\documentclass[11pt,a4paper]{article}
\usepackage{jcappub}
\usepackage{graphicx,epsfig}

\title{Primordial bispectrum from inflation with background gauge fields}

\author[a]{Hiroyuki Funakoshi}

\author[a,b]{and Kei Yamamoto}

\affiliation[a]{DAMTP, University of Cambridge, \\
Wilberforce Road, Cambridge, CB3 9AL United Kingdom}

\affiliation[b]{Institute of Theoretical Astrophysics, University of Oslo, \\
P.O. Box 1029, Blindern, N-0315 Oslo, Norway}

\emailAdd{K.Yamamoto@damtp.cam.ac.uk}
\emailAdd{H.Funakoshi@damtp.cam.ac.uk}

\abstract{We study the primordial bispectrum of curvature perturbation in the uniform-density 
slicing generated by the interaction between the inflaton and isotropic background gauge fields. 
We derive the action up to cubic order in perturbation and take into account all the relevant 
effects in the leading order of slow-roll expansion. We first treat the quadratic vertices
perturbatively and confirm the results of past studies, while identifying their regime of validity.
We then extend the analysis to include the effect of the quadratic vertices 
to all orders
by introducing exact linear mode functions, allowing us to make 
accurate predictions long
after horizon crossing where the features of both the power spectrum and the bispectrum are
drastically different. It is shown that the spectra become constant and scale-invariant in
the limit of large e-folding. As a result, we are able to impose reliable constraints on the
parameters of our theory using the recent observational data coming from Planck.} 

\keywords{Non-Gaussianity, In-in formalism}

\begin{document}

\maketitle

\section{Introduction}
The prediction on the primordial density fluctuation from inflation offers an exciting opportunity
to test the physics at high energy that is inaccessible for ground-based experiments. The advent 
of the Planck satellite, which is expected to improve the constraint on the three-point and higher 
correlations of the density perturbation at recombination by a factor of $10 $ to $100$, has prompted 
detailed theoretical investigations into the interaction of the inflaton \cite{Bartolo2004}. So far, the efforts 
have been focused on the scalar self-interactions and interactions among multiple scalar fields. 
It is found that single-scalar models with a canonical kinetic term generically predict an undetectable level of
non-Gaussian signals \cite{Maldacena2002} while a scalar with the DBI action or multi-scalar dynamics,
such as hybrid inflation or curvaton scenarios, can lead to significant higher-order 
correlations \cite{Seery2005, Langlois2008, Wands2010}. These models being based on the string theories in
their origin of the inflaton, a detection of significant bispectrum or trispectrum can give us 
a clue for understanding the high-energy physics.

In the context of unified theories of fundamental interactions, however, scalar fields cannot be 
the only ingredients of the universe. Most of the proposed theories such as superstring theories
and M-theory rely on gauge symmetries, and gauge fields are indispensable
to mediate interactions among the fields and in some cases to preserve supersymmetries.
Even when they are absent in the fundamental Lagrangians, it is a generic prediction of
dimensional reduction that a typical scalar field is coupled to some gauge fields \cite{Hull1995, Gibbons2010}.
Earlier attempts to drive inflation with vector fields \cite{Ford1989, Koivisto2008, Golovnev2008, Bamba2008} turned out to 
be largely unsuccessful since one needs to abandon gauge symmetries, which results in the 
introduction of additional degrees of freedom and various 
instabilities \cite{Himmetoglu2008, Koivisto2009, Himmetoglu2009, Golovnev2009, Esposito-Farese2010}. More recently, interactions between 
the inflaton and gauge fields, motivated by those unified theories of interactions, have been 
taken into account in the context of 
preheating \cite{Yokoyama2008, Bartolo2009, Bartolo2009a, Kanno2009, Dimopoulos2009, Dimopoulos2010}. 
In addition to interesting phenomenologies including non-Gaussianity, primordial magnetic fields
and gravitational waves \cite{Valenzuela-Toledo2009, Valenzuela-Toledo2010, Karciauskas2011, Shiraishi2011, Koivisto2012, Barnaby2012, Anber2012, Namba2012, Urban2012, Jain2012}, it was realized that the back reaction of gauge fields on 
the inflaton can effectively act as an extra friction term so that they slow down the rolling of 
the scalar field and help causing an accelerated expansion \cite{Anber2010, Dimopoulos2010a, Dimastrogiovanni2010, Barnaby2010, Barnaby2011, Barnaby2011a, Dimopoulos2011}. In fact, this back reaction can be
so strong that it may generate a significant vacuum expectation value of the gauge fields and
violate the isotropy of the universe. On the other hand, there has been a growing interest 
to maintain a small, but non-vanishing amplitude of classical gauge 
fields during inflation in order to explain the reported statistical anisotropy of cosmic microwave
background radiation (CMBR) in WMAP 7-year \cite{Jaffe2005, Jaffe2005a, Jaffe2006}. 
By taking into account the aforementioned
back reaction classically, the same types of scalar-gauge interactions 
arising from the high-energy particle theories have been found to enable acceleration of
the cosmic expansion without requiring a sufficiently flat potential for the inflaton \cite{Watanabe2009, Kanno2010}. 
This scenario turned out to be free of any classical instabilities or 
fine-tuning \cite{Moniz2010,  Emami2010, Murata2011, Hervik2011, Do2011, Do2011a}. 
There have also been extensive
studies on its potential imprints on CMBR and it was revealed that even a very small amplitude
of background energy density of the gauge fields could result in a significant statistical
anisotropy in the curvature fluctuation \cite{Dulaney2010, Gumrukcuoglu2010, Watanabe2010, Watanabe2011, Bartolo2012}. While it implies such an anisotropic vacuum expectation
value of gauge fields must be severely constrained, a recent study suggests that their effect
on primordial bispectum is as drastic as its linear counterpart and the resulting non-Gaussianity
may still be observable. In another recent development, it has been shown that multiple vector
degrees of freedom generically suppress the residual anisotropy of the background
space-time through a dynamical attractor mechanism. In particular, when three or more
gauge fields are coupled to a scalar field via a common gauge-kinetic function, the
final state of the universe is completely isotropic regardless of initial conditions \cite{Yamamoto2012}. Such
a circumstance may naturally be realized by non-Abelian gauge fields since the equal
coupling is guaranteed by the symmetry \cite{Maeda2012}. There are other instances of isotropic inflation
involving non-Abelian gauge fields which also exhibit similar attractor behaviours \cite{Maleknejad2011, Maleknejad2012, Adshead2012, Adshead2012a, Sheikh-Jabbari2012, Martinec2012}. The
linear perturbation of this isotropic inflation with background gauge fields has been
studied, which has found that the primordial power spectrum is not strongly constrained
by the current observations since the spectrum is perfectly isotropic and almost
scale-invariant \cite{Yamamoto2012a}.

In this paper, we investigate the second-order perturbation of an isotropic
universe containing three $U(1)$ gauge fields and a scalar inflaton. We compute the
bispectrum of curvature perturbation by deploying the in-in formalism and compare the
results with the corresponding work in an anisotropic background \cite{Bartolo2012}, which is expected
to be qualitatively similar. There are theoretical, phenomenological, and technical reasons 
for this particular model to be studied:
\begin{enumerate}
\item The isotropy maintained by a triad configuration of gauge fields appears to be 
a generic feature of multiple vector degrees of freedom according to \cite{Yamamoto2012}. This model 
serves as a prototype of the more complicated instances with non-Abelian gauge fields, 
for which similar features are expected.
\item Because of its isotropy, the model cannot be effectively constrained by the power
spectrum. As we anticipate a strong signal in the bispectrum given the analogy to
the anisotropic models, it is important to quantify it from a phenomenological point
of view.
\item While perturbation around anisotropic backgrounds is extremely involved,
one can make a transparent perturbative expansion in the present isotropic model and 
identify all the relevant contributions. 
\end{enumerate}
Besides, it is worth emphasizing that the interactions under discussion frequently appear
in supergravity theories that are low-energy effective theories of superstring theories and 
M-theory. It is therefore of great interest to study observational consequences of these 
models as they are beyond the reach of any ground-based experiments. 

Our results reproduce the previous studies when the e-folding number is relatively small
while extending the analysis so that it is applicable to the period long after the horizon exit. 
A full perturbative expansion
of the Lagrangian up to cubic order is carried out and several interaction terms that are 
not suppressed by any of the slow-roll parameters are identified. We first treat
all the interaction terms, both quadratic and cubic, as perturbation and compute the 
three-point function for the curvature $\zeta $. The amplitude is solely controlled 
by the parameter $\mathcal{I}^2 $ that represents the ratio of background energy density 
of the gauge fields to the scalar kinetic energy density. We explicitly show that 
the leading contribution comes from the vertex involving a scalar field and two gauge fields, 
which confirms the claim of \cite{Bartolo2012}. The three-point function scales as 
$\propto \mathcal{I}^2 N_k^3$ where $N_k$ is the e-folding number after horizon exit 
for a mode with wavenumber $k$. The shape is local as has been shown in the previous studies. 
This results in a large $f_{NL}$ when the modification to the power spectrum is assumed 
to be small. However, we find that this conclusion is valid only if $\mathcal{I}^2 \ll N_k^{-2} $, 
which is not satisfactory for this isotropic model since $\mathcal{I}^2 $ is not necessarily 
that small in contrast to the anisotropic cases where this is required to keep the background 
anisotropy within the range allowed by the observations. A reason for the limited
applicability is the quadratic vertices that generate an infinite number of Feynman
diagrams in the perturbative expansion even at tree level.
In the second half of this paper, we take into account this fact by introducing exact
linear mode functions. It turns out that one can solve the linear evolution equations 
analytically at superhorizon scales. By exploiting their general features, we shall prove
that both power spectrum and bispectrum are convergent in the limit $N_k \rightarrow \infty $, 
determining the late-time value of $f_{NL}$.
In order
to obtain more quantitative estimates and handle the intermediate regime, we also 
solve the linear equations numerically from deep inside the horizon and use them
in the integrand of the three-point correlators. We confirm the initial logarithmic
behaviours in both power spectrum and bispectrum and their convergence 
at late times. It turns out that the time evolution of $f_{NL}$ (squeezed) displays some
interesting features. It first peaks at $N_k \sim 0.3 \mathcal{I}^{-1}$
where the peak value scales as $\mathcal{I}^{-1}$; thus for certain small values of $\mathcal{I}$, 
the latest Planck data appear to rule out the possibily of the observable
modes in the CMBR arising from this intermediate phase. 
Then, $f_{NL}$ monotonically
decreases and converges to a negative $\mathcal{I}$-independent constant, 
$-5/3$.

The paper is organized as follows. In the next section, after sketching the dynamics
of the background evolution and introducing relevant parameters, we derive the
perturbed Lagrangian up to cubic order. Section 3 gives the detailed procedure
of computing the three-point function by perturbative expansion with respect to free
de-Sitter mode functions. We calculate all the relevant contributions and identify
the leading order term. Section 4 discusses the importance of the deviation from
the de-Sitter mode functions on superhorizon scales. 
In the end we provide estimates for the late-time values for the power spectrum and bispectrum. 
In section 5, we numerically confirm these analytical results and make the prediction on 
non-Gaussianity more quantitative. Concluding remarks are given in section 6.

\section{Perturbative expansion up to cubic order}

Our model contains a scalar field and several gauge fields minimally coupled to gravity:
\begin{equation*}
S = \int d^4 x \sqrt{-^4 \! g}\left( \frac{1}{16\pi G} R - \frac{1}{2}\partial _{\mu }\varphi \partial ^{\mu }\varphi  
-V(\varphi ) -\frac{f(\varphi )^2 }{4}F^a_{\ \mu \nu }F^{a \mu \nu } \right).
\end{equation*}
$R$ is the Ricci scalar curvature and $F^a_{\mu \nu } = \partial _{\mu }A^a_{\nu } -\partial _{\nu }
A^a_{\mu } ; a = 1,2,3$ are three copies of $U(1)$ gauge field strengths. These types of
actions have been well studied in the context of magnetogenesis, preheating and
anisotropic inflation. It has been realised that the coupling between the scalar field,
which is identified to be the inflaton, and gauge fields enables an accelerated phase
of expansion even with a relatively steep potential, as we will see later. We note that
the energy density of the gauge fields stays constant in the first approximation in the
inflating universe, violating the cosmic-no-hair conjecture. It has also been
shown that the isotropic configuration of the gauge fields is a dynamical attractor of
the system. Based on this result, we study the perturbation of this theory around the
isotropic background with a non-vanishing triad of the gauge fields.

We set $8\pi G =1$ and follow the ADM formalism \cite{Arnowitt2004} and parametrize the metric as
\begin{equation*}
^4 \! g_{\mu \nu } =  \left(   \begin{array}{cc}
    -N^2 + N_k N^k  & N_j \\ 
    N_i & g_{ij} \\ 
  \end{array}\right)
\end{equation*}
where 
\begin{equation*}
N^i = g^{ij}N_j  , \ \ \ \ \ g^{ik}g_{kj} = \delta ^i_{\ j} .
\end{equation*}
The normalized extrinsic curvature of the constant time slice is given by
\begin{equation*}
E_{ij} = -\frac{1}{2}\left( \dot{g}_{ij}-2N_{(i|j)} \right)
\end{equation*}
and its intrinsic scalar curvature is
\begin{equation*}
^3 \! R = \left( g_{ij,kl}+g_{mn}\Gamma ^m_{\ ij} \Gamma ^n_{\ kl} \right) \left( g^{ik}g^{jl}-g^{ij}g^{kl} \right) .
\end{equation*}
Electric fields are defined to be
\begin{equation*}
E^a_{i} = F^a_{\ 0i}   .
\end{equation*}

\subsection{Gravity and the scalar field}
The Einstein-Hilbert action in ADM formalism is given by
\begin{equation*}
\mathcal{L}_g = \frac{\sqrt{g}}{2N}\left( E_{ij}E^{ij}-E^2  \right) + \frac{N\sqrt{g}}{2} \ ^3 \! R .
\end{equation*}
We assume that the background is a flat Friedmann-Lema\^itre-Robertson-Walker space-time 
\begin{equation*}
ds^2 = a(\eta )^2 \left( -d\eta ^2 +  \delta _{ij} dx^i dx^j \right)
\end{equation*}
and write the perturbed metric components as
\begin{equation*}
N = a(1+\phi ), \ \ \ \ \ N_i = a^2 \beta _i, \ \ \ \ \ g_{ij} = a^2 (\delta _{ij} + 2\gamma _{ij} ) .
\end{equation*}
When the problem concerns perturbation beyond linear order, one has to be careful in choosing 
the small quantities with respect to which the order of perturbation is determined. In the present case,
we will solve the constraint equations so that $\phi $ and $\beta _i$ are expressed in terms of 
$\gamma _{ij}$ and the other matter variables. Thus, their order in perturbative expansion is 
subject to the equations to be solved and we should distinguish different orders as
\begin{eqnarray*}
\phi &=& \phi _{(1)}  + \frac{1}{2}\phi _{(2)} + \cdots , \\
\beta _i &=& \beta ^{(1)}_i + \frac{1}{2}\beta ^{(2)}_i + \cdots .
\end{eqnarray*}
On the other hand, we avoid a similar expansion for $\gamma _{ij}$ since it will hardly appear 
in the following analysis as we are primarily working in the flat gauge, where the perturbation is 
set to be zero at each order by the choice of gauge. The only exception is the curvature perturbation 
on the uniform-density slice that will be expanded in terms of the dynamical variables in the flat gauge . 
As usual, the scalar-vector-tensor decomposition is made in order to decouple the 
linear-order equations. It is defined by
\begin{eqnarray*}
&& \gamma _{ij} = -\psi \delta _{ij} + E_{,ij} + F_{(i,j)} + \frac{1}{2}h_{ij} , \ \ \ \ \ \beta _i^{(n)} = B^{(n)}_{,i} -S^{(n)}_i ,\\
&& S^{(n)}_{i,i} = F_{i,i} = 0 , \ \ \ \ \ h_{ii} = 0 , \ \ \ \ \ h_{ij,j} =0 .
\end{eqnarray*}
In the uniform-density gauge, we denote the curvature perturbation $- \zeta = \psi $ and expand it as
\begin{equation*}
\zeta = \zeta _{(1)} + \frac{1}{2}\zeta _{(2)} + \cdots .
\end{equation*}

The $1+3$ decomposition of the action for the scalar field is given by
\begin{equation*}
\mathcal{L}_{\varphi } = \frac{\sqrt{g}}{2N} \left[ \varphi ^{\prime 2} - 2 \varphi ^{\prime }\varphi _{,i}N^i + \left( \varphi _{,i}N^i \right) ^2 \right] -N\sqrt{g}\left[ \frac{1}{2}g^{ij}\varphi _{,i} \varphi _{,j} +V(\varphi ) \right] ,
\end{equation*}
where primes denote derivatives with respect to the conformal time $\eta $.
We split $\varphi $ into the background and the perturbation:
\begin{equation*}
\varphi = \bar{\varphi } + \pi .
\end{equation*}
$\pi $ will be treated as the dynamical variable in terms of which the perturbative expansion is
defined so that we do not need to expand it further. 

\subsection{Gauge field perturbations}
The Maxwell Lagrangian in the ADM formalism reads 
\begin{equation*}
\mathcal{L}_{M} = \frac{\sqrt{g}}{2N}f^2 g^{ij}\left( E^a_{\ i}+F^a_{\ ik}N^k \right) \left( E^a_{\ j}+F^a_{\ jl}N^l \right) -\frac{N\sqrt{g}}{4}f^2 g^{ik}g^{jl}F^a_{\ ij}F^a_{\ kl} .
\end{equation*}
The perturbative expansion of the vector potentials yields
\begin{equation*}
A^a_0 = \sigma ^a , \ \ \ \ \ A^a_i = A(\eta ) \delta ^a_{\ i} + \chi ^a_{\ i} ,
\end{equation*}
where the background quantity $A(\eta )$ behaves effectively as a second scalar field. The background
values of $A^a_0 $ are taken to be zero by a gauge choice. As in the gravity sector, we should
in principle distinguish the different orders of perturbation for the variables that are expanded in terms of 
the dynamical ones. However, after the adoption of flat slicing and $U(1)$ gauge fixing, we are
left only with the dynamical variables from this sector. Hence we suppress this distinction and 
the scalar-vector-tensor decomposition is carried out as follows:
\begin{eqnarray*}
&& \sigma ^{a} = \mu _{,a} + \nu _{a} , \ \ \ \ \ \chi ^a_{\ i} = \alpha \delta _{ai} + \theta _{,ai} + \epsilon _{aij} \left( \tau _{,j} + \lambda _j \right) + \kappa _{(a,i)} + \omega _{ai} , \\
&& \nu _{i,i} = \lambda _{i,i} = \kappa _{i,i} = 0 , \ \ \ \ \  \omega _{ii} = 0 , \ \ \ \ \ \omega _{ij,j} =0.
\end{eqnarray*}

\subsection{Background dynamics and parameters}
Before going into the perturbative analysis, we briefly review the background evolution of the system and identify the relevant parameters. The Maxwell's equation can be trivially integrated to give
\begin{equation*}
A^{\prime } = \frac{c}{f^2}
\end{equation*}
where $c$ is an integration constant. As usual, we introduce the "slow-roll" parameters
\begin{equation}
\epsilon _H = 1-\frac{\mathcal{H}^{\prime }}{\mathcal{H}^2}, \ \ \ \ \ \eta _H = \frac{\epsilon _H^{\prime }}{\mathcal{H}\epsilon _H}, \ \ \ \ \ \left( \mathcal{H} = \frac{a^{\prime }}{a} \right) , \label{eq:epsilons}
\end{equation}
which characterize the evolution of the scale factor $a(\eta )$.
The Raychaudhuri equation
\begin{equation}
2 \mathcal{H}^{\prime }+\mathcal{H}^2 = -\frac{1}{2}\bar{\varphi }^{\prime 2} - \frac{c^2}{2a^2 f^2 } + a^2 V 
\end{equation}
tells that the potential energy has to dominate over the scalar kinetic energy and the energy of gauge fields in order to have an accelerated expansion. This suggests the introduction of another parameter
\begin{equation}
\epsilon _{\varphi } = \frac{\bar{\varphi }^{\prime 2}}{2\mathcal{H}^2} 
\end{equation}
which controls the evolution of the inflaton. Combined with the Friedmann equation
\begin{equation}
3\mathcal{H}^2 = \frac{1}{2}\bar{\varphi }^{\prime 2} + a^2 V + \frac{3c^2}{2a^2 f^2} ,
\end{equation}
one derives
\begin{equation*}
\frac{c^2}{a^2 f^2} = (\epsilon _H -\epsilon _{\varphi })\mathcal{H}^2 ,
\end{equation*}
which is the representative of the energy density for the gauge fields.
Note that $\epsilon _{\varphi } \leq \epsilon _H$ where equality holds when the gauge fields vanish. Since this deviation from the single-scalar inflation plays a central role, we define the parameter
\begin{equation}
\mathcal{I} = \sqrt{\frac{\epsilon _H - \epsilon _{\varphi }}{\epsilon _{\varphi }}}\,,  \label{eq:scri}
\end{equation}
which measures the ratio between the energy density of the gauge fields and the scalar kinetic energy. 
We note that $\mathcal{I}$ does not have to be small as far as the background dynamics and the
power spectrum are concerned. Without loss of generality, 
we can assume $\bar{\varphi }^{\prime } >0$ and use $\bar{\varphi }^{\prime } = \sqrt{2\epsilon _{\varphi }}\mathcal{H}$.
Now the equation of motion for $\bar{\varphi }$ gives
\begin{equation}
\bar{\varphi }^{\prime \prime } = \sqrt{\frac{\epsilon _{\varphi }}{2}}\left( 2-2\epsilon _H + \eta _{\varphi } \right) \mathcal{H}^2 
\end{equation}
where 
\begin{equation}
\eta _{\varphi } = \frac{\epsilon _{\varphi }^{\prime }}{\mathcal{H}\epsilon _{\varphi }} .
\end{equation}
In principle, this quantity does not have to be small as long as $\eta _H \ll 1$, but we do assume that it is in order to control the perturbative expansion. Now by differentiating 
\begin{equation*}
\frac{c^2}{f^2} = \left( \epsilon _H -\epsilon _{\varphi } \right) \mathcal{H}^2 a^2 ,
\end{equation*}
one obtains
\begin{equation}
\frac{(f^2)_{,\varphi }}{f^2} = -\sqrt{\frac{2}{\epsilon _{\varphi }}}\left( 2-\epsilon _H +\frac{\epsilon _H \eta _H - \epsilon _{\varphi } \eta _{\varphi }}{2(\epsilon _H -\epsilon _{\varphi }) }\right) ,
\end{equation}
and using the equation of motion for the scalar field yields
\begin{equation}
\frac{a^2 V_{,\varphi }}{\mathcal{H}^2} = -\frac{1}{\sqrt{2\epsilon _{\varphi }}}\left( 6\epsilon _H -3\epsilon _H^2 + \epsilon _H \epsilon _{\varphi } + \frac{3}{2}\epsilon _H \eta _H - \frac{1}{2}\epsilon _{\varphi } \eta _{\varphi } \right) .
\end{equation}
The first expression tells that the slope of $f(\varphi )$ must be steep in order to maintain the amplitude of gauge fields during inflation. The second implies that the gradient of potential is not necessarily small if $\epsilon _{\varphi } \ll \epsilon _H$, or equivalently, if $\mathcal{I} \gg 1$. The reason is that the slow roll of the inflaton can be achieved by transferring the scalar kinetic energy to the gauge fields through the coupling
$f(\varphi )$. It later turns out that the perturbative approach breaks down when $\mathcal{I}>1$ anyway, so we assume that $\mathcal{I} <1$, where the usual intuition from single-scalar model works well. The higher order derivatives of $V$ and $f$ take complicated forms in general, but assuming the constancy of $\eta _{H,\varphi }$ and keeping only the leading-order terms in the small parameters, we obtain
\begin{equation}
\frac{a^2 V_{,\varphi \varphi }}{\mathcal{H}^2} \sim \frac{3\epsilon _H}{2\epsilon _{\varphi }}\left( 4\epsilon _H -2\eta _H +\eta _{\varphi } \right) , \ \ \ \ \ 
\end{equation}
\begin{equation}
\frac{a^2 V_{,\varphi \varphi \varphi }}{\mathcal{H}^2} \sim \frac{3}{2\sqrt{2\epsilon _{\varphi }}} \frac{\epsilon _H}{\epsilon _{\varphi }} \left( 8\epsilon _H \eta _H - 4\epsilon _H \eta _{\varphi } -2\eta _H^2 + 3\eta _H \eta _{\varphi }-\eta _{\varphi }^2 \right) ,
\end{equation}
and
\begin{equation}
\frac{(f^2)_{,\varphi \varphi }}{f^2} \sim \frac{8}{\epsilon _{\varphi }}, \ \ \ \ \ \frac{(f^2)_{,\varphi \varphi \varphi }}{f^2} \sim -\left( \frac{8}{\epsilon _{\varphi }} \right) ^{\frac{3}{2}} . \label{eq:fgrads}
\end{equation}
Finally, we emphasize that this regime of accelerated expansion aided by gauge fields is a dynamical 
attractor for a wide range of potential and coupling. The readers are referred to ref. \cite{Yamamoto2012}.

\subsection{The cubic action for scalar perturbations}
Since the curvature perturbation does not receive any contribution from vector or tensor modes at
the linear order, we can eliminate them from the tree-level calculations of three-point correlation functions
arising from cubic interactions. Since their power spectra are known to be small, the higher order 
contribution is expected to be negligible. Hence, we focus on scalar perturbations hereafter. 

For the scalar perturbation, the gauge field variables are given by
\begin{eqnarray*}
\sigma ^a_0 &=& \mu _{,a} , \\
\chi ^a_{\ i} &=& \alpha \delta _{ai}+\theta _{,ai}+\epsilon _{aij}\tau _{,j} .
\end{eqnarray*} 
We use the $U(1)$ gauge freedom to set $\theta ^{\prime } +\mu = 0$. It follows that
\begin{eqnarray*}
\delta F^a_{ij} &=&  \alpha _{,j}\delta _{ia} - \alpha _{,i} \delta _{ja}+\epsilon _{aik}\tau _{,kj} -\epsilon _{ajk}\tau _{,ki}  , \\
\delta E^a_i &=& \alpha ^{\prime } + \epsilon _{aij}\tau ^{\prime }_{,j} ,
\end{eqnarray*}
where $\delta $ indicates the perturbation of the following variables.
For the metric, we adopt the flat slicing $\psi = E =0$. Focusing on scalar modes, we can completely ignore $\gamma _{ij}$. The gravitational action drastically simplifies up to cubic order to become
\begin{equation*}
a^{-2}\mathcal{L}_g = (1-\phi _{(1)}) \left[ -3\mathcal{H}^2 \phi _{(1)}^2 + \frac{1}{2}\left( B^{(1)}_{,ij}B^{(1)}_{.ij} -B^{(1)}_{,ii}B^{(1)}_{,jj} \right) -2\mathcal{H}\phi _{(1)}B^{(1)}_{,ii} \right] .
\end{equation*}
The scalar part is the same as the standard:
\begin{eqnarray*}
\mathcal{L}_{\varphi } &=& (1-\phi _{(1)})\left( \frac{1}{2}\bar{\varphi }^{\prime 2}\phi _{(1)}^2 - \phi _{(1)}\left( \bar{\varphi }^{\prime }\pi ^{\prime }+a^2 V_{,\varphi }\pi \right) - \bar{\varphi }^{\prime }\pi _{,i}B^{(1)}_{,i}+\frac{1}{2}\pi ^{\prime 2}-\frac{1}{2}\pi _{,i}\pi _{,i}-\frac{1}{2}a^2 V_{,\varphi \varphi }\pi ^2 \right) \\
&& -a2 V_{,\varphi }\phi _{(1)}^2 \pi - \pi ^{\prime }\pi _{,i}B^{(1)}_{,i}-\frac{1}{6}a^2 V_{,\varphi \varphi \varphi }\pi ^3 .
\end{eqnarray*}
After some straightforward algebra, one obtains the gauge Lagrangian as
\begin{eqnarray*}
\mathcal{L}_{M} &=& (1-\phi _{(1)}) \mathcal{L}^{(2)}_{MS} -2f^2 \alpha _{,i}B^{(1)}_{,i}\left( \alpha ^{\prime } + \frac{c(f^2)_{,\varphi }}{f^4}\pi \right)  + f^2 B^{(1)}_{.i} \left( \epsilon _{ijk}\alpha _{,j}\tau ^{\prime }_{,k} - \tau _{,ij}\tau ^{\prime }_j - \tau ^{\prime }_i \tau _{,jj} \right) \\
&& + \frac{c^2(f^2)_{,\varphi \varphi \varphi }}{4f^4} \pi ^3 +\frac{3c(f^2)_{,\varphi \varphi }}{2f^2} \pi ^2 \alpha ^{\prime } + (f^2)_{,\varphi }\pi \left( \frac{3}{2} \alpha ^{\prime 2} -\alpha _{,i}\alpha _{,i} \right) \\
&& +(f^2)_{,\varphi }\pi \left( \tau ^{\prime }_{,k} \tau ^{\prime }_{,k} -\frac{1}{2}\tau _{,ij}\tau _{,ij}-\frac{1}{2}\tau _{,ii}\tau _{,jj} \right) ,
\end{eqnarray*}
where
\begin{eqnarray*}
\mathcal{L}^{(2)}_{MS} &=& \frac{3c^2}{2f^2}\phi _{(1)}^2 -3c\phi \left( \alpha ^{\prime } + \frac{c(f^2)_{,\varphi }}{2f^4} \pi \right) -2c\alpha _{,i}B^{(1)}_{,i} + \frac{3c^2 (f^2)_{,\varphi \varphi }}{4f^4}\pi ^2 + \frac{3c(f^2)_{,\varphi }}{f^2}\alpha ^{\prime }\pi \\
&& + \frac{f^2}{2} \left( 3\alpha ^{\prime 2} -2\alpha _{,i} \alpha _{,i}+ 2\tau ^{\prime }_{,i}\tau ^{\prime }_{,i} -\tau _{,ij}\tau _{,ij}-\tau _{,ii}\tau _{,jj} \right) .
\end{eqnarray*}
Therefore, the total Lagrangian up to cubic order is written as
\begin{eqnarray}
\mathcal{L} &=& (1-\phi _{(1)}) \mathcal{L}^{(2)} - a^4 V_{,\varphi }\phi _{(1)}^2 \pi - a^2 \pi ^{\prime }\pi _{,i}B^{(1)}_{,i} -2f^2 \left( \alpha ^{\prime } +\frac{c(f^2)_{,\varphi }}{f^4}\pi \right) \alpha _{,i}B^{(1)}_{,i} \nonumber \\
&& +\left( \frac{c^2 (f^2)_{,\varphi \varphi \varphi }}{4f^4} -\frac{1}{6}a^4 V_{,\varphi \varphi \varphi } \right) \pi ^3 + \frac{3c(f^2)_{,\varphi \varphi }}{2f^2} \pi ^2 \alpha ^{\prime } + (f^2)_{,\varphi }\pi \left( \frac{3}{2}\alpha ^{\prime 2}-\alpha _{,i}\alpha _{,i} \right)  \label{eq:cubicaction} \\
&& + f^2 \left( \epsilon _{ijk}\alpha _{,j}\tau ^{\prime }_{,k} - \tau _{,ij}\tau ^{\prime }_{,j}-\tau ^{\prime }_{,i}\tau _{,jj} \right) B^{(1)}_{,i} + (f^2)_{,\varphi }\pi \left( \tau ^{\prime }_{,k}\tau ^{\prime }_{,k}-\frac{1}{2}\tau _{,ij}\tau _{,ij}-\frac{1}{2}\tau _{,ii}\tau _{,jj} \right) , \nonumber 
\end{eqnarray}
where the quadratic Lagrangian is given by
\begin{eqnarray}
a^{-2}\mathcal{L}^{(2)} &=& \left( -3\mathcal{H}^2 +\frac{1}{2}\bar{\varphi }^{\prime 2}+\frac{3c^2}{2a^2 f^2} \right) \phi _{(1)}^2 + \frac{1}{2}\left( B^{(1)}_{,ij}B^{(1)}_{ij}-B^{(1)}_{,ii}B^{(1)}_{,jj} \right) \nonumber \\
&& - \phi _{(1)}\left( 2\mathcal{H}B^{(1)}_{,ii} + \bar{\varphi }^{\prime }\pi ^{\prime } + \left( a^2 V_{,\varphi }+ \frac{3c^2 (f^2)_{,\varphi }}{2a^2 f^4} \right) \pi + \frac{3c}{a^2}\alpha ^{\prime } \right) \nonumber \\
&& -\bar{\varphi }^{\prime }\pi _{,i}B^{(1)}_{,i} -\frac{2c}{a^2}\alpha _{,i}B^{(1)}_{,i} + \frac{1}{2}\pi ^{\prime 2}-\frac{1}{2}\pi _{,i}\pi _{,i} -\frac{1}{2}a^2 V_{,\varphi \varphi }\pi ^2 \\
&& + \frac{3c^2 (f^2)_{,\varphi \varphi }}{4a^2 f^4}\pi ^2 + \frac{3c(f^2)_{,\varphi }}{a^2 f^2}\alpha ^{\prime }\pi + \frac{f^2}{2a^2}\left( 3\alpha ^{\prime 2}-2\alpha _{,i}\alpha _{,i} \right) \nonumber \\
&& +\frac{f^2}{2a^2}\left( 2\tau ^{\prime }_{,i}\tau ^{\prime }_{,i}-\tau _{,ij}\tau _{,ij}-\tau _{,ii}\tau _{,jj} \right) . \nonumber 
\end{eqnarray}
It should be mentioned that we dropped the terms involving $\phi _{(2)}$ and $B^{(2)}$ from 
the beginning since they multiply the background and linear-order constraint equations, 
which would be automatically satisfied in our formulation.

\subsection{Solving the linear constraints}
Using the background equations and parameters, the quadratic Lagrangian can be rewritten as
\begin{eqnarray*}
a^{-2} \mathcal{L}^{(2)} &=& -a^2 V \phi _{(1)}^2 - \mathcal{H} \phi _{(1)}\left( 2\nabla ^2 B^{(1)} + \sqrt{2\epsilon _{\varphi }} \pi ^{\prime } + \frac{3f}{a}\sqrt{\epsilon _{\varphi }} \mathcal{I} \alpha ^{\prime } \right)  + q_{\phi } \mathcal{H}^2 \phi _{(1)}\pi \\
&& + \mathcal{H} B^{(1)} \left( \sqrt{ 2\epsilon _{\varphi }}\nabla ^2 \pi + \frac{2f}{a}\sqrt{\epsilon _{\varphi }} \mathcal{I}\nabla ^2 \alpha \right) + \frac{1}{2}\pi ^{\prime 2} -\frac{1}{2}\pi _{,i} \pi _{,i} \\
&& + \frac{f^2}{2a^2}\left( 3\alpha ^{\prime 2} -2\alpha _{,i}\alpha _{,i} + 2\tau ^{\prime }_{,i}\tau ^{\prime }_{,i} -\tau _{,ij} \tau _{,ij} -\tau _{,ii}\tau _{,jj} \right)   \\
&& -\frac{1}{2}\left( a^2 V_{,\varphi \varphi } -\frac{3}{2}\left( \epsilon _H - \epsilon _{\varphi } \right) \frac{(f^2)_{,\varphi }}{f^2 }\mathcal{H}^2 \right) \pi ^2 \\
&&  -3\sqrt{2}\frac{f}{a}\mathcal{I} \left( 2 -\epsilon _H +\frac{\epsilon _H \eta _H -\epsilon _{\varphi }\eta _{\varphi }}{2(\epsilon _H -\epsilon _{\varphi })} \right) \mathcal{H} \alpha ^{\prime } \pi 
\end{eqnarray*}
where we discarded the surface term and defined
\begin{equation*}
q_{\phi } = \frac{1}{\sqrt{2\epsilon _{\varphi }}} \left( 6\epsilon _{\varphi }(1+2\mathcal{I}^2 ) -6\epsilon _H^2 + 4\epsilon _H \epsilon _{\varphi }+3\epsilon _H \epsilon _{\varphi }-2\epsilon _{\varphi }\eta _{\varphi } \right) .
\end{equation*}
Varying $B^{(1)} $ determines $\phi _{(1)}$ as
\begin{equation}
\phi _{(1)}=\sqrt{ \frac{\epsilon _{\varphi }}{2}} \left( \pi + \sqrt{2}\frac{f}{a}\mathcal{I} \alpha \right) . \label{eq:phi} 
\end{equation}
Using this, variation of $\phi _{(1)}$ leads to
\begin{eqnarray}
\sqrt{\frac{2}{\epsilon _{\varphi }}}\nabla ^2 B^{(1)} &=& - \pi ^{\prime } + \left( 6 \mathcal{I}^2 +\frac{\epsilon _{\varphi }}{2}\mathcal{I}^2 \left( 5+6\mathcal{I}^2 \right) +\frac{3}{2}\eta _H (1+\mathcal{I}^2 ) -\eta _{\varphi } \right) \mathcal{H} \pi \label{eq:B} \\
&& -\frac{f}{\sqrt{2}a}\mathcal{I} \left( 3 \alpha ^{\prime } + (6-3\epsilon _H +\epsilon _{\varphi } )\mathcal{H}  \alpha \right) . \nonumber 
\end{eqnarray}
These relations will be substituted into the Lagrangian derived in the previous subsection and 
the curvature perturbation in the uniform-density gauge introduced in the following.

\subsection{Curvature of the uniform-density surface}
For the purpose of quantum field theory calculations in the multi-field dynamics, the most convenient 
gauge is the flat gauge where $\gamma _{ij} = h_{ij}$ \cite{Seery2005}. However, the observationally 
relevant quantity is the curvature perturbation in the comoving gauge $\mathcal{R}_c$ that coincides
with the curvature in the uniform-density gauge $\zeta $ beyond the horizon scale. The latter is more
often picked up as done here since it possesses a desirable mathematical property. Hence, 
we need the transformation law between flat gauge and uniform-density gauge, which we cite from
\cite{Malik2008} as
\begin{equation}
-\zeta _{(1)} = \mathcal{H}\frac{\delta \rho _{(1)} }{\bar{\rho }^{\prime }} \label{eq:curvature1}
\end{equation}
and 
\begin{equation}
-\zeta _{(2)} = \frac{\mathcal{H}}{\bar{\rho }^{\prime }} \left( \delta \rho _{(2)} - \frac{\delta \rho _{(1)}^{\prime }}{\bar{\rho }^{\prime }}\delta \rho _{(1)} \right) - \frac{1}{4}\Xi _{kk}+ \frac{1}{4}\nabla ^{-2}\Xi _{ij ,ij} \label{eq:curvature2}
\end{equation}
where the right-hand sides are evaluated in the flat gauge. We defined the perturbative expansion 
of the energy density 
\begin{equation*}
\rho = \bar{\rho } + \delta \rho _{(1)} + \frac{1}{2}\delta \rho _{(2)} + \cdots 
\end{equation*}
and a quadratic expression
\begin{eqnarray*}
\Xi _{ij  } &= & =-\frac{2\mathcal{H}}{\bar{\rho }^{\prime }}\left( \mathcal{H}(1+3c_s^2 ) \left( \frac{\delta \rho_{(1)}^2}{\bar{\rho }^{\prime }} \right) - \frac{\delta \rho _{(1)} ^{\prime }}{\bar{\rho }^{\prime }}\delta \rho _{(1)} \right) \delta _{ij} \\
&& -\frac{2}{\bar{\rho }^{\prime }}\left( \delta \rho _{(1),i}B^{(1)}_{,j} + \delta \rho _{(1),j}B^{(1)}_{,i} \right) - \frac{2}{\bar{\rho }^{\prime 2}}\delta \rho _{(1),i}\delta \rho _{(1),i} .
\end{eqnarray*}
The background sound speed $c_s^2$ in the present setting is
\begin{equation*}
c_s^2 = \frac{\bar{p }^{\prime }}{\bar{\rho }^{\prime }} = -1+\frac{2}{3}\epsilon _H + \frac{1}{3} \eta _H 
\end{equation*}
with $\bar{p}$ being the background pressure.
At the linear order, the energy density in the flat gauge is neatly written as
\begin{equation*}
a^2 \delta \rho _{(1)} = -6\mathcal{H}^2 \phi _{(1)}-2\mathcal{H} \nabla ^2 B^{(1)} .
\end{equation*}
The background energy density satisfies
\begin{equation*}
\bar{\rho } = 3\frac{\mathcal{H}^2}{a^2} ,
\end{equation*}
thus
\begin{equation*}
\bar{\rho }^{\prime } = -6\epsilon _H \frac{\mathcal{H}^3}{a^2} .
\end{equation*}
Therefore, the first-order curvature perturbation is given by
\begin{equation}
\zeta _{(1)} = -\frac{1}{3\mathcal{H}\epsilon _H} \left( 3\mathcal{H} \phi _{(1)}+ \nabla ^2 B^{(1)} \right) . \label{eq:zetag} 
\end{equation}
The second-order part will be discussed later.

\section{Analytical estimate of the bispectrum in the limit of small $\mathcal{I}$}\label{sec:analytic}

In this section, we apply the standard methods of the in-in formalism to the Lagrangian obtained in 
the previous section. In order to render the problem tractable, we keep only the leading-order 
contributions in the small parameters $\epsilon _{H,\varphi } ,\eta _{H,\varphi }$. An interesting
point is that even in the limit of de-Sitter space-time, the key parameter $\mathcal{I}$ does not
necessarily vanish. Using equations
(\ref{eq:epsilons}) - (\ref{eq:fgrads}) and substituting (\ref{eq:phi}) and (\ref{eq:B}), the cubic 
Lagrangian (\ref{eq:cubicaction}) becomes
\begin{eqnarray}
a^{-2} \mathcal{L} &=& \frac{1}{2} \left( \pi ^{\prime 2} - ( \nabla \pi )^2 \right) + \frac{f^2}{2a^2} \left( 3\alpha ^{\prime 2} - 2(\nabla \alpha )^2 \right) + \frac{6\mathcal{I}^2}{\eta ^2} \pi ^2 + \frac{6\sqrt{2}\mathcal{I}}{\eta } \frac{f}{a} \pi \alpha ^{\prime } \label{eq:lagrangian} \\
&& - \frac{4\sqrt{2}\mathcal{I}^2}{\sqrt{\epsilon _{\varphi }}\eta ^2} \pi ^3 -\frac{12\mathcal{I}}{\sqrt{\epsilon _{\varphi }}\eta }\frac{f}{a}\pi ^2 \alpha ^{\prime } -\sqrt{\frac{2}{\epsilon _{\varphi }}} \frac{f^2}{a^2} \pi \left( 3\alpha ^{\prime 2} -2(\nabla \alpha )^2 \right) . \nonumber 
\end{eqnarray}
We dropped $\tau $ for a reason that becomes clear soon.
We assume the background is close to de-Sitter, which implies
\begin{equation}
a = -\frac{1}{H\eta } , \ \ \ \ \ f = f_0 \eta ^2 
\end{equation}
and discarded all the terms higher order in slow roll. One  can rescale $\alpha $ to set $f_0 =1$ 
without loss of generality. We further demand $\mathcal{I} < 1$ since
we would like to treat all but kinetic terms perturbatively. The factors of $\sqrt{\epsilon _{\varphi }}^{-1}$ appearing in the cubic terms might look worrying for the validity of the perturbative approach. But 
when the action is written in terms of $\zeta $, they are of the same order as the quadratic kinetic terms
and the perturbative expansion should be marginally applicable. The following analysis is expected to be valid for $\epsilon _H \ll \mathcal{I} <1$. We are concerned with the three-point correlation function of the curvature perturbation
\begin{equation}
\zeta _{(1)}= -\frac{\eta }{3\epsilon _H} \sqrt{\frac{\epsilon _{\varphi }}{2}} \left( \pi ^{\prime } + \frac{3(1+2\mathcal{I}^2)}{\eta }\pi - \frac{3}{\sqrt{2}} H \mathcal{I}\eta ^3  \alpha ^{\prime }\right) , \label{eq:zeta1}
\end{equation}
which is obtained from (\ref{eq:zetag}) with (\ref{eq:phi}) and (\ref{eq:B}), neglecting all the higher order 
terms in slow roll. The absence of $\tau $ at this linear order justifies its omission from the Lagrangian.

 \subsection{Notations}
We take the free massless part of the action to be the background and treat all the other terms perturbatively. In the interaction picture, we set
\begin{eqnarray}
\pi _I (\eta , \mathbf{x}) &=& \int \frac{d^3k}{(2\pi )^3} \frac{H}{\sqrt{2k^3}} \left( u_k (\eta ) a_{\mathbf{k}} e^{i\mathbf{k}\cdot \mathbf{x}} + u^{\ast }_k (\eta ) a_{\mathbf{k}}^{\dagger } e^{-i\mathbf{k}\cdot \mathbf{x} } \right) , \\
\alpha _I (\eta , \mathbf{x}) &=& \frac{1}{\eta ^3} \int \frac{d^3 k}{(2\pi )^3} \frac{1}{\sqrt{6c_s^3 k^3}} \left( v_k (\eta ) b_{\mathbf{k}} e^{i\mathbf{k}\cdot \mathbf{x}} + v^{\ast }_k (\eta ) b_{\mathbf{k}}^{\dagger } e^{-i\mathbf{k}\cdot \mathbf{x} } \right) ,
\end{eqnarray}
where the de-Sitter mode functions are defined to be
\begin{eqnarray*}
u_k (\eta ) &=& (i-k\eta ) e^{-ik\eta } , \\
v_k (\eta ) &=& (c_s k \eta -i ) e^{-ic_s k \eta } , \ \ \ \ \ c_s \ =\  \sqrt{\frac{2}{3}} .
\end{eqnarray*}
The interaction Hamiltonian is given by
\begin{equation*}
H_I = \frac{6H^2 \mathcal{I}^2}{\eta ^4} \int d^3 w \pi _I^2 + H_I^q + H_I^A + H_I^B +H_I^C 
\end{equation*}
where
\begin{eqnarray*}
&& H_I^q = \frac{6\sqrt{2}\mathcal{I}}{H} \int d^3 w \ \pi _I \alpha _I^{\prime } , \ \ \ \ \ H_I^A = \sqrt{\frac{2}{\epsilon _{\varphi }}} \frac{4\mathcal{I}^2}{H^2 \eta ^4 } \int d^3 w \ \pi _I^3 , \\
&& H_I^B = -\frac{12\mathcal{I}}{\sqrt{\epsilon _{\varphi }}H} \int d^3 w \ \pi _I^2 \alpha _I^{\prime } , \ \ \ \ \ H_I^C = 3 \sqrt{\frac{2}{\epsilon _{\varphi }}}\eta ^4 \int d^3 w \ \pi _I \alpha _I^{\prime 2} .
\end{eqnarray*}
The term with higher spatial derivatives has been omitted. We often drop the subscript $I$.
Note that it was claimed in \cite{Bartolo2012} that $H_I^C$ gives the
leading contribution to the bispectrum. We shall explicitly confirm that it is the case as long as we remain
within the regime of validity for the perturbative treatment of the quadratic vertices (i.e. $H_I^q$ and the mass term for $\pi $).

We are going to compute the three-point correlation function in Fourier space defined by
\begin{equation*}
\langle \zeta (\eta ,\mathbf{x}) \zeta (\eta ,\mathbf{y}) \zeta (\eta ,\mathbf{z}) \rangle = \iiint \frac{d^3 k_1}{(2\pi )^3} \frac{d^3 k_2}{(2\pi )^3} \frac{d^3 k_3}{(2\pi )^3} \langle \zeta _{k_1} \zeta _{k_2} \zeta _{k_3}\rangle (2\pi )^3 \delta (\mathbf{k}_1 +\mathbf{k}_2 +\mathbf{k}_3 ) e^{i(\mathbf{k}_1 \cdot \mathbf{x}+ \mathbf{k}_2 \cdot \mathbf{y} + \mathbf{k}_3 \cdot \mathbf{z})} .
\end{equation*}
We often abbreviate it as $\langle \zeta _k^3 \rangle $. 

\subsection{The outline of the calculation}
Introducing an auxiliary function
\begin{equation}
\xi (\eta , \mathbf{x}) = \pi ^{\prime } + \frac{3}{\eta }\pi ,
\end{equation}
the three-point function for $\zeta $ can be written as
\begin{eqnarray}
- \frac{54\sqrt{2}\epsilon _H^3}{\sqrt{\epsilon _{\varphi }^3}} \langle \zeta (\eta ,\mathbf{x}) \zeta (\eta ,\mathbf{y}) \zeta (\eta ,\mathbf{z}) \rangle &=& \eta ^3 \langle \xi (\eta ,\mathbf{x} ) \xi (\eta ,\mathbf{y})\xi (\eta ,\mathbf{z}) \rangle \label{eq:zeta} \\
&& - \frac{3H\mathcal{I}\eta ^6}{\sqrt{2}} \left( \langle \xi (\eta , \mathbf{x} ) \xi (\eta , \mathbf{y} ) \alpha ^{\prime }(\eta , \mathbf{z}) \rangle +  {\rm 2 \ perms}\right) \nonumber \\
&& + \frac{9H^2 \mathcal{I}^2 \eta ^9}{2} \left( \langle \xi (\eta ,\mathbf{x}) \alpha ^{\prime }(\eta ,\mathbf{y} )\alpha ^{\prime }(\eta , \mathbf{z}) \rangle + {\rm 2 \ perms} \right) \nonumber .
\end{eqnarray}
Despite the appearance of lower powers of $\mathcal{I}$ in the Lagrangian, the leading-order contribution to $\langle \zeta ^3 \rangle $ turns out to be quadratic.\footnote{This is true for tree-level calculations, but
loop contributions may contain terms without any factor of $\mathcal{I}$.} Then, the $\pi ^2$ term in the interaction Hamiltonian is clearly irrelevant. However, we do have to keep $H_I^q$ since it affects, for example $\langle \xi ^3 \rangle $ with $H_I^B$ at this order. More specifically, $\langle \xi ^3  \rangle $ can be written as follows:
\begin{eqnarray}
\langle \xi ^3 \rangle &=& i \int ^{\eta }d\eta _1 \langle \left[ H_I^A (\eta _1 ) , \xi _I^3 \right]  \rangle \label{eq:xi1} \\ 
&& - \int ^{\eta }d\eta _1 \int ^{\eta _1}d\eta _2 \langle \left[ H_I^B (\eta _2 ) , \left[ H_I^q (\eta _1) , \xi _I^3 \right] \right] \rangle  \label{eq:xi2A} \\
&& - \int ^{\eta }d\eta _1 \int ^{\eta _1}d\eta _2 \langle \left[ H_I^q (\eta _2 ) ,  \left[ H_I^B (\eta _1 ) ,\xi _I^3 \right] \right] \rangle \label{eq:xi2B} \\
&&  -i \int ^{\eta }d\eta _1 \int ^{\eta _1 }d\eta _2 \int ^{\eta _2}d\eta _3 \langle \left[ H_I^C (\eta _3 ) , \left[ H_I^q (\eta _2 ) , \left[ H_I^q (\eta _1 ) , \xi _I^3 \right] \right] \right] \rangle \label{eq:xi3A} \\
&&  -i \int ^{\eta }d\eta _1 \int ^{\eta _1 }d\eta _2 \int ^{\eta _2}d\eta _3 \langle \left[ H_I^q (\eta _3 ) , \left[ H_I^C (\eta _2 ) , \left[ H_I^q (\eta _1 ) , \xi _I^3 \right] \right] \right] \rangle \label{eq:xi3B} \\
&&  -i \int ^{\eta }d\eta _1 \int ^{\eta _1 }d\eta _2 \int ^{\eta _2}d\eta _3 \langle \left[ H_I^q (\eta _3 ) , \left[ H_I^q (\eta _2 ) , \left[ H_I^C (\eta _1 ) , \xi _I^3 \right] \right] \right] \rangle  + O(\mathcal{I}^3 )  . \label{eq:xi3C} 
\end{eqnarray}
In a similar way, $\langle \xi ^2 \alpha ^{\prime } \rangle $ contains the quadratic term in the Hamiltonian given as
\begin{eqnarray}
\langle \xi ^2 \alpha ^{\prime } \rangle  &=& i \int ^{\eta }d\eta _1 \langle \left[ H_I^B (\eta _1 ) , \xi _I^2 \alpha ^{\prime }_I \right] \rangle  \label{eq:xia1} \\
&& - \int ^{\eta }d\eta _1 \int ^{\eta _1}d\eta _2 \langle \left[ H_I^C (\eta _2 ) , \left[ H_I^q (\eta _1 ) , \xi _I^2 \alpha _I^{\prime } \right] \right] \rangle \label{eq:xia2A} \\
&& - \int ^{\eta }d\eta _1 \int ^{\eta _1}d\eta _2 \langle \left[ H_I^q (\eta _2 ) , \left[ H_I^C (\eta _1 ) , \xi _I^2 \alpha _I^{\prime } \right] \right] \rangle + O(\mathcal{I}^2 ) \label{eq:xia2B}  .
\end{eqnarray}
Finally, $\langle \xi \alpha ^{\prime 2} \rangle $ receives no contribution from the quadratic interaction and becomes
\begin{equation}
\langle \xi \alpha ^{\prime 2} \rangle = i\int ^{\eta } d\eta _1 \langle \left[ H_I^C (\eta _1 ) , \xi _I \alpha _I^{\prime 2} \right] \rangle + O(\mathcal{I})  \label{eq:xiaa} .
\end{equation}
Hence, we have to compute ten distinct integrations to fully work out $\langle \zeta ^3 \rangle $ at the quadratic order in $\mathcal{I}^2 $. We focus on the superhorizon limit of the spectrum, i.e.
$-k_i \eta \ll 0 $ for all $i=1,2,3$. 
 
 \subsection{Summary of the results}
We list the contributions from each of the ten integrations in the limit of $- k_i \eta \rightarrow 0$.
The detailed calculations are presented in the appendix. \newline

\noindent {\bf 1-vertex contributions}
\begin{align}
\begin{split}
& i \int ^{\eta } d\eta _1 \langle  \left[ H_I^A (\eta _1 ),\xi _k^3 \right]  \rangle  \rightarrow \sqrt{\frac{2}{\epsilon _{\varphi }}} \frac{9H^4 \mathcal{I}^2 \left( k_1^3 + k_2^3 + k_3^3 \right) }{k_1^3 k_2^3 k_3^3 \eta ^3} \int ^{\eta } \frac{d\eta _1}{\eta } \cos \left[ (k_1+k_2 +k_3 ) (\eta - \eta _1 ) \right]  , 
\end{split} \nonumber \\
\begin{split}
& i \int ^{\eta } d\eta _1 \langle \left[ H_I^B (\eta _1 ) , \xi _{k_1}\xi _{k_2} \alpha _{k_3}^{\prime } \right]  \rangle  \rightarrow -\frac{27H^3 \mathcal{I}(k_1^3 + k_2^3 )}{\sqrt{\epsilon _{\varphi }}\eta ^6 c_s^3 k_1^3 k_2^3 k_3^3 }\int ^{\eta } \frac{d\eta _1}{\eta _1} \cos \left[ (k_1 + k_2 +c_s  k_3 )(\eta - \eta _1 ) \right]  , 
\end{split} \nonumber \\
\begin{split}
& i \int ^{\eta } d\eta _1 \langle  \left[ H_I^C (\eta _1 ) , \xi _{k_1}\alpha ^{\prime }_{k_2} \alpha ^{\prime }_{k_3} \right]  \rangle  \rightarrow \frac{27H^2 }{2\sqrt{2\epsilon _{\varphi }}\eta ^9 c_s^6  k_2^3 k_3^3}\int ^{\eta } \frac{d\eta _1}{\eta _1} \cos \left[ (k_1 +c_s k_2 + c_s k_3 ) (\eta - \eta _1 ) \right]  .
\end{split} \nonumber 
\end{align}

\noindent {\bf 2-vertex contributions}
\begin{eqnarray*}
&&- \int ^{\eta }d\eta _1 \int ^{\eta _1}d\eta _2 \langle  \left[ H_I^B (\eta _2 ) , \left[ H_I^q (\eta _1 ) , \xi _k^3 \right] \right]  \rangle  = -\sqrt{\frac{2}{\epsilon _{\varphi }}} \frac{12H^4 \mathcal{I}^2}{c_s^3 k_1^3 k_2^3 k_3^3 \eta ^3} \left( \frac{\mathcal{A}_{2a}}{k_1^3} + 2 \ {\rm perms} \right) ,  \\
&& \mathcal{A}_{2a} \rightarrow 27 k_1^3 (k_2^3 +k_3^3 ) \int ^{\eta }\frac{d\eta _1}{\eta _1} \int ^{\eta _1}\frac{d\eta _2}{\eta _2} \cos \left( k_1 (\eta - \eta _1 ) \right) \cos \left(  c_s k_1 (\eta _1 -\eta _2 )+(k_2 +k_3 )(\eta -\eta _2 ) \right) , \\
&& - \int ^{\eta }d\eta _1 \int ^{\eta _1}d\eta _2 \langle  \left[ H_I^q (\eta _2 ) , \left[ H_I^B (\eta _1 ) , \xi _k^3 \right] \right]  \rangle = -\sqrt{\frac{2}{\epsilon _{\varphi }}} \frac{12H^4 \mathcal{I}^2 }{c_s^3 k_1^3 k_2^3 k_3^3  \eta ^3} \left( \frac{\mathcal{A}_{2b}}{k_1^3 } + 2 \ {\rm perms} \right)  , \\
&& \mathcal{A}_{2b} \rightarrow 27 k_1^3 (k_2^3 +k_3^3 ) \int ^{\eta } \frac{d\eta _1}{\eta _1}\int ^{\eta _1}\frac{d\eta _2}{\eta _2} \cos \left( (k_1 +k_2 )(\eta -\eta _1 ) \right)  \cos \left( k_1 (\eta - \eta _2 ) +c_s k_1 (\eta _1 -\eta _2 ) \right) , \\
&& -\int^{\eta } d\eta _1 \int ^{\eta _1}d\eta _2 \langle  \left[ H_I^C (\eta _2 ) , \left[ H_I^q (\eta _1 ) , \xi _{k_1}\xi _{k_2} \alpha ^{\prime }_{k_3} \right] \right]  \rangle = \frac{2H^3 \mathcal{I}}{\sqrt{\epsilon _{\varphi }}c_s^6 k_1^6 k_2^3 k_3^3 \eta ^6} \mathcal{B}_{2a} + (1 \leftrightarrow 2 )  ,  \\
&& \mathcal{B}_{2a} \rightarrow 81 k_1^3 k_2^3 \int ^{\eta }\frac{d\eta _1}{\eta _1} \int ^{\eta _1}\frac{d\eta _2}{\eta _2 } \cos \left( k_1 (\eta  -\eta _1 ) \right) \cos \left( c_s k_1 (\eta _1 -\eta _2 ) + (k_2 + c_s k_3 ) (\eta - \eta _2 ) \right) , \\
&& -\int^{\eta } d\eta _1 \int ^{\eta _1}d\eta _2 \langle  \left[ H_I^q (\eta _2 ) , \left[ H_I^C (\eta _1 ) , \xi _{k_1}\xi _{k_2} \alpha ^{\prime }_{k_3} \right] \right]  \rangle = \frac{2H^3 \mathcal{I}}{\sqrt{\epsilon _{\varphi }}c_s^6 k_1^6 k_2^3 k_3^3 \eta ^6} \mathcal{B}_{2b} + (1 \leftrightarrow 2)  , \\
&& \mathcal{B}_{2b} \rightarrow 81 k_1^3 k_2^3 \int ^{\eta }\frac{d\eta _1}{\eta _1} \int ^{\eta _1}\frac{d\eta _2 }{\eta _2 } \cos \left( (k_2 +c_s k_3 ) (\eta -\eta _1 ) \right) \cos \left( c_s k_1 (\eta _1 -\eta _2 ) + k_1 (\eta -\eta _2 ) \right) .
\end{eqnarray*}

\noindent {\bf 3-vertex contributions}
\begin{eqnarray*}
&& -i \int ^{\eta }d\eta _1 \int ^{\eta _1}d\eta _2 \int ^{\eta _2}d\eta _3 \langle  \left[ H^C_I (\eta _3) , \left[ H_I^q  (\eta _2 ) , \left[ H_I^q (\eta _1 ) , \xi _k^3 \right] \right] \right]  \rangle \\
&& \qquad \qquad \qquad \qquad = - \sqrt{\frac{2}{\epsilon _{\varphi }}} \frac{12H^4 \mathcal{I}^2 }{c_s^6 k_1^3 k_2^3 k_3^3 \eta ^3} \left( k_3^3 \mathcal{A}_{3a}+ 5 \ {\rm perms} \right)  , \\
&& \mathcal{A}_{3a} \rightarrow -81 k_1^3 k_2^3 k_3^3 \int ^{\eta }\frac{d\eta _1}{\eta _1} \cos \left( k_1 (\eta -\eta _1) \right) \int ^{\eta _1}\frac{d\eta _2}{\eta _2} \cos \left( k_2 (\eta - \eta _2 ) \right) \\
&& \qquad  \qquad \times  \int ^{\eta _2}\frac{d\eta _3}{\eta _3} \cos \left( c_s k_1 (\eta _1 -\eta _3 ) +c_s k_2 (\eta _2 -\eta _3 ) +k_3 (\eta - \eta _3 ) \right) , \\
&& -i \int ^{\eta }d\eta _1 \int ^{\eta _1}d\eta _2 \int ^{\eta _2}d\eta _3 \langle  \left[ H^q_I (\eta _3) , \left[ H_I^C  (\eta _2 ) , \left[ H_I^q (\eta _1 ) , \xi _k^3 \right] \right] \right]  \rangle \\
&& \qquad \qquad \qquad \qquad = - \sqrt{\frac{2}{\epsilon _{\varphi }}} \frac{12H^4 \mathcal{I}^2 }{c_s^6 k_1^3 k_2^3 k_3^3 \eta ^3} \left( k_2^3 \mathcal{A}_{3b} + 5 \ {\rm perms} \right)  , \\
&& \mathcal{A}_{3b} \rightarrow -81 k_1^3 k_2^3 k_3^3 \int ^{\eta }\frac{d\eta _1}{\eta _1 } \cos \left( k_1 (\eta -\eta _1 ) \right) \int ^{\eta _1}\frac{d\eta _2}{\eta _2} \cos \left( c_s k_1 (\eta _1 -\eta _2 ) + k_2 (\eta - \eta _2 ) \right) \\
&& \qquad \qquad \times \int ^{\eta _2 }\frac{d\eta _3}{\eta _3} \cos \left( k_3 (\eta - \eta _3 ) +c_s k_3 (\eta _2 -\eta _3 ) \right) , 
\end{eqnarray*}
\begin{eqnarray*}
&& -i \int ^{\eta ,}d\eta _1 \int ^{\eta _1}d\eta _2 \int ^{\eta _2}d\eta _3 \langle  \left[ H^q_I (\eta _3) , \left[ H_I^q  (\eta _2 ) , \left[ H_I^C (\eta _1 ) , \xi _k^3 \right] \right] \right]  \rangle \\
&& \qquad \qquad \qquad  \qquad = - \sqrt{\frac{2}{\epsilon _{\varphi }}} \frac{12H^4 \mathcal{I}^2 }{c_s^6 k_1^3 k_2^3 k_3^3 \eta ^3} \left( k_2^3 \mathcal{A}_{3c} + 5 \ {\rm perms} \right)  , \\
&& \mathcal{A}_{3c} \rightarrow -81 k_1^3 k_2^3 k_3^3 \int ^{\eta }\frac{d\eta _1}{\eta _1} \cos \left( k_1 (\eta -\eta _1 ) \right) \int ^{\eta _1}\frac{d\eta _2}{\eta _2} \cos \left( c_s k_1 (\eta _1 -\eta _2 ) +k_2 (\eta - \eta _2 ) \right) \\
&& \qquad \qquad \times \int ^{\eta _2}\frac{d\eta _3}{\eta _3}\cos \left( k_3 (\eta - \eta _3 ) +c_s k_3 (\eta _1 -\eta _3 ) \right) .
\end{eqnarray*}

Note that all of the remaining integrals can be carried out in the limit $-k_i \eta \rightarrow 0$,
which result in a logarithm of $- \eta $ for each integration. Therefore, 3-vertex contributions
dominate over the others in superhorizon limit, as claimed in \cite{Bartolo2012}. 

In addition, there is a contribution to bispectrum arising from second- and higher order
perturbations of $\zeta $ in terms of the field variables $\pi $ and $\alpha $. It is evaluated
for the second-order term in the appendix and shown to be of order $\mathcal{I}^2$, hence
subdominant compared to the logarithms from the integrations listed above. 

In the end, our result is summarized as follows. At the order of $\mathcal{I}^2$, the tree-level
amplitude of the three-point function in the super horizon limit becomes
\begin{align}
\begin{split}
\langle \zeta _{k_1}\zeta _{k_2}\zeta _{k_3} \rangle \ \rightarrow \ & \frac{3\epsilon _{\varphi }H^4 \mathcal{I}^2}{2\epsilon _H^3 k_1^6 k_2^6 k_3^6} \left( k_3^3 \mathcal{A}_{3a} + k_2^3 \mathcal{A}_{3b} +k_2^3 \mathcal{A}_{3c} + 2 \ {\rm perms}  \right) \\
 \ \sim \ & \frac{243\epsilon _{\varphi } H^4 \mathcal{I}^2}{4\epsilon _H^3 k_1^3 k_2^3 k_3^3 } \left( k_1^3 +k_2^3 +k_3^3 \right) \left( \ln \left( -K\eta \right) \right) ^3 
 \end{split} \label{eq:bispectrum}
\end{align}
where $K$ is a reference momentum, say $K = \frac{1}{3}(k_1 +k_2 +k_3 )$. While the ambiguity
of $K$ arising from the lower limits of the integrations leads to errors of order $\ln ( k_i / k_j ) , i \neq j$,
for the wavelengths of interests, this should be of order $10$. Since there are many other
contributions of similar order which we have already ignored, it does not make sense to overly
worry about this reference momentum. As we can see, the bispectrum is of local shape. 
In order to estimate the $f_{NL}$ in the squeezed limit, which is defined as
\begin{equation}
f_{NL} = \frac{5}{6} \frac{\langle \zeta _{k_1} \zeta _{k_2} \zeta _{k_3} \rangle }{\langle \zeta _{k_1} \zeta _{k_2} \rangle + \langle \zeta _{k_2} \zeta _{k_3} \rangle + \langle \zeta _{k_3} \zeta _{k_1} \rangle } \ , 
\end{equation}
we quote the result from
\cite{Yamamoto2012a} for the power spectrum
\begin{equation}
 \langle  \zeta _k ^2 \rangle \rightarrow \frac{\epsilon _{\varphi }}{\epsilon _H^2}\frac{H^2}{4k^3} \left( 1 + 18\sqrt{6}\mathcal{I}^2 \left( \ln (-k\eta ) \right) ^2 \right) . \label{eq:powerPerturbative}
\end{equation}
Under the condition
\begin{equation}
\mathcal{I}^2 \ll 1 , \label{eq:limitation}
\end{equation}
which implies we can replace $\epsilon _{\varphi } $ with $\epsilon _H$, we obtain
\begin{equation}
f_{NL} \sim \frac{810 \mathcal{I}^2 N_K^3}{(1+18\sqrt{6}\mathcal{I}^2 N_K^2)^2} , \label{eq:result1}
\end{equation}
where $N_K$ is the number of efoldings experienced by the relevent modes after horizon crossing.
This result qualitatively agrees with the one derived in \cite{Bartolo2012} if the correction term
in the denominator is ignored. 

However, there are a few unsatisfactory features in this result. The first is the limitation
arising from our perturbative approach. Since we are sticking to perturbative expansion in
terms of $\mathcal{I}$, the formula (\ref{eq:result1}) can be trusted only for 
\begin{equation}
\mathcal{I}^2 \left( \ln (-K\eta ) \right) ^2 \ll 1 \label{eq:limitation2}
\end{equation}
since otherwise we would have to take into account the higher order terms from the Taylor
expansion of the denominator. However, the condition (\ref{eq:limitation2}) is much more
strict than the generic one (\ref{eq:limitation}) considering that $N_K = -\ln (-K\eta )$ for the modes
relevant in CMBR are of order $50$. Namely, the applicability of the analysis so far is limited
to $\mathcal{I}^2 \lesssim 10^{-4}$ and we are unable to say anything about $f_{NL}$ for
the range $10^{-4} \lesssim \mathcal{I}^2 \lesssim 1$.
Furthermore, the fact that $\langle \zeta _k^3 \rangle $ may grow indefinitely as long
as inflation continues sounds unpleasant considering the classical stability of the
quasi-de-Sitter background. It is distinct from the infrared divergence discussed in \cite{Bartolo2012}
which concerns the back reaction of the quantum fluctuations and loop corrections
which is beyond the scope of the present article.
The divergence is already there at the tree-level calculation.
Motivated by this, in the next section, we shall give a more careful analysis on the 
superhorizon dynamics of the fluctuations.

\section{Non-perturbative treatment of the quadratic vertices}

It is clear that the above approach based on the perturbative expansion in terms
of $\mathcal{I}$ bares a limited applicability even if $\mathcal{I} \ll 1$. From the point
of view of the classical stability of this inflationary regime shown in \cite{Yamamoto2012}, the 
apparent indefinite growth of the correlation functions after horizon exit should halt
sooner or later if all the relevant effects are taken into account. In the Lagrangian
(\ref{eq:lagrangian}), we have regarded the quadratic interaction terms as 
perturbative corrections along with the cubic ones. In this way, the proper
tree-level amplitude involves an infinite number of Feynman diagrams
generated by those quadratic vertices. While we have avoided this issue
by focusing on the leading-order contribution in $\mathcal{I}$, one should
expect a convergent result if the higher order corrections are treated
appropriately. For this purpose, we investigate the linear perturbation
more closely and show that both the power spectrum and the bispectrum become 
constant in the limit of $\eta \rightarrow 0$ despite the appearance of logarithmic 
divergence $\ln (-k\eta )$ in the perturbative analysis.

\subsection{Linear evolution equations and their superhorizon solutions}
The equations of motion at linear order are given by
\begin{eqnarray}
&& \frac{1}{H^2 \eta ^2 } \left( \pi _k^{\prime \prime } +k^2 \pi _k^2 \right) -\frac{2}{H^2 \eta ^3} \pi _k^{\prime } -\frac{12\mathcal{I}^2}{H^2 \eta ^4} \pi _k = -\frac{6\sqrt{2}\mathcal{I}}{H} \alpha _k^{\prime }  \label{eq:evpi} \\
&& \left( 3\eta ^4 \alpha _k^{\prime } \right) ^{\prime } +2 k^2  \eta ^4 \alpha _k = \frac{6\sqrt{2}\mathcal{I}}{H} \pi _k^{\prime }   . \label{eq:evalpha}
\end{eqnarray}
It turns out that one can write down analytic expressions for the solutions in the 
superhorizon limit. Ignoring the spatial gradients, the second immediately integrates to give
\begin{equation*}
\alpha _k^{\prime } = \frac{c_0}{\eta ^4} +\frac{2\sqrt{2}\mathcal{I}}{H\eta ^4} \pi _k
\end{equation*}
where $c_0 $ is an integration constant. We suppress its $k$-dependence since there should be
no confusion as far as the linear theory is concerned. The same applies to the rest of the
integration constants. Plugging this into the first equation, we derive
\begin{equation*}
\pi _k^{\prime \prime } -\frac{2}{\eta } \pi _k^{\prime } +\frac{12\mathcal{I}^2}{\eta ^2 } \pi _k= -\frac{6\sqrt{2}H\mathcal{I}c_0}{\eta ^2 }
\end{equation*}
whose general solution can be written as
\begin{equation}
\pi _k= -\frac{H}{\sqrt{2}\mathcal{I}} \left( c_0 + c_{+}\left( -k\eta \right) ^{p_{+}} +c_- \left( -k\eta \right) ^{p_{-}} \right) \label{eq:solpi}
\end{equation}
with two arbitrary constant $c_{\pm }$. 
The power exponents are given by
\begin{equation*}
p _{\pm } = \frac{3\pm \sqrt{9-48\mathcal{I}^2}}{2} .
\end{equation*}
The corresponding $\alpha $ is
\begin{equation}
\alpha _k= \frac{c_0 }{3\eta ^3} +2(-k)^3 \left( c_1 + \frac{c_+}{p_-} \left( -k\eta \right) ^{-p_-} + \frac{c_-}{p_+} \left( -k\eta \right) ^{-p_+} \right)  \label{eq:solalpha}
\end{equation}
with the fourth integration constant $c_1$. We used the relations
\begin{equation*}
p_{\pm }-3 = -p_{\mp } .
\end{equation*}

\subsection{Canonical mode functions}
When the off-diagonal terms in the quadratic Lagrangian are taken into account, the introduction
of mode functions is not so straightforward as with independent free fields. In this subsection, 
we look into the canonical formulation of the field theory. From the Lagrangian (\ref{eq:lagrangian}), 
we read off
\begin{equation*}
\hat{\pi } = a \pi \ \ \ \ \ {\rm and} \ \ \ \ \ \hat{\alpha } = \sqrt{3}f \alpha 
\end{equation*}
as the canonically normalised field variables. Their conjugate momenta are given by
\begin{eqnarray*}
\hat{p}_{\pi } &=& \hat{\pi }^{\prime } , \\
\hat{p}_{\alpha } &=& \hat{ \alpha }^{\prime } +\frac{2\sqrt{6}\mathcal{I}}{\eta } \hat{\pi } 
\end{eqnarray*}
and we impose the canonical commutation relations
\begin{equation*}
\left[ \hat{\pi }(\tau ,\mathbf{x}) , \hat{p}_{\pi }(\tau , \mathbf{y}) \right] = i\delta \left( \mathbf{x} -\mathbf{y} \right) , \ \ \ \ \ \left[ \hat{\alpha } (\tau , \mathbf{x}) , \hat{p}_{\alpha }(\tau , \mathbf{y}) \right] = i \delta \left( \mathbf{x} -\mathbf{y} \right) 
\end{equation*}
with all the cross commutators being zero.
To diagonalize the Hamiltonian, we introduce the creation and annihilation operators
\begin{equation*}
\left[ \hat{a}_{ a\mathbf{p}} , \hat{a}_{b\mathbf{q}}^{\dagger } \right] = \delta _{ab} \delta \left( \mathbf{p}- \mathbf{q} \right) , \ \ \ \ a, b = 1,2 
\end{equation*}
and expand the field operators in terms of the mode functions:
\begin{eqnarray*}
\hat{\pi }\left( \eta ,\mathbf{x} \right)  &=& a \sum _{a=1,2} \int \frac{d^3 k}{(2\pi )^3} \left( \pi _k^a (\eta ) \hat{a}_{a\mathbf{k}} e^{i\mathbf{k}\cdot \mathbf{x}} + \pi _k^{a\ast }(\eta ) \hat{a}_{a\mathbf{k}}^{\dagger }e^{-i\mathbf{k}\cdot \mathbf{x} } \right)  , \\
\hat{\alpha } \left( \eta, \mathbf{x} \right) &=& \sqrt{3} f \sum _{a=1,2} \int \frac{d^3 k}{(2\pi )^3 } \left(  \alpha _k^a (\eta )\hat{a}_{a\mathbf{k}}e^{i\mathbf{k}\cdot \mathbf{x}}  +\alpha _k^{a \ast } (\eta )\hat{a}_{a\mathbf{k}}^{\dagger } e^{-i\mathbf{k}\cdot \mathbf{x}} \right) .
\end{eqnarray*}
Here, $( \pi _k^a ,\alpha _k^a) , a =1,2 $ are two independent solutions of equations (\ref{eq:evpi}) and (\ref{eq:evalpha}). As an example, for the superhorizon solutions derived in the previous 
subsection, the mode functions become
\begin{eqnarray}
\pi _k^a &=& -\frac{H}{\sqrt{2}\mathcal{I} } \left( c_0^a + c_+^a \left( -k\eta \right) ^{p_+} + c_-^a \left( -k\eta \right) ^{p_-} \right) , \label{eq:modeu} \\
\alpha _k^a &=&  \frac{c_0^a}{3\eta ^3 } +2(-k)^3 \left(  c_1^a  + \frac{c_+^a}{p_-} \left( -k\eta \right) ^{-p_-} + \frac{c_-^a}{p_+} \left( -k\eta \right) ^{-p_+} \right) , \label{eq:modev} 
\end{eqnarray}
which are characterized by eight complex constants.
In this way, we see that each field operator may excite two different particles. Conversely,
for each particle species $a$, there are two associated mode functions $\hat{u}^a_k$ and $\hat{v}_k^a$.
This simply reflects the fact that the fields themselves do not define particles when there is a 
quadratic mixing term. This formulation is consistent as long as the mode functions satisfy the
following conditions arising from the canonical commutators (from here on, the summation
convention for indices $a,b,\cdots $ is assumed):
\begin{eqnarray*}
&& \pi ^a \pi ^{a\ast \prime } - \pi ^{a\ast } \pi ^{a\prime } = \frac{i}{a^2}  , \quad  \alpha ^a \alpha ^{a\ast \prime } - \alpha ^{a\ast } \alpha ^{a\prime } = \frac{i}{3f^2} ,\\
&& \pi ^a \alpha ^{a\ast } - \pi ^{a\ast } \alpha ^a = 0 , \quad \pi ^a \alpha ^{a\ast \prime } - \pi ^{a\ast } \alpha ^{a\prime } = 0 , \\
&& \pi ^{a\prime } \alpha ^{a\ast } - \pi ^{a\ast \prime } \alpha ^a = 0 , \quad \pi ^{a\prime } \alpha ^{a\ast \prime } - \pi ^{a\ast \prime } \alpha ^{a\prime } = \frac{2\sqrt{2}\mathcal{I}}{a^3 f \eta } i  .
\end{eqnarray*}
It can be checked that they are preserved by the evolution equations (\ref{eq:evpi}) and (\ref{eq:evalpha})
if they are satisfied at an initial time.

Later, it proves to be useful to write down these conditions specifically for the superhorizon
mode functions. Those six equations translate into algebraic conditions on the integration
constants:
\begin{eqnarray}
&& c_+^a c_-^{a\ast } - c_+^{a\ast }c_-^a = \frac{2i \mathcal{I}^2 }{k^3 \left( p_+ -p_- \right)} , \label{eq:con1} \\
&& c_0^a c_1^{a\ast } -c_0^{a\ast } c_1^a = \frac{i}{6k^3} , \\
&& c_0^a c_{\pm }^{a\ast } - c_0^{a\ast } c_{\pm }^a = c_1^a c_{\pm }^{a\ast } -c_1^{a\ast } c_{\pm }^a = 0 . \label{eq:con3} 
\end{eqnarray}

\subsection{Matching with the de-Sitter mode functions}
From the solutions (\ref{eq:solpi}) and (\ref{eq:solalpha}), it is almost obvious that the power
spectrum should converge to a constant proportional to $ |c_0^a |^2 $. 
In order to estimate its magnitude, however, one needs to determine the 
constants $c_{\alpha }^a , \alpha = 0,1,\pm $ from appropriately initial conditions set
deep inside the horizon. As a first approximation, we can match the de-Sitter mode functions:
\begin{eqnarray*}
\pi _k^a &=&  \frac{H}{\sqrt{2k^3}}\delta ^a_1 \left( i -k\eta \right)  e^{-ik\eta } , \\
\alpha _k^a  &=& \frac{\eta ^{-3}}{\sqrt{6c_s^3 k^3}}\delta ^a_2 \left( c_s k \eta - i  \right) e^{-ic_s k\eta } 
\end{eqnarray*}
with the superhorizon counterparts (\ref{eq:modeu}) and (\ref{eq:modev}) at 
the horizon crossing $k\eta =-1$. We evaluate them and their first derivatives and equate each other.
This leads to the following eight equations:
\begin{eqnarray*}
&& c_0^a + c_+^a +c_-^a = -\frac{\mathcal{I}\delta ^a_{\ 1} }{\sqrt{k^3}} \left( 1+i\right) e^i , \\
&& \frac{c_0^a }{6} +c_1^a + \frac{c_+^a}{p_-} +\frac{c_-^a}{p_+} = -\frac{\delta ^a_{\ 2}}{\sqrt{24c_s k^3}}\left( 1 + \frac{i}{c_s} \right) e^{c_s i},   \\
&& -c_0^a + \left( p_+ -1 \right) c_+^a + \left( p_- -1 \right) c_-^a = \frac{\mathcal{I}\delta ^a_{\ 1}}{\sqrt{k^3}} e^i , \\
&& -\frac{c_0^a }{6} + 2c_1^a + \frac{p_+ -1}{p_-} c_+^a + \frac{p_- -1}{p_+} c_-^a = \frac{\delta ^a_{\ 2}}{\sqrt{24c_s k^3}} \left( 1 + \frac{c_s}{2}i \right) e^{c_s i} .
\end{eqnarray*}
These easily solve as
\begin{eqnarray*}
&& c_0^1 = -\frac{2\mathcal{I}}{\sqrt{k^3}} \left( 1+i \right) e^i , \ \ \ \ \ c_1^1 = \frac{\mathcal{I}}{3\sqrt{k^3}}e^i \left( 1 + i + \frac{3i}{p_+ p_-} \right)  ,  \\
&& c_0^2 = \frac{e^{c_s i}}{2\sqrt{6 c_s k^3 }} \left( 6 + 7c_s i \right) , \ \ \ \ \ c_1^2 = -\frac{e^{c_s i}}{3\sqrt{6c_s k^3}} \left( 3 + 4c_s i \right) , \\
&& c_{\pm }^1 = \frac{\mathcal{I}}{\sqrt{k^3}} \frac{e^i }{p_{\mp } -p_{\pm } }\left( i + (1+i) p_{\mp } \right)  , \ \ \ \ \ c_{\pm }^2 = \frac{p_{\mp }}{p_{\pm } -p_{\mp }} c_0^2 .
\end{eqnarray*}

We can now estimate the final amplitude of the two-point function. Clearly, the dominant
contribution at late time comes from $c_0^a$s. The mode functions asymptotically approach
\begin{equation*}
\pi _k^a (\eta ) \sim - \frac{H c_0^a }{\sqrt{2}\mathcal{I} } , \ \ \ \ \ \alpha _k^a (\eta ) \sim \frac{c_0^a}{3\eta ^3 } .
\end{equation*}
Using
\begin{equation*}
\zeta = \frac{H\eta ^2 }{3\epsilon _H } \sqrt{\frac{\epsilon _{\varphi }}{2}} \left[ \hat{\pi }^{\prime } +\frac{4+6\mathcal{I}^2 }{\eta } \hat{\pi } +\sqrt{\frac{3}{2}} \mathcal{I} \left( \hat{\alpha }^{\prime } -\frac{2}{\eta } \hat{\alpha } \right) \right] ,
\end{equation*}
we derive
\begin{equation}
\langle \zeta _k^2 \rangle \rightarrow \frac{H^2 \epsilon _{\varphi }}{4\epsilon _H^2 \mathcal{I}^2} \left( 1+\mathcal{I}^2 \right) \left| c_0^a \right| ^2 = \frac{H^2 \epsilon _{\varphi }}{\epsilon _H^2 k^3 } \left( 1+\mathcal{I}^2 \right) \left( 2+ \frac{103}{144 c_s \mathcal{I}^2} \right) . \label{eq:powerresult}
\end{equation}
Although the exact numerical factors should
not be trusted due to the errors coming from the matching, the dependence on $\mathcal{I}^2$
is generic (we will confirm this numerically). The fact that the final amplitude is divergent in the
limit small $\mathcal{I}^2 $ might look worrying. But it is also the case in the standard
single-scalar inflation where the power spectrum is formally infinite when the background is 
exactly de-Sitter. The same nature can be seen taking the limit of $\mathcal{I} \rightarrow 0 $ 
and large e-folding number ($-k\eta  \rightarrow 0$), rewriting the amplitude of the power spectrum
as
\begin{equation}
\langle \zeta _k^2  \rangle \sim O(1) \frac{H^2 \epsilon _{\varphi }}{\mathcal{I}^2 \epsilon _H^2 k^3 } 
\sim \frac{H^2}{(\epsilon _H -\epsilon _{\varphi }) k^3 }.
\end{equation}
In the last approximation, we used (\ref{eq:scri}) and 
\begin{equation*}
\mathcal{I} \ll 1 \Leftrightarrow \frac{\epsilon _H}{ \epsilon _{\varphi }} \sim 1  . \label{eq:scriapprox}
\end{equation*}
This result is reasonable: the dependence of the power spectrum on the parameters is the
same as the single-scalar inflation except that $\epsilon _{\varphi }$ is now replaced by
$\epsilon _H- \epsilon _{\varphi }$. Recalling that it represents the energy density of the
gauge fields in the unit of $H^2$, the power spectrum is inversely proportional 
to the energy density of the background gauge fields instead of the background scalar kinetic energy. 
We emphasise that the only approximation we used to derive (\ref{eq:powerresult}) was the
matching with de-Sitter mode functions. Hence, we expect the expression is valid even
for $\mathcal{I} \gtrsim 1$ in the superhorizon limit. 

Comparing the expression (\ref{eq:powerresult}) with the perturbative result 
(\ref{eq:powerPerturbative}), one can estimate the time when the power spectrum settles
down to a constant value after horizon exit. We simply equate these two in the limit of small 
$\mathcal{I}$ and infer that
\begin{equation}
N_k \sim \mathcal{I}^{-2} . \label{eq:saturatevalue}
\end{equation}
Beyond this point, $\langle \zeta _k^2 \rangle $ is conserved as it is in the usual adiabatic
perturbation.

\subsection{Estimating the superhorizon contribution to the late-time bispectrum}
Under the general conditions (\ref{eq:con1}) - (\ref{eq:con3}) arising from the requirement
of canonical commutation relations, one can show that the tree-level amplitude of the
three-point function is convergent in the superhorizon limit too.

First of all, let us introduce the mode functions for $\zeta $ by
\begin{equation}
\zeta _k^a (\eta ) = -\frac{\eta }{3\epsilon _H} \sqrt{\frac{\epsilon _{\varphi }}{2}} \left( \pi _k^{a\prime } + \frac{3(1+2\mathcal{I}^2)}{\eta } \pi ^a_k - \frac{3}{\sqrt{2}}H\mathcal{I}\eta ^3 \alpha ^{a\prime }_k \right) \ ,
\end{equation}
whose superhorizon limit becomes
\begin{equation}
\zeta _k^a \rightarrow \frac{H\sqrt{\epsilon _{\varphi }}}{6\epsilon _H \mathcal{I}} \left[ 3 (1+ \mathcal{I}^2 ) c_0^a (k)+ (p_+ + 3) c_+^a (k) \left( -k\eta \right) ^{p_+} + (p_- + 3 ) c_-^a (k) \left( -k\eta \right) ^{p_-} \right] \ .
\end{equation}
We restored the $k$-dependence of the coefficients.
Now the tree-level amplitude does not involve any multiple integrals and we derive
\begin{eqnarray*}
\langle \zeta _k^3 \rangle &=& i \int ^{\eta } d\eta _1 \langle \left[ H^A_I (\eta _1 ) + H_I^B (\eta _1 ) + H_I^C (\eta _1 ) , \zeta _k^3 (\eta ) \right] \rangle \\
&=& - \sqrt{\frac{2}{\epsilon _{\varphi }}} \frac{48 \mathcal{I}^2}{H^2} \int \frac{d\eta _1}{\eta _1^4} \Im \left( \zeta _{k_1}^{a\ast } (\eta ) \zeta _{k_2}^{b\ast } (\eta ) \zeta _{k_3}^{c\ast }(\eta ) \pi ^a_{k_1}(\eta _1 ) \pi ^a_{k_2}(\eta _1 ) \pi ^a_{k_3} (\eta _1 )\right) \\
&& + \frac{48\mathcal{I}}{\sqrt{\epsilon _{\varphi }}H} \int d\eta _1 \left[ \Im \left( \zeta _{k_1}^{a\ast } (\eta ) \zeta _{k_2}^{b\ast }(\eta ) \zeta _{k_3}^{c\ast }(\eta ) \pi _{k_1}^a (\eta _1 ) \pi _{k_2}^b (\eta _1 )\alpha _{k_3}^{c\prime } (\eta _1 ) \right) + 2 \ {\rm prems} \right] \\
&& - 12 \sqrt{\frac{2}{\epsilon _{\varphi }}} \int d\eta _1 \eta _1^4 \left[ \Im \left( \zeta _{k_1}^{a\ast }(\eta ) \zeta _{k_2}^{b\ast } (\eta ) \zeta _{k_3}^{c\ast } (\eta ) \pi _{k_1}^a (\eta _1) \alpha _{k_2}^{b\prime } (\eta _1 ) \alpha _{k_3}^{c\prime } (\eta _1 ) \right) + 2 \ {\rm perms } \right] \ .
\end{eqnarray*}
Note that
\begin{align}
\begin{split}
- \sqrt{\frac{2}{\epsilon _{\varphi }}} \frac{6\epsilon _H \mathcal{I}^2}{H^2} \zeta _k^{a\ast }(\eta ) \pi _k^a (\eta _1) =& 3(1+\mathcal{I}^2 ) \left[ |c_0^a (k) |^2 + c_0^{a\ast }(k) c_+^a (k) (-k\eta _1 )^{p_+} + c_0^{a\ast }(k) c_-^a (k) (-k\eta _1 )^{p_-} \right]  \\
& + (p_+ + 3) \left[ c_0^a (k) c_+^{a\ast }(k) (-k\eta )^{p_+} + |c_+^a (k) |^2 (-k\eta )^{p_+} (-k\eta _1 )^{p_+} \right]  \\
& + (p_- + 3) \left[ c_0^a (k) c_-^{a\ast }(k) (-k\eta )^{p_-} + |c_-^a (k) |^2 (-k\eta )^{p_-} (-k\eta _1 )^{p_-} \right]  \\
& + (p_- +3) c_+^a (k) c_-^{a\ast }(k) (-k\eta )^{p_-}(-k\eta _1 )^{p_+}  \\
& + (p_+ +3 ) c_+^{a\ast }(k) c_-^a (k) (-k\eta )^{p_+} (-k\eta _1 )^{p_-} \ .
\end{split}\notag
\end{align}
Because of the conditions (\ref{eq:con3}), we have
\begin{align}
\begin{split}
& \Im \left( \zeta _k^{a\ast }(\eta ) \pi _k^a (\eta _1 ) \right) =  - \sqrt{\frac{\epsilon _{\varphi }}{2}} \frac{H^2}{6\epsilon _H \mathcal{I}^2} \Im \left( c_+^a (k) c_-^{a\ast }(k) \right) \\
& \qquad \qquad \times \left[ (p_- +3 )(-k\eta )^{p_-}(-k\eta _1 )^{p_+}  - (p_+ +3 )(-k\eta )^{p_+}(-k\eta _1 )^{p_-}  \right] \ .
\end{split}
\end{align}
Thus, we see that the lowest power of the integrand for the first term must come from
\begin{equation*}
\frac{1}{\eta _1^4} \Im \left( \zeta _{k_1}^{a\ast }(\eta ) u_{k_1}^a (\eta _1 ) \right) \Re \left( \zeta _{k_2}^{b\ast }(\eta )  u_{k_2}^b (\eta _1 ) \right) \Re \left( \zeta _{k_3}^{c\ast }(\eta ) u_{k_3}^c (\eta _1 ) \right) + 2 \ {\rm perms} .
\end{equation*}
The time dependence of its dominant contribution is given by 
\begin{equation*}
\eta ^{p_-} \eta _1^{p_+ -4} \quad {\rm or} \quad \eta ^{p_+} \eta _1^{p_- -4} \ ,
\end{equation*}
both of which have the total power of $-1$ and contain a positive power of $\eta $,
which implies the integration in the limit $-\eta \rightarrow 0 $ is convergent. The bispectrum
generated by $\pi ^3$ vertex long after horizon exit is therefore 
\begin{equation}
i\int ^{\eta }d\eta _1 \langle \left[ H_I^A (\eta _1 ) , \zeta _k^3 (\eta ) \right] \rangle \sim  \frac{\epsilon _{\varphi }H^4 (1+\mathcal{I}^2 )^2}{4\epsilon _H^3 \mathcal{I}^4} \left( | c_0^a (k_2) |^2 |c_0^a (k_3 ) |^2 + 2 \ {\rm perms } \right) \ .
\end{equation}
Similarly,
using 
\begin{align}
\begin{split}
-\frac{6\epsilon _H \mathcal{I}\eta _1^4}{H\sqrt{\epsilon _{\varphi }}} \zeta _k^{a\ast }(\eta ) \alpha _k^{a\prime } (\eta _1 ) = & 3(1+\mathcal{I}^2 ) \left[ |c_0^a (k)|^2 + 2c_0^{a\ast }(k) c_+^a (k) (-k\eta _1 )^{p_+} + 2c_0^{a\ast }(k) c_-^a (k) (-k\eta _1 )^{p_-} \right] \\
& + (p_+ +3) \left[ c_0^a (k) c_+^{a\ast }(k) (-k\eta )^{p_+} + 2|c_+^a (k)|^2 (-k\eta )^{p_+} (-k\eta _1 )^{p_+} \right]  \\
& + (p_- +3) \left[ c_0^a (k) c_-^{a\ast }(k) (-k\eta )^{p_-} + 2|c_-^a (k)|^2 (-k\eta )^{p_-} (-k\eta _1 )^{p_-} \right] \\
& +2(p_- +3) c_+^a (k) c_-^{a\ast }(k) (-k\eta )^{p_-} (-k\eta _1 )^{p_+} \\
& + 2(p_+ + 3) c_+^{a\ast }(k) c_-^a (k) (-k\eta )^{p_+} (-k\eta _1 )^{p_-} \ , 
\end{split}\notag
\end{align}
one can show that the second and third integrals give convergent results as
\begin{equation}
i\int ^{\eta } d\eta _1 \langle \left[ H_I^B (\eta _1 ) , \zeta _k^3 (\eta ) \right] \rangle \sim - \frac{\epsilon _{\varphi }H^4 (1+\mathcal{I}^2 )^2}{\epsilon _H^3 \mathcal{I}^4} \left( |c_0^a (k_2 )|^2 |c_0^a (k_3 )|^2 + 2 \ {\rm perms } \right) 
\end{equation}
and
\begin{equation}
i\int ^{\eta } d\eta _1 \langle \left[ H_I^C (\eta _1 ) , \zeta _k^3 (\eta ) \right] \rangle \sim \frac{5\epsilon _{\varphi } H^4 (1+\mathcal{I}^2 )^2 }{8\epsilon _H^3 \mathcal{I}^4} \left( |c_0^a (k_2 )|^2 |c_0^a (k_3 )|^2 + 2 \ {\rm perms } \right) 
\end{equation}
respectively. In the end, the late-time contribution to the three-point function becomes
\begin{equation}
\langle \zeta _k^3 \rangle \sim -\frac{\epsilon _{\varphi }H^4 (1 +\mathcal{I}^2 )^2}{8\epsilon _H^3 \mathcal{I}^4} \left( |c_0^a (k_1 )|^2 |c_0^a (k_2) |^2 + |c_0^a (k_2 ) |^2 |c_0^a (k_3) |^2 + |c_0^a (k_3 ) |^2 |c_0^a (k_1 ) |^2 \right) \ .
\end{equation}
 
Assuming the final bispectrum is dominated by the superhorizon contribution, which appears
to be the case in the evidence of the numerical study in the next section, one can now
estimate the final value of $f_{NL}$ in the squeezed limit $k_1 \ll k_2 \sim k_3$. Note that
the dependence of $c_0^a (k)$ on $k$ derived by matching is rather generic. Then, in this limit,
we have
\begin{equation}
\langle \zeta _k^3 \rangle \rightarrow - \frac{\epsilon _{\varphi }H^4 (1+\mathcal{I}^2 )^2}{4\epsilon _H^3 \mathcal{I}^4} |c_0^a (k_1 )|^2 |c_0^a (k_2 )|^2 \ .
\end{equation}
Combined with (\ref{eq:powerresult}), the appropriately normalised $f_{NL}$ is computed as
\begin{equation}
f_{NL} \rightarrow  -\frac{5}{3}\frac{\epsilon _{\varphi }}{\epsilon _H} \rightarrow -\frac{5}{3}  \ ,
\end{equation}
where  the last limit was taken for $\mathcal{I}\rightarrow 0 \Leftrightarrow \epsilon _{\varphi } \rightarrow \epsilon _H $. This beautiful result will be confirmed in the following section.

\section{Numerical calculation of exact tree-level amplitude}

Following from the previous section, here we treat the quadratic vertices non-pertubatively 
with the only difference being that now we calculate most of the contributions numerically. 
The aim is to negate the need for making any approximations and therefore to make 
our result more quantatively accurate. In the analytic results from section 4, we derived
the qualitative features of power spectrum and bispectrum by assuming that at horizon crossing 
the mode functions are those of the free de-Sitter case and applying the superhorizon 
approximation ($k\eta=0$) as soon as the mode crosses the horizon $(k\eta = -1$). 
We have been able to estimate
the final amplitude of power spectrum, the time of transition from the perturbative regime
discussed in section 3 to the one dictated by the superhorizon mode functions, and calculate
the superhorizon contribution to the bispectrum in the limit $k\eta \rightarrow 0$. 
However, we are yet to have a reliable estimate for the time evolution (or equivalently scale dependence)
of the bispectrum.

Here, we calculate the exact mode function, first setting the $\pi$ and $\alpha$ fields in the 
Bunch-Davies vacuum deep inside the horizon, solve the coupled linear equations of motion
numerically until the modes are far into the superhorizon regime. At this point we switch to 
using the superhorizon equations of motion and use the analytic solution - this is simply to 
avoid numerical instabilities encountered in this calculation. Now we only use the analytic 
superhorizon solution for $-k\eta\ll 1$, so the error introduced by doing so is negligible. 

Ultimately, we will be interested in the value of $f_{NL}$ (squeezed) here. 
Factors of $f_0H$ present in the Lagrangian (\ref{eq:lagrangian}) will be absorbed into 
the definition of the $\alpha$ field, and the overall multiplicative factor of $H^{-2}$ in 
front of the Lagrangian will not affect the value of $f_{NL}$. We therefore set $H=1$; for 
quantities such as the power spectrum or bispectrum, reintroducing $H$ will be a matter
of an overall multiplicative factor which will be included in the plots. When reintroducing $H$, 
$\mathcal{I}$ will need to be replaced with $\mathcal{I}/H$.

This leaves the factors of $\epsilon _H$ and $\epsilon _{\varphi }$ in the 
Lagrangian and the definition of the curvature perturbation; in the numerical calculation below, 
they will be set to 1. It can be easily seen that these two parameters can be reintroduced 
at the end as an overall multiplicative factor of $\epsilon _H /\epsilon _{\varphi }=1+\mathcal{I}^2$ for the value of $f_{NL}$ computed.
The $\zeta$ mode functions, power spectrum and bispectrum will need to be multiplied by
$H(1+\mathcal{I}^2)^{-1}\epsilon_{\varphi }^{-\frac{1}{2}}$, $H^2(1+\mathcal{I}^2)^{-2}\epsilon_{\varphi}^{-1}$ 
and $H^4(1+\mathcal{I}^2)^{-3}\epsilon_{\varphi }^{-2}$
respectively, to restore the dependence on these constants.

\subsection{Subhorizon linear evolution and initial conditions}
\begin{figure}[htpb]
 \begin{center}
\includegraphics[width=0.24\linewidth]{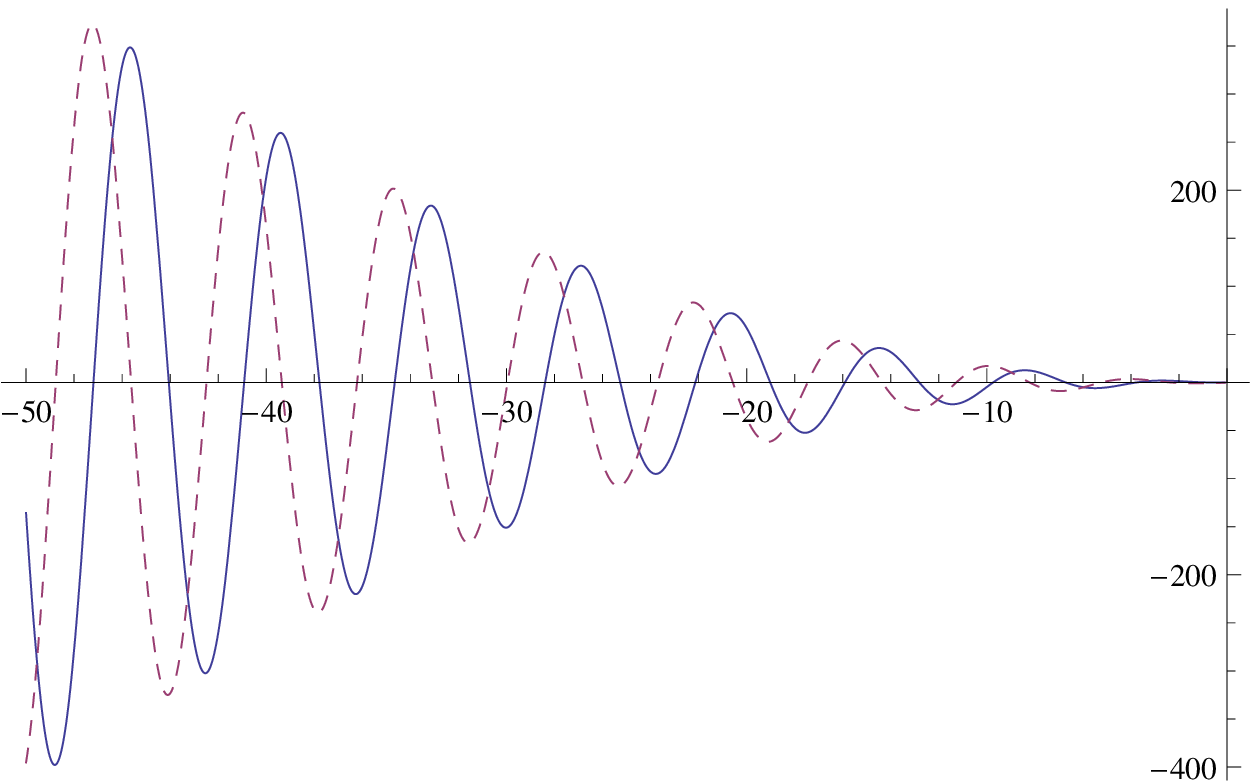}
\includegraphics[width=0.24\linewidth]{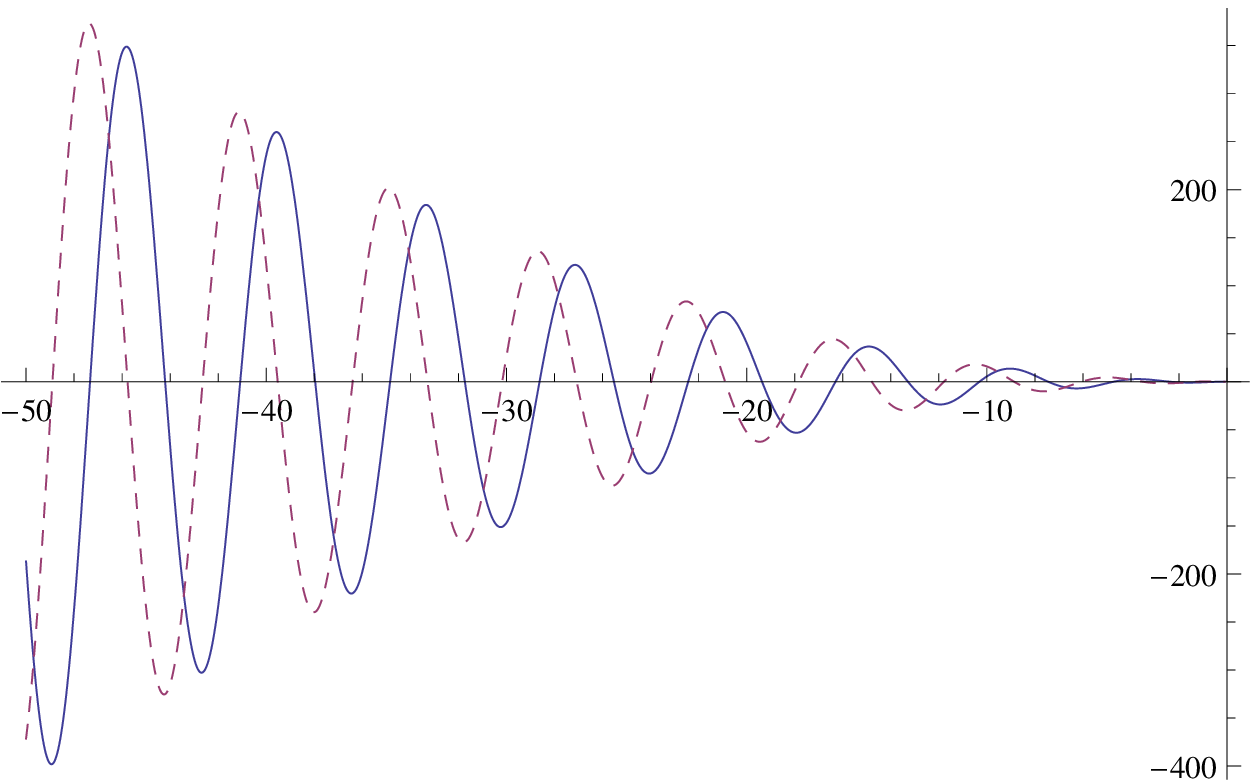}
\includegraphics[width=0.24\linewidth]{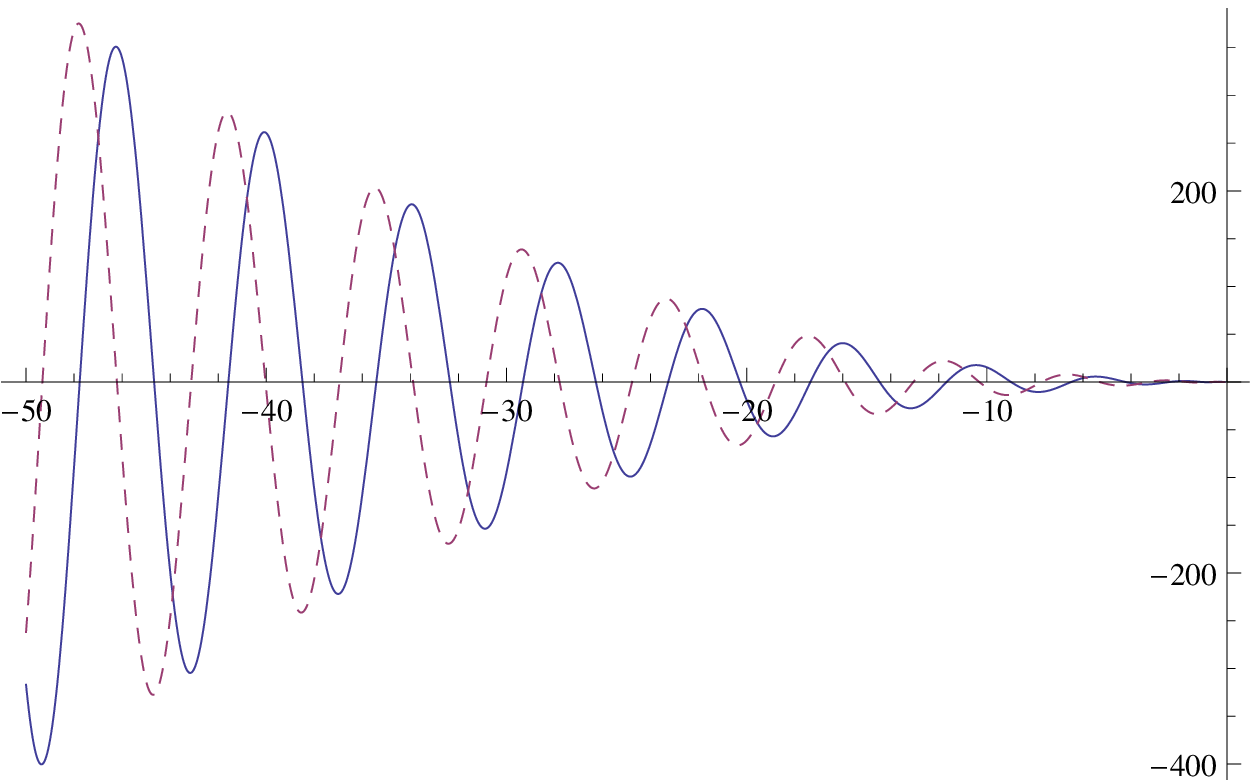}
\includegraphics[width=0.24\linewidth]{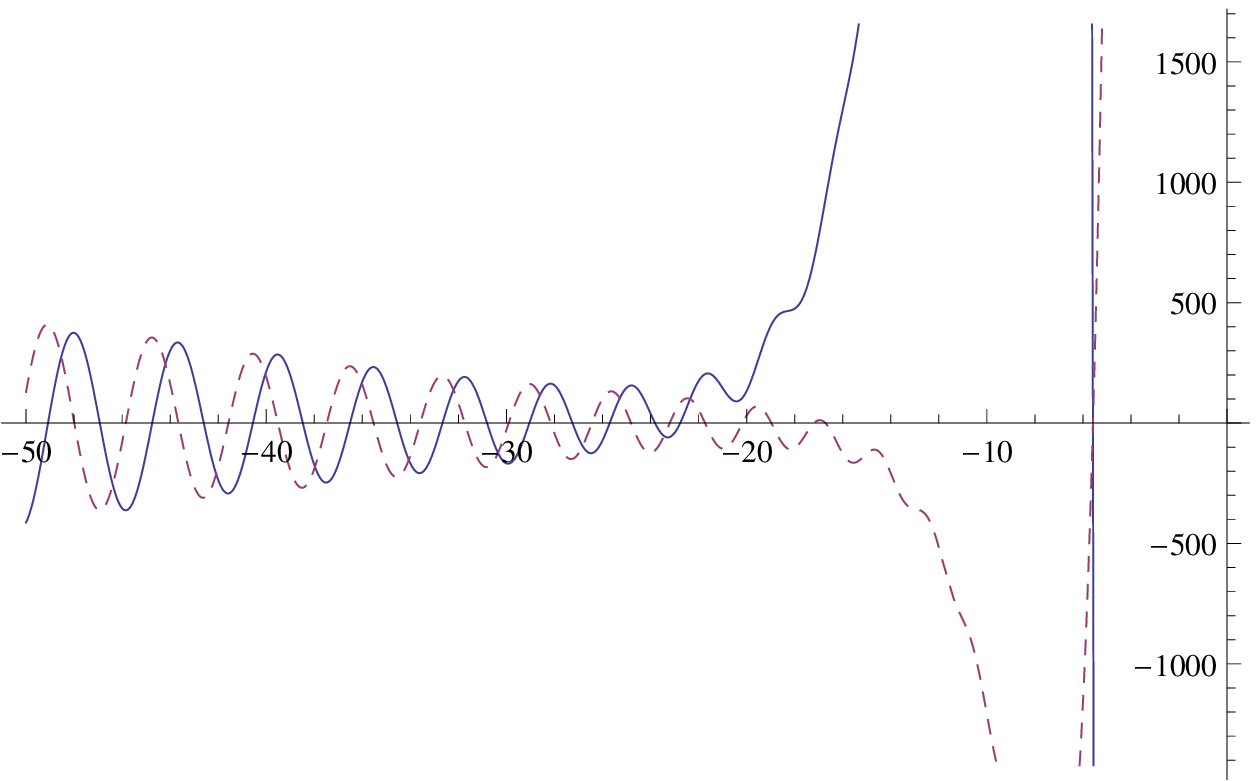}
\caption{Plots of the $\zeta^1_k$ mode functions during subhorizon, with $k\eta$ along 
the $x$-axis. For all plots on this page, the solid line represents the real part and the dashed 
line represents the imaginary part. The different plots are for different values of $\mathcal{I}$; 
from left to right: $\mathcal{I}$ = 0.1, 0.5, 1 and 10 respectively. For smaller values of 
$\mathcal{I}$, one can observe the characteristic oscillation and its damping towards 
horizon exit of de-Sitter mode functions, while the behaviour near the horizon is
significantly different for $\mathcal{I} = 10$.}
\label{fig:zetamf1}
\includegraphics[width=0.24\linewidth]{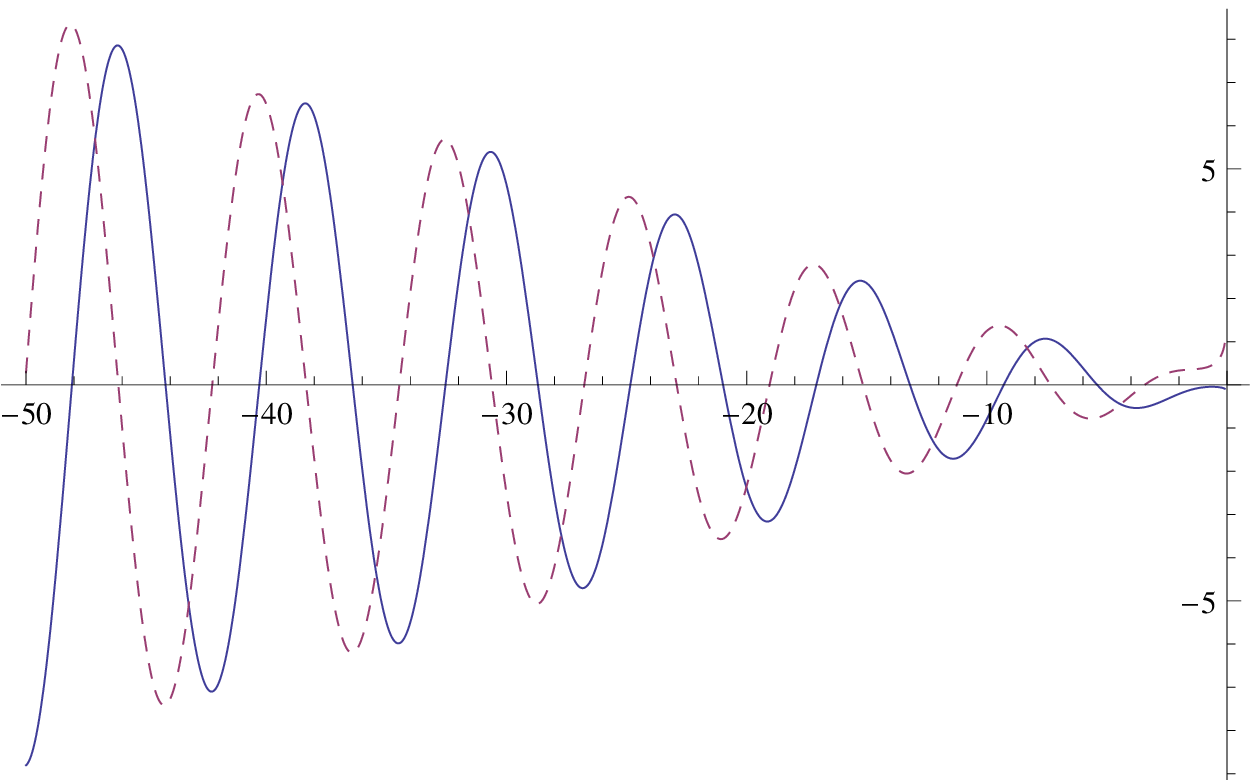}
\includegraphics[width=0.24\linewidth]{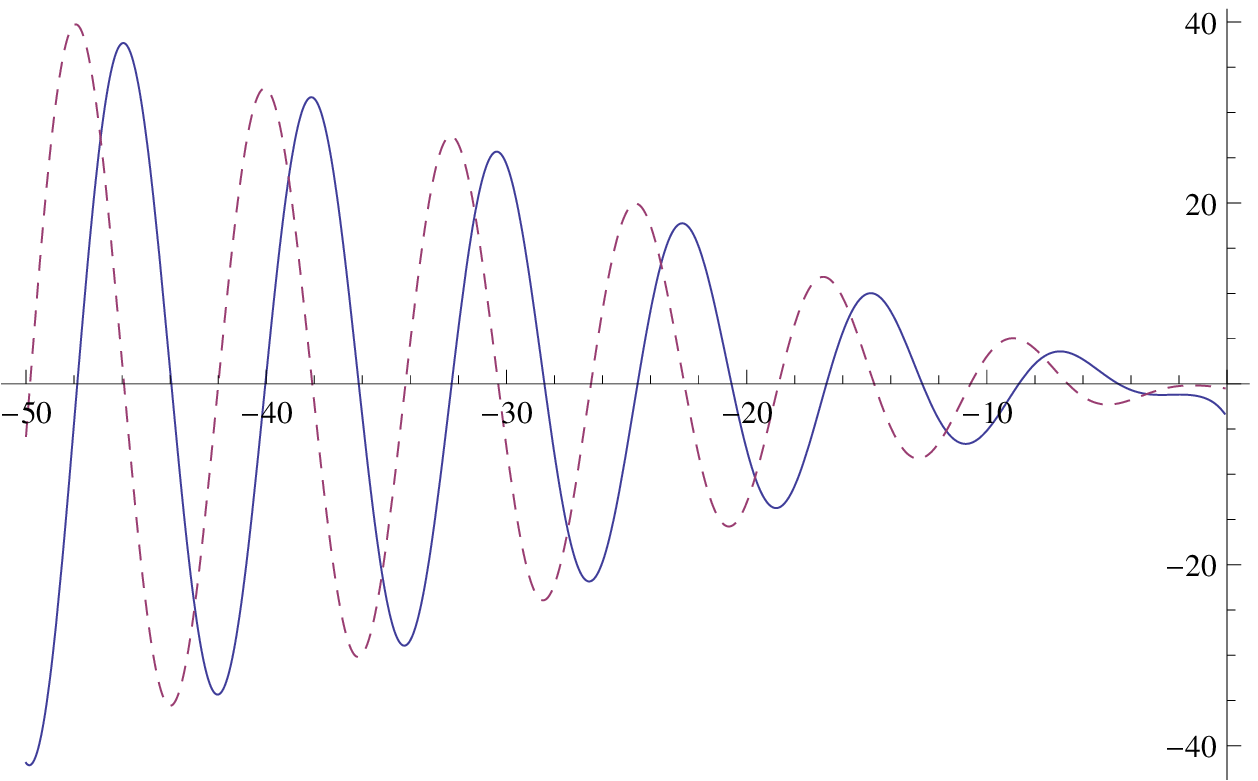}
\includegraphics[width=0.24\linewidth]{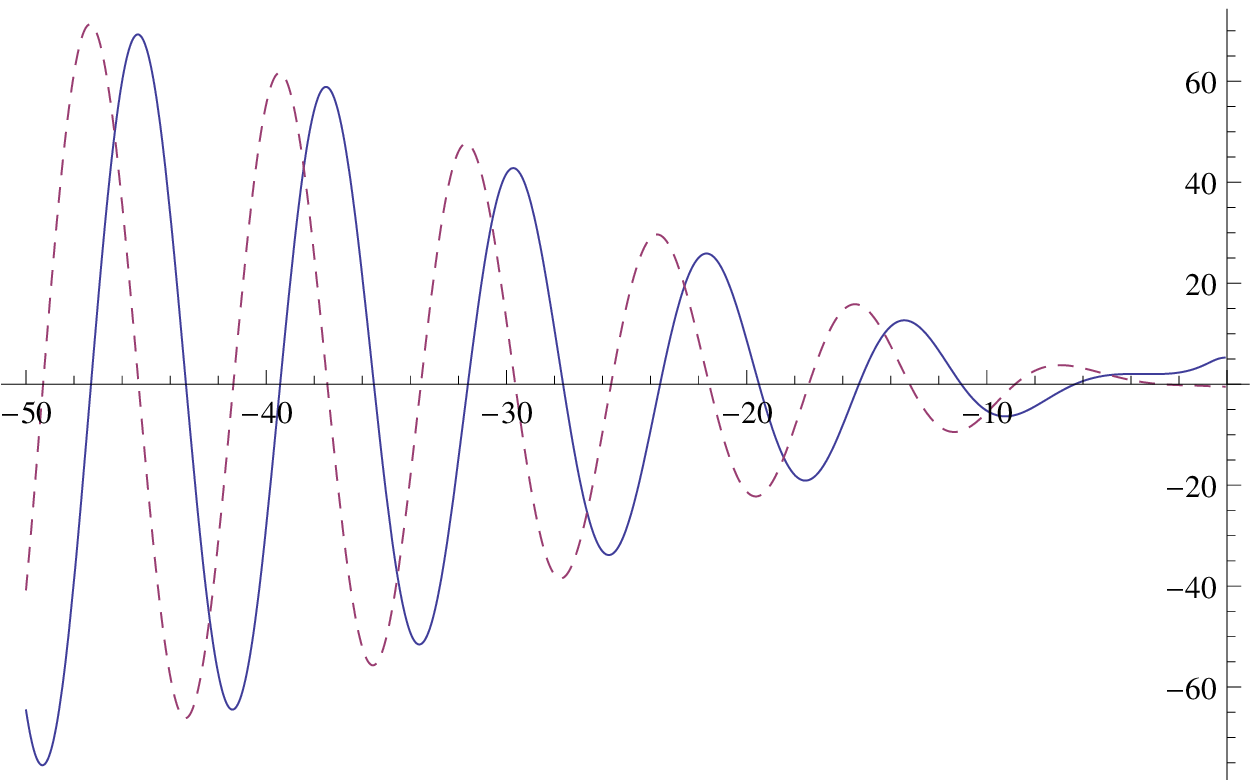}
\includegraphics[width=0.24\linewidth]{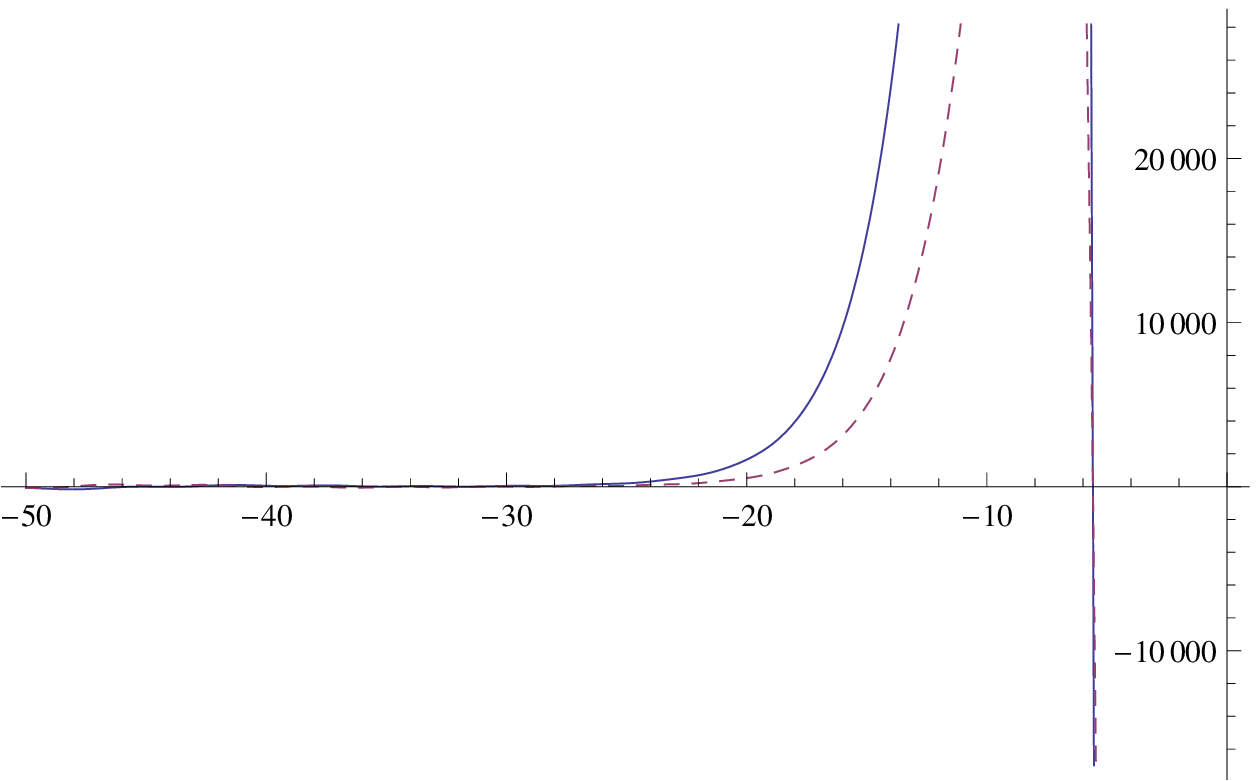}
\caption{Plots of the $\zeta^2_k$ mode functions during subhorizon, with $k\eta$ along the $x$-axis.
The subhorizon dynamics appears to be similar between $\zeta ^1_k $ and $\zeta ^2_k$.} 
\label{fig:zetamf2}
\includegraphics[width=0.24\linewidth]{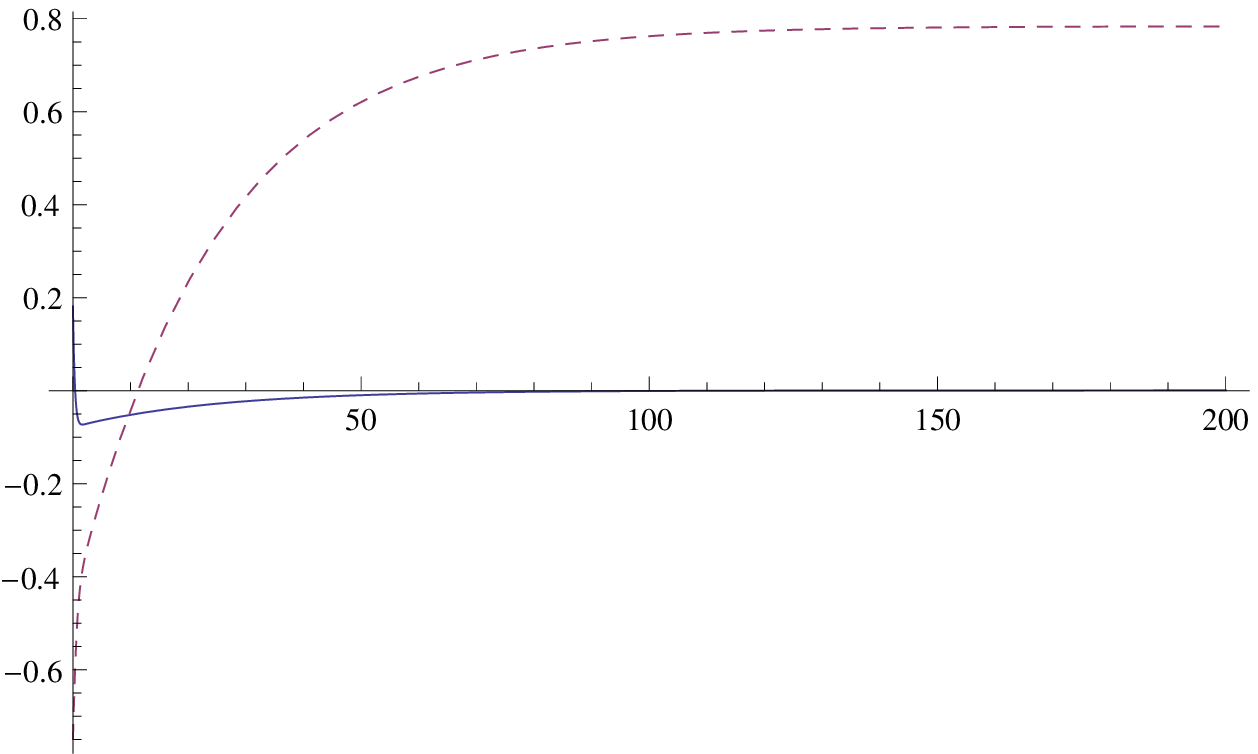}
\includegraphics[width=0.24\linewidth]{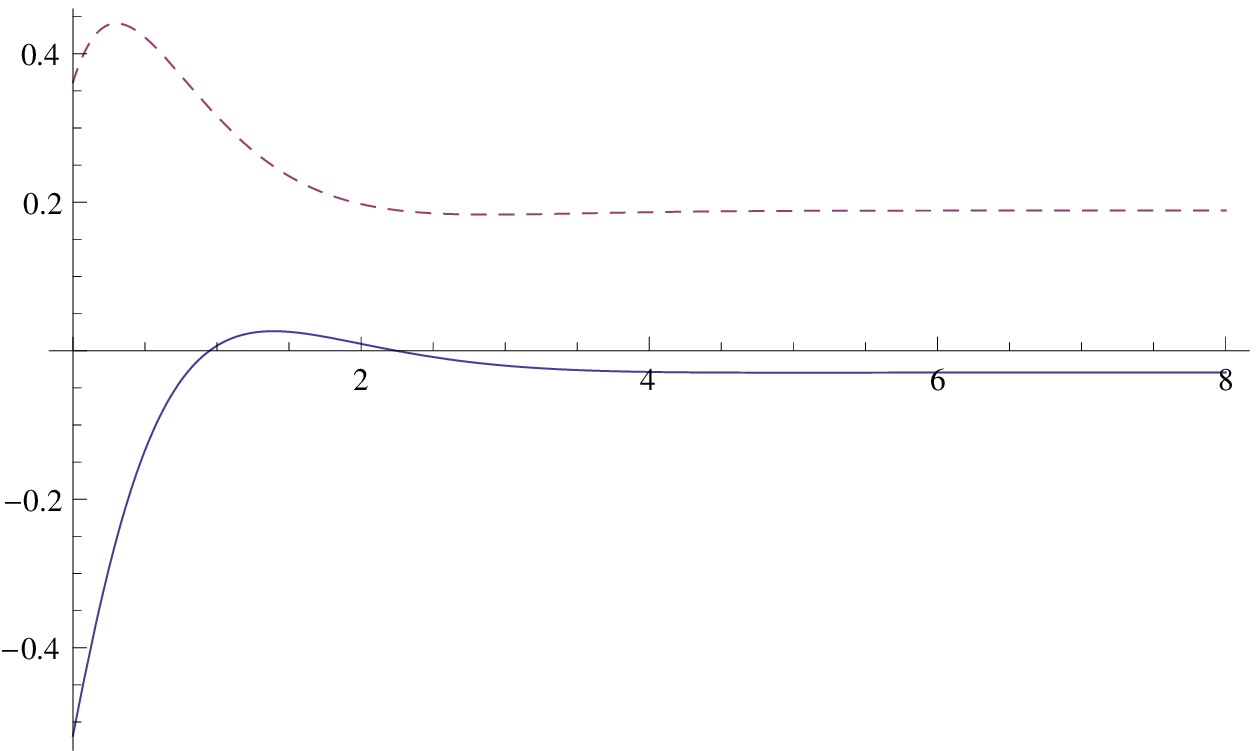}
\includegraphics[width=0.24\linewidth]{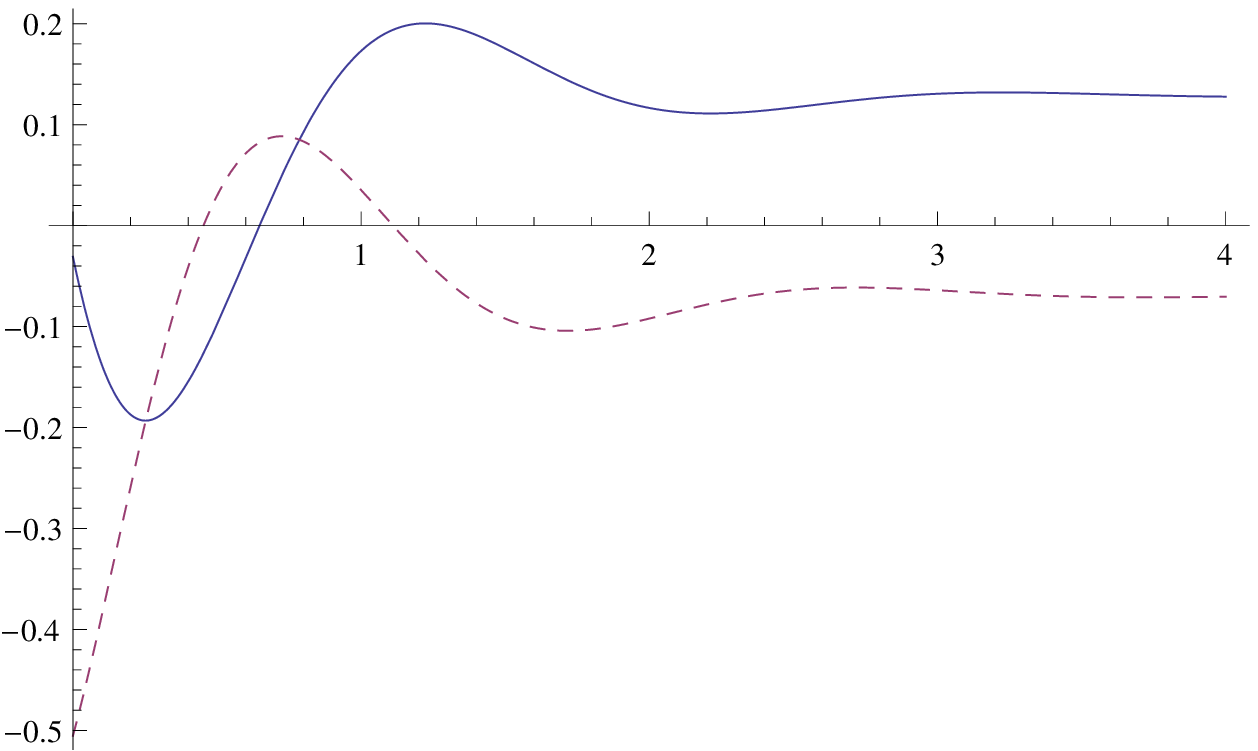}
\includegraphics[width=0.24\linewidth]{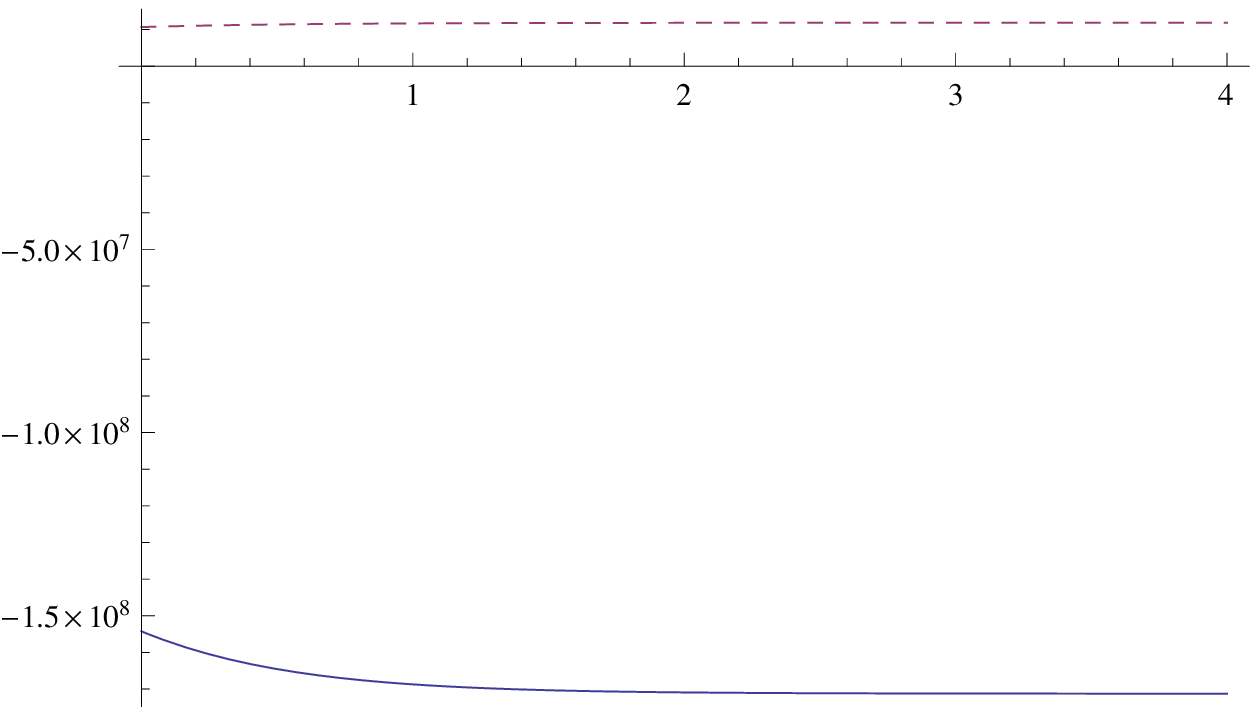}
\caption{Plots of the $\zeta^1_k$ mode functions during superhorizon, with cosmic time 
({\bf $-\ln (-k\eta )$}) along the $x$-axis. While the evolution of individual mode functions 
significantly depends on the value of $\mathcal{I}$, all of them settle down to constant 
in agreement with the analytical results.} 
\label{fig:zetamf3}
\includegraphics[width=0.24\linewidth]{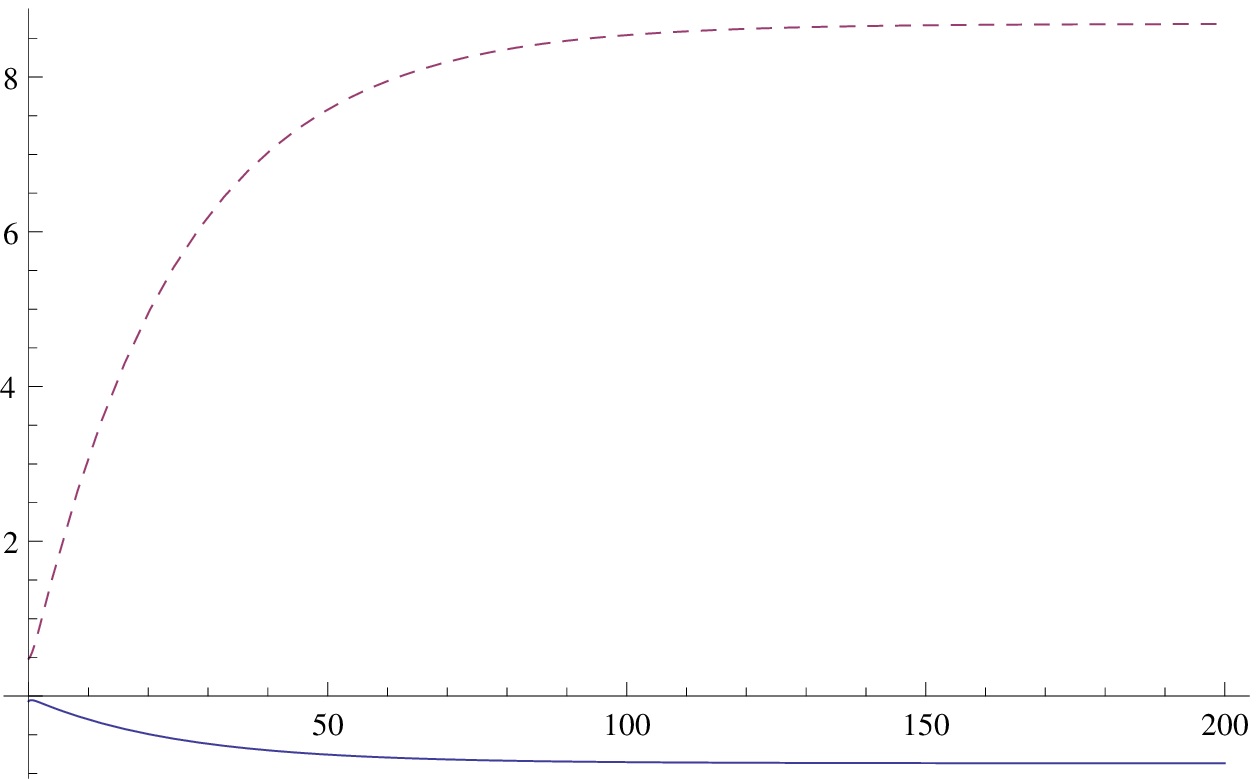}
\includegraphics[width=0.24\linewidth]{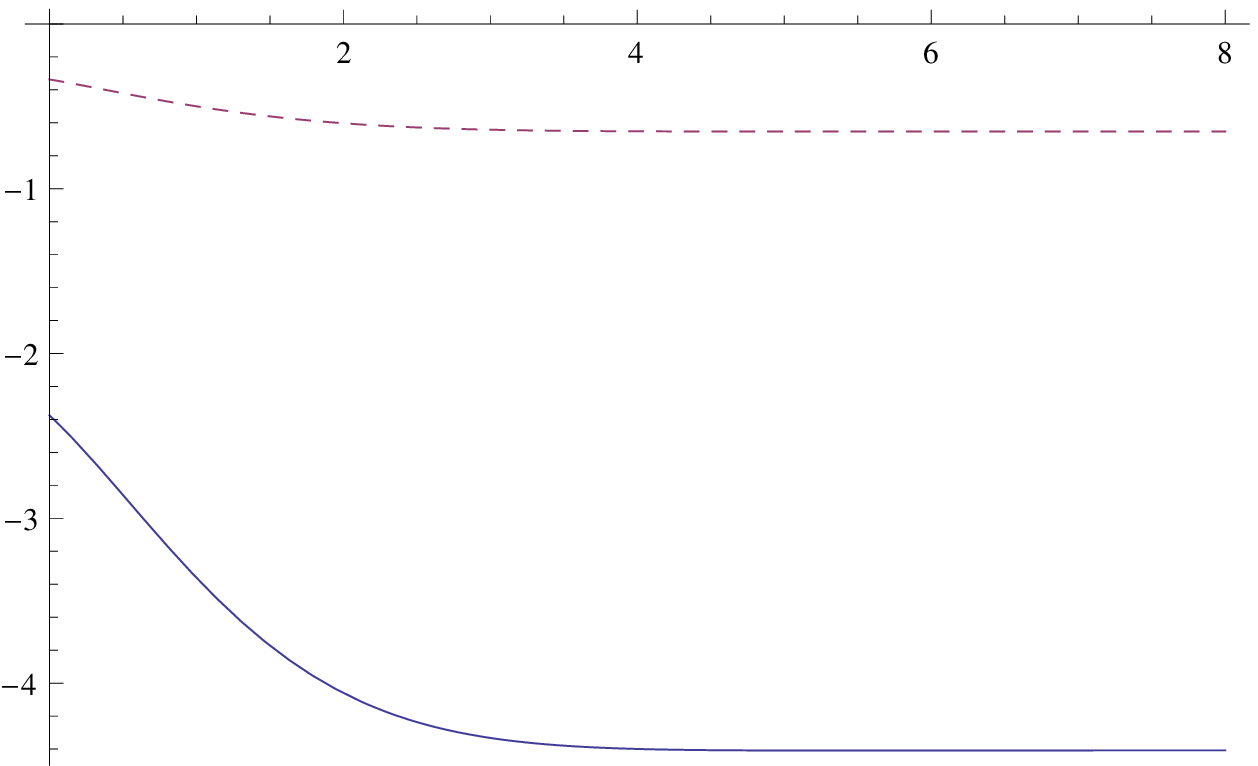}
\includegraphics[width=0.24\linewidth]{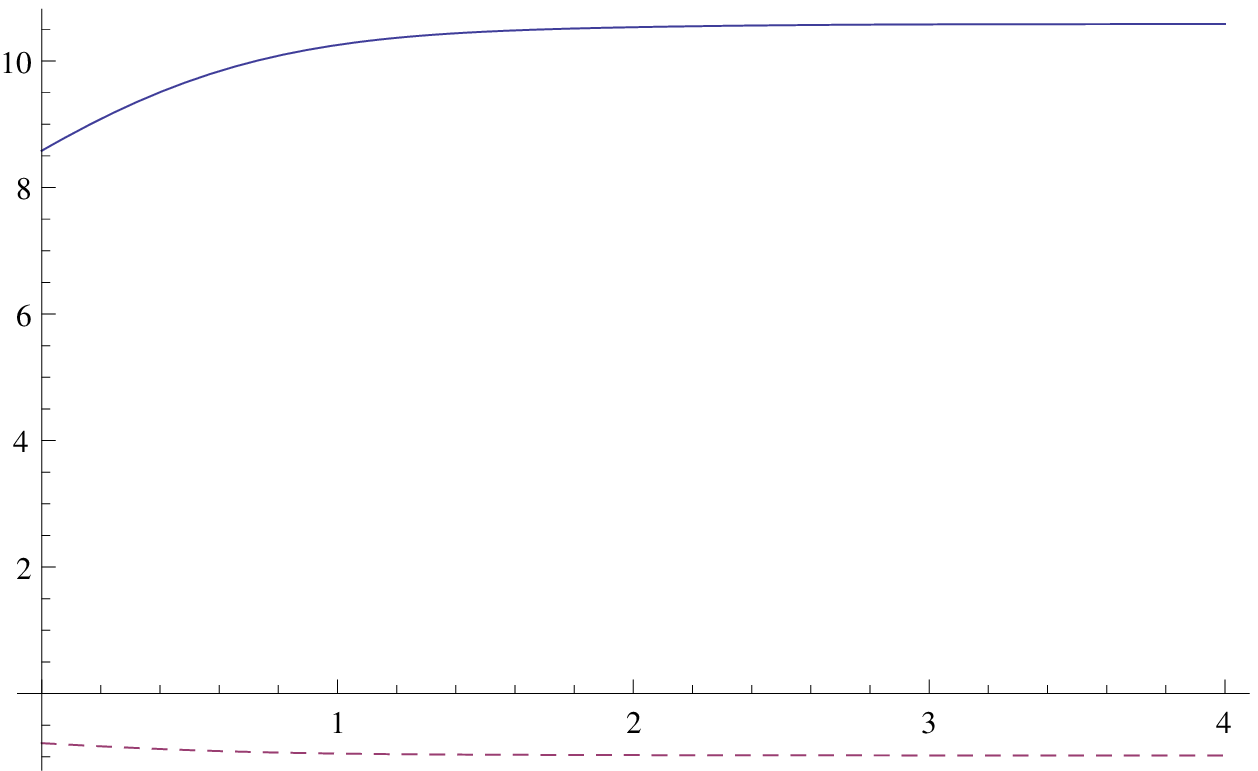}
\includegraphics[width=0.24\linewidth]{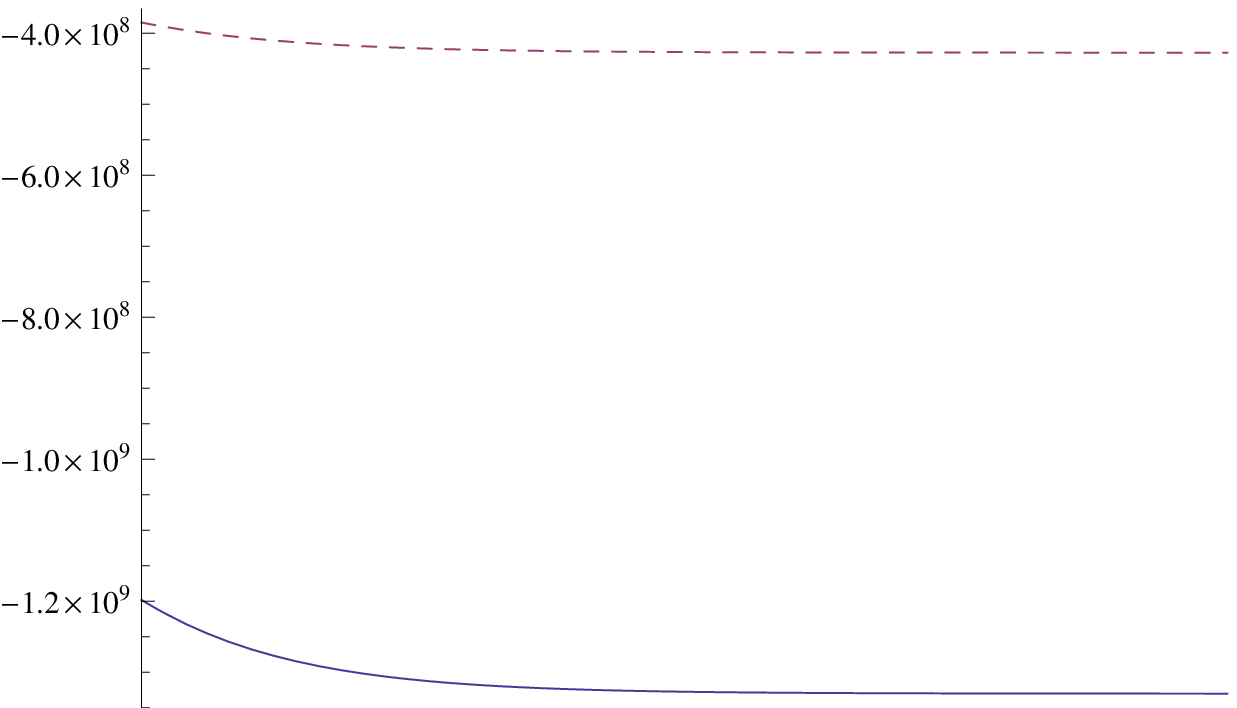}
\caption{Plots of the $\zeta^2_k$ mode functions during subhorizon, with cosmic time along the $x$-axis. On superhorizon scales, $\zeta ^1_k$ and $\zeta ^2_k$ evolve differently.} 
\label{fig:zetamf4}
\end{center}
\end{figure}
Again, the linear equations of motion (\ref{eq:evpi}) and (\ref{eq:evalpha}) are used, 
this time keeping the gradient terms. Since the evolution equations do not admit 
an analytic solution, we will find solutions numerically. For computational convenience, 
the canonical variables used in this section are $\pi$ and $\alpha$, and 
so their conjugate momenta are given by
\begin{eqnarray*}
p_\pi&=&a^{2}\pi'\,,\nonumber\\
p_\alpha&=&3a^{-4}\alpha'-6\sqrt{2}\mathcal{I}\pi\,.\nonumber
\end{eqnarray*}

The initial conditions for the mode functions are given by the Bunch-Davies condition, 
expressed here in terms of the canonical variables for each $k$ mode as:
\begin{eqnarray}
\pi^a&=&\frac{\delta^a_1}{\sqrt{2k^3}}(i-k\eta)e^{-ik\eta}\hbox{,}\quad p^a_\pi \ = \ \delta^a_1a^2\sqrt{\frac{k}{2}}i\eta e^{-ik\eta}\,, \label{eq:initial1} \\
\alpha^a&=&\frac{\delta^a_2}{\sqrt{6c_s^3k^3}}(c_sk\eta^{-2}-i\eta^{-3})e^{-ic_sk\eta}\hbox{,} \label{eq:initial2} \\
 p^a_\alpha &=& \frac{3\delta^a_2}{a^4\sqrt{6c_s^3k^3}}(-ic_s^2k^2\eta^{-2}-3c_sk\eta^{-3}-3i\eta^{-4})e^{-ic_sk\eta}\,. \label{eq:initial3}
\end{eqnarray}
The point here is that these are the conditions required on the mode functions for the fields to be in the Bunch-Davies vacuum deep inside the horizon, and for the canonical commutation relations to hold. Given the definitions of the conjugate momenta, the initial conditions for solving the linear evolution equations will then be given by (\ref{eq:initial1}) and (\ref{eq:initial2}) along with
\begin{equation}
\alpha_k^{a\prime} = \frac{\delta^a_2}{\sqrt{6c_s^3k^3}}(-ic_s^2k^2\eta^{-2}-3c_sk\eta^{-3}-3i\eta^{-4})e^{-ic_sk\eta}+\frac{2\mathcal{I}\delta^a_1}{\sqrt{k^3}}(-k\eta^{-3}+i\eta^{-4})e^{-ik\eta}
\end{equation}
in place of (\ref{eq:initial3}) for some $-k\eta\gg1$. 
For the results in this section, the initial conditions for the modefunctions were set at 
$(\eta)_{\hbox{init}}=-1000$.

\begin{figure}[htbp]
 \begin{center}
 \includegraphics[width=0.24\linewidth]{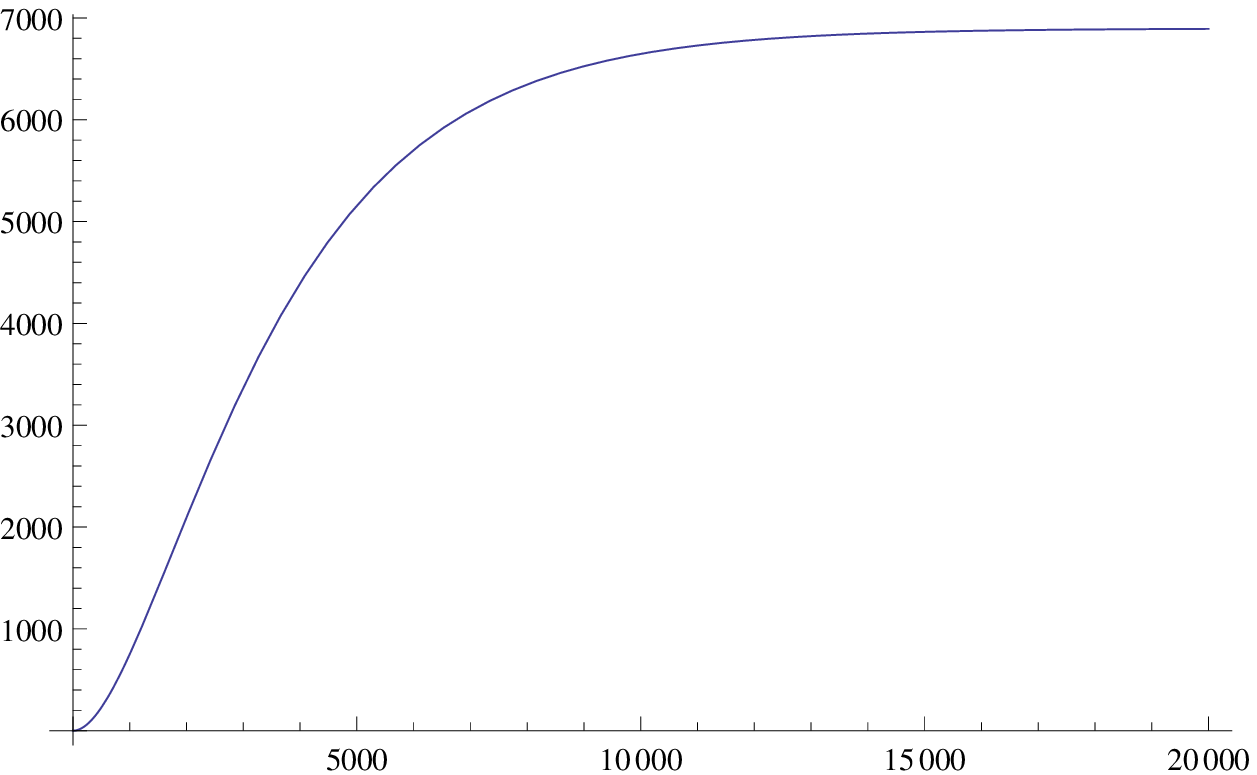}
\includegraphics[width=0.24\linewidth]{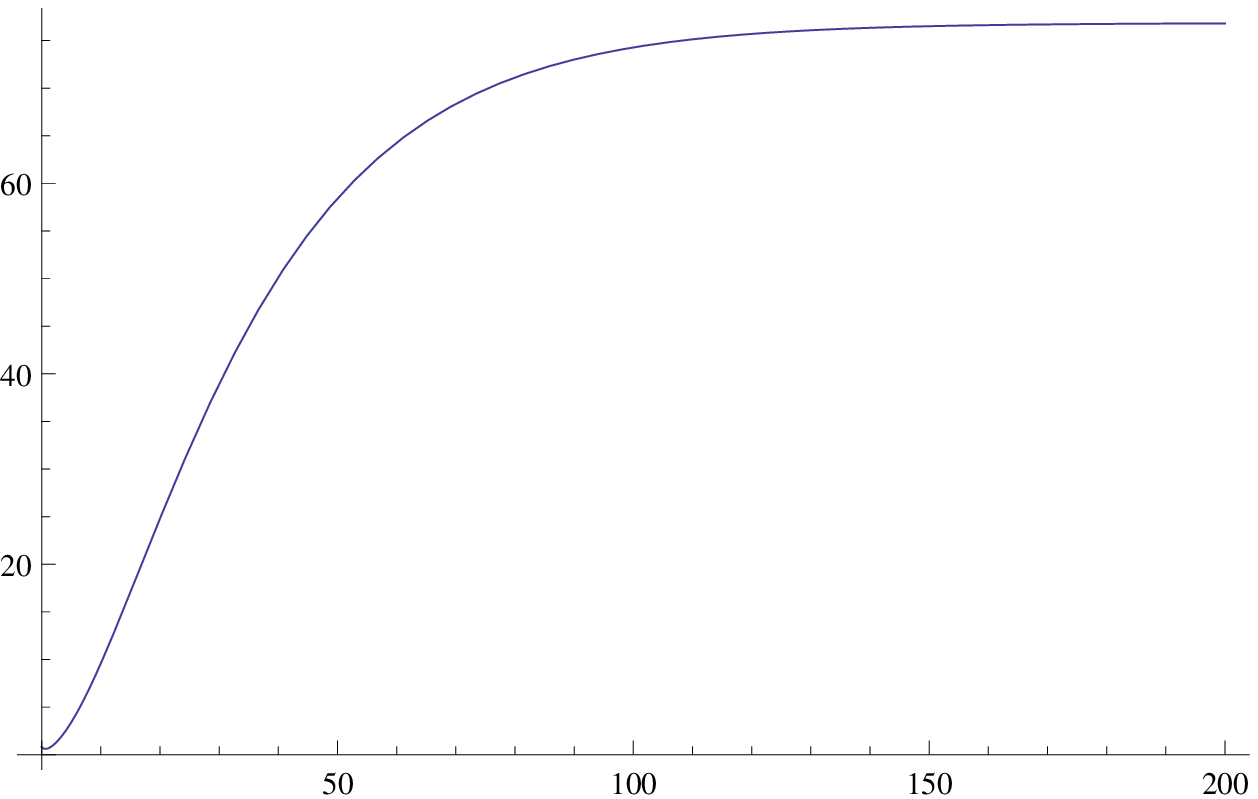}
\includegraphics[width=0.24\linewidth]{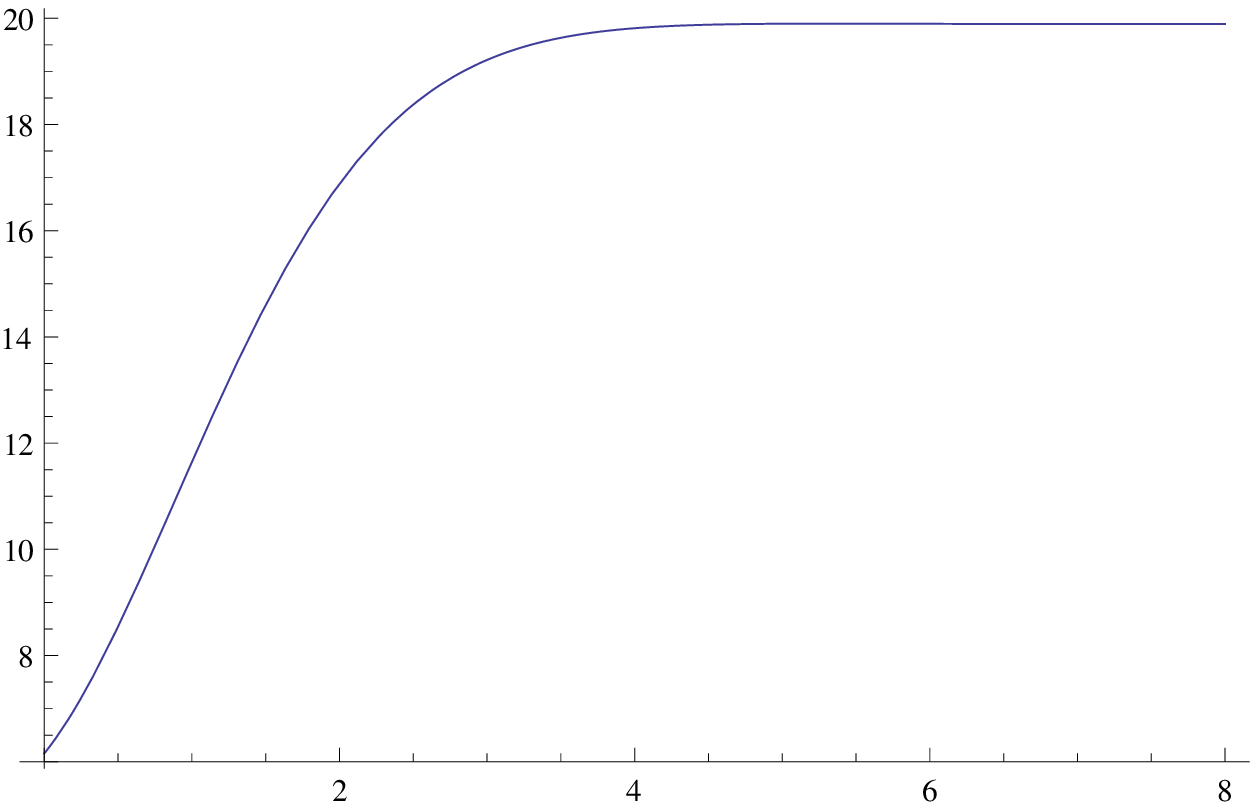}
\includegraphics[width=0.24\linewidth]{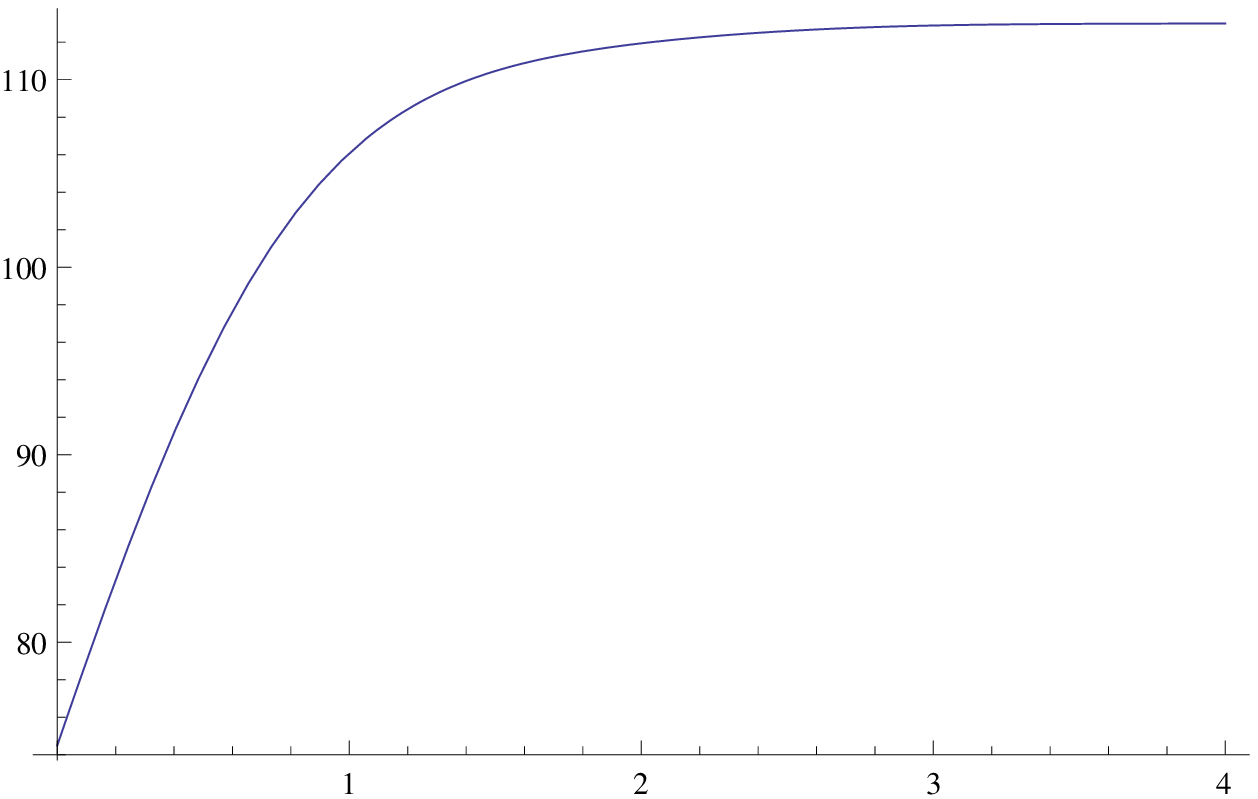}
\includegraphics[width=0.24\linewidth]{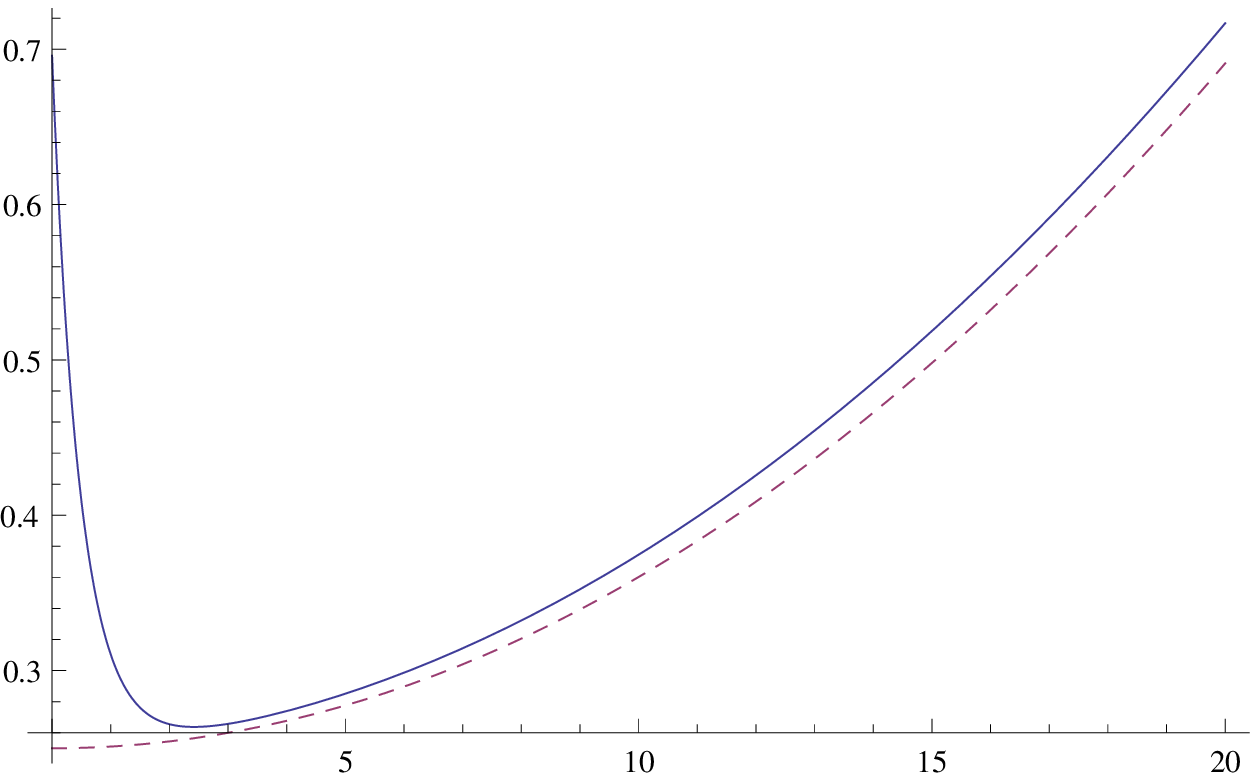}
\includegraphics[width=0.24\linewidth]{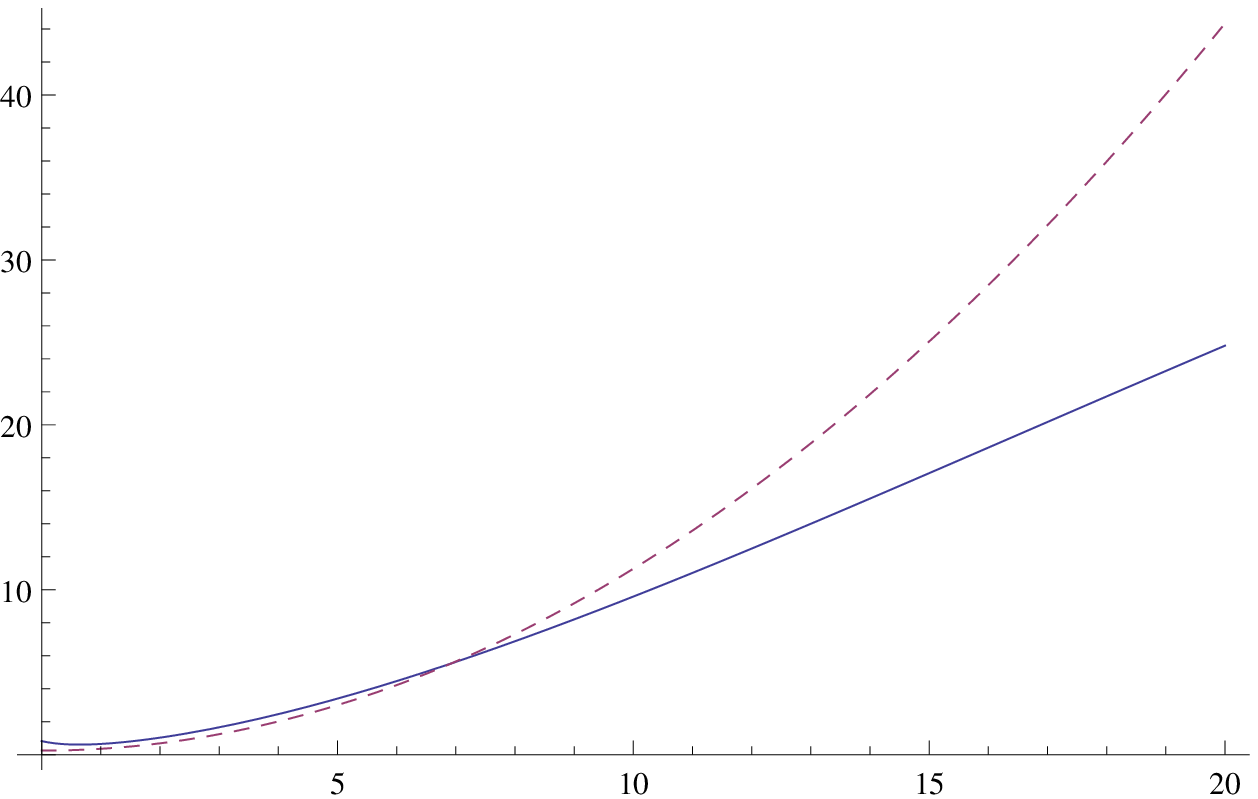}
\includegraphics[width=0.24\linewidth]{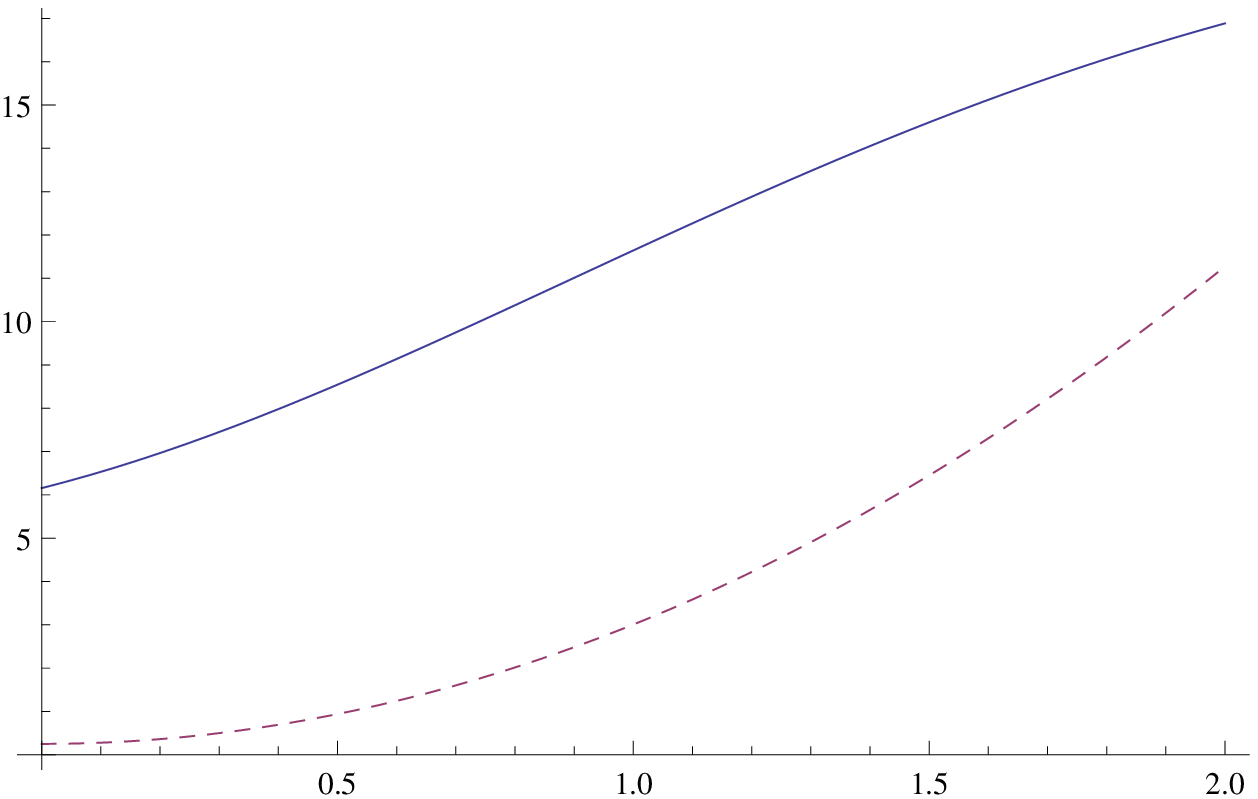}
\includegraphics[width=0.24\linewidth]{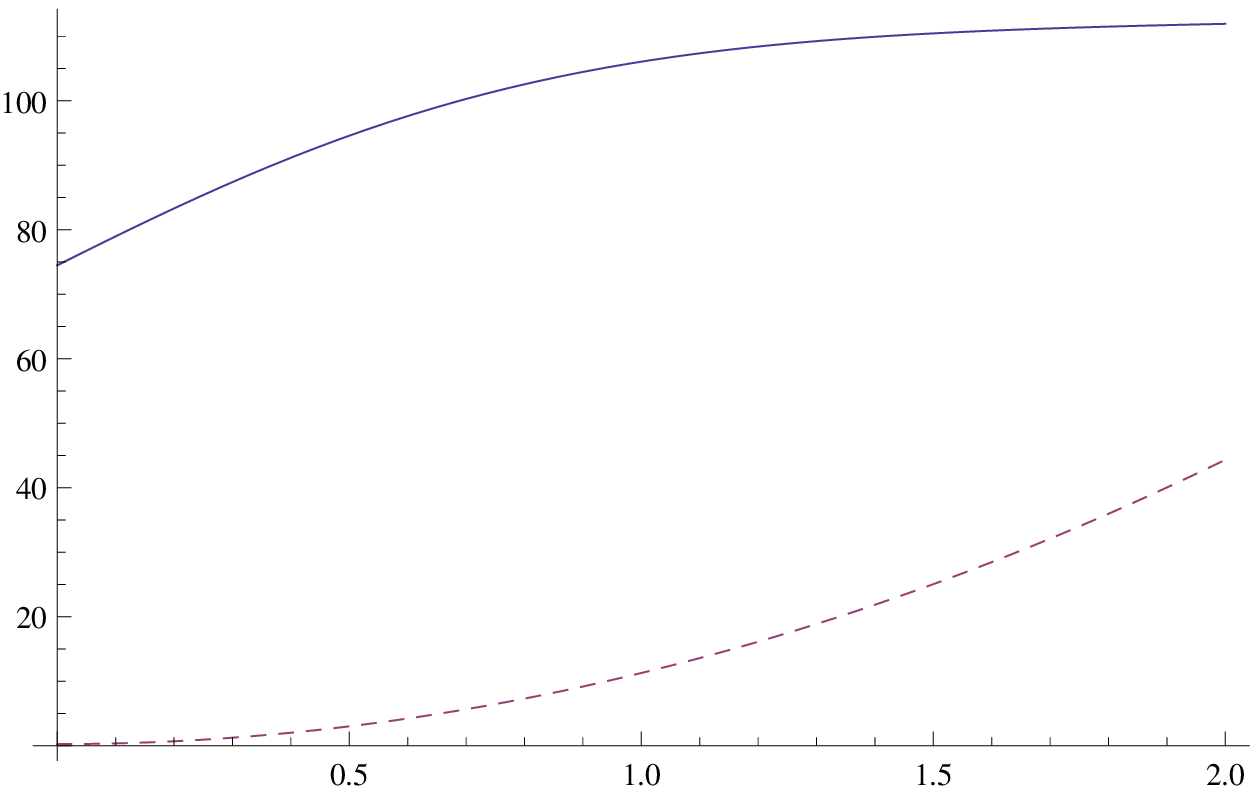}
\caption{Time evolution for the power spectrum $\langle\zeta_k{}^2\rangle$. The first line 
represents the time evolution until the power spectrum becomes constant and the second line 
represents the time evolution for a some time after horizon crossing, both plotted against cosmic time. 
The second line provides a comparison between the perturbative result from 
\cite{Yamamoto2012} and the numerical results from section 5 for the time evolution of 
$\langle\zeta_k{}^2\rangle$, with the numerical results in solid lines and the analytic ones 
in dashed lines. From left to right: $\mathcal{I}$ = 0.01, 0.1, 0.5 and 1. While the analytical results
fit the initial growth fairly well for $\mathcal{I} = 0.01, 0.1$, the discrepancy is significant for
larger $\mathcal{I}$ or late time.}
\label{fig:zac}
\end{center}
\end{figure}
\begin{figure}
\begin{center}
\includegraphics[width=0.5\linewidth]{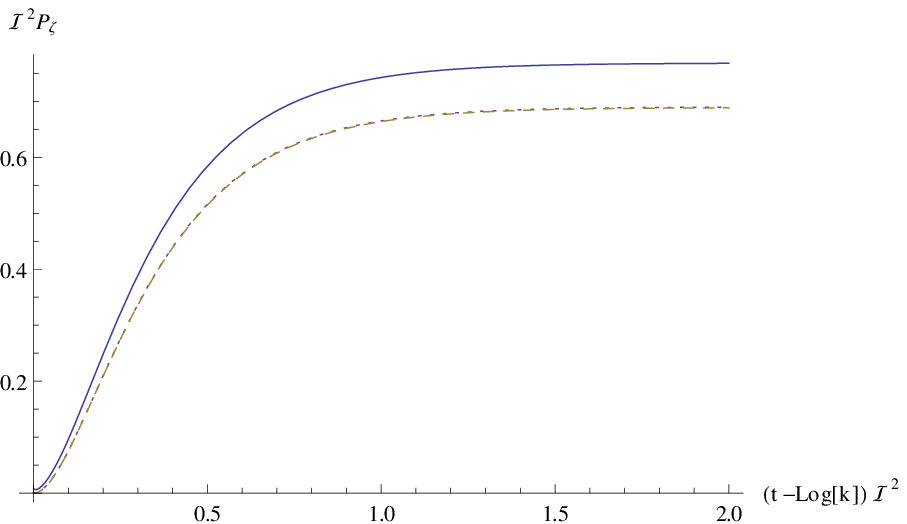}
\caption{Time evolution of $\langle\zeta_k{}^2\rangle$ rescaled according to analytic expectation; the (overlapping) dotted and dashed lines are $\mathcal{I}$ = 0.001 and 0.01, and the solid line is $\mathcal{I}$ = 0.1. The time coordinate has been rescaled by $\mathcal{I}^{2}$ and the power spectrum amplitude by $\mathcal{I}^2$. It clearly indicates the time of transition to the constant
regime $- \ln (-k\eta ) \sim \mathcal{I}^{-2}$.}
\label{fig:ssev}
\includegraphics[width=0.5\linewidth]{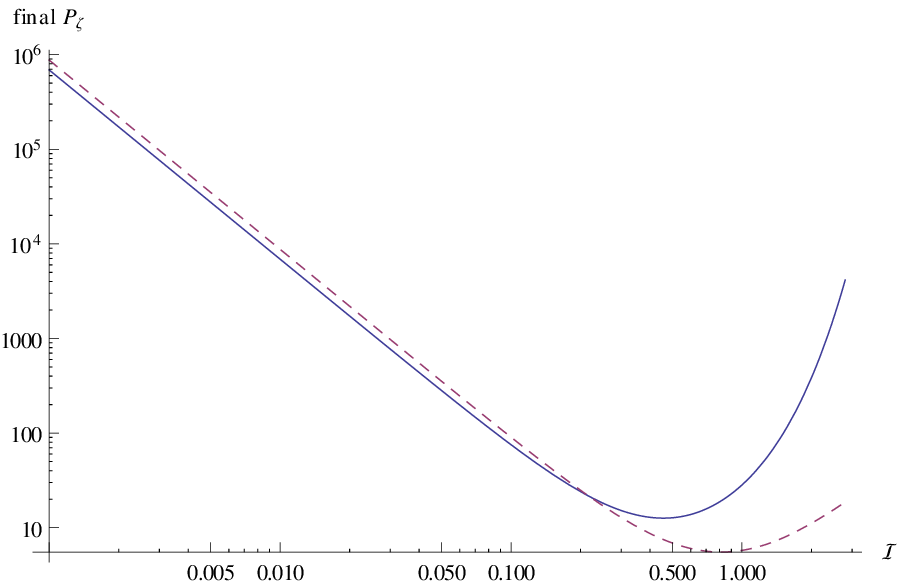}
\caption{Values for the power spectrum $\langle \zeta^2_k \rangle $ after settling down to a constant, plotted as a function of $\mathcal{I}$. The solid line represents the numerical result, and the dotted line represents the analytic expression from section 4. There is a good agreement for $\mathcal{I} \lesssim 0.2$.}
\label{fig:finalps}
\end{center}
\end{figure}
\subsection{Numerical calculation of the $\zeta$ power spectrum}
Now we calculate the power spectrum for the curvature perturbation. A similar analysis has been carried out in \cite{Gumrukcuoglu2010}, so the results in this section are to recap these results, and to verify that these results are consistent with the perturbative expression for the curvature power spectrum \cite{Yamamoto2012}.
Using the $\pi$ and $\alpha$ mode functions we are able to define the $\zeta$ mode function as
\begin{equation}
\zeta^a_k= -\frac{\eta }{3\sqrt{2}}\left( \pi_k ^{a\prime } + \frac{3(1+2\mathcal{I}^2)}{\eta }\pi_k^a - \frac{3}{\sqrt{2}} H \mathcal{I}\eta ^3  \alpha_k ^{a\prime }\right) ,
\end{equation}
and hence
\begin{equation}
\langle\zeta_k{}^2\rangle=\zeta_k^a\zeta_k^a\,.
\end{equation}
The $\zeta$ mode functions are plotted in figures \ref{fig:zetamf1}, \ref{fig:zetamf2}, \ref{fig:zetamf3} 
and \ref{fig:zetamf4}, while the numerical and analytic results for the power spectrum are shown in 
figures \ref{fig:zac}, \ref{fig:ssev} and \ref{fig:finalps}. The perturbative solution for the time evolution 
of $\langle\zeta_k{}^2\rangle$ is shown to be useful only for small $\mathcal{I}$ (figure \ref{fig:zac}), 
while the analytic estimate for the final value of the power spectrum, derived in section 4, 
is valid for values of $\mathcal{I}$ up to around 0.2 (figure \ref{fig:finalps}). The lack of
quantitative agreement beyond $\mathcal{I} \sim 0.2$ is presumably due to the error
arising from the matching since for larger values of $\mathcal{I}$, the numerical calculations (figrues
\ref{fig:zetamf1}, \ref{fig:zetamf2}, \ref{fig:zetamf3}, \ref{fig:zetamf4}) show a significant deviation 
from the de-Sitter mode functions around horizon crossing. The characteristic timescale 
for the time evolution for $\langle\zeta_k{}^2\rangle$ before it reaches constant is shown to be 
$\mathcal{I}^{-2}$ for $\mathcal{I}\lesssim 0.1$, in agreement with the analytical estimate
(\ref{eq:saturatevalue}) from the previous section (figure \ref{fig:ssev}). This result has a
significant implication on the validity of the perturbative treatment of quadratic vertices 
discussed at the end of section \ref{sec:analytic}. The transition to constant regime
occurs around 
\begin{equation}
N_k \sim \mathcal{I}^{-2} 
\end{equation}
which is much later than the time at which the correction term to the power spectrum
becomes comparable to the leading-order term $N_k \sim \mathcal{I}^{-1}$. In fact,
the numerical evidence suggests that the perturbative formula (\ref{eq:powerPerturbative})
is valid right up to $ \mathcal{I}^{2} \lesssim N_k^{-1} $, or for CMBR scale ($N_k \sim 50$),
 $\ \mathcal{I} \lesssim O(0.1)$. This observation plays a key role in imposing
the observational constraint from Planck later.

\subsection{Numerical calculation of the $\zeta$ bispectrum}

By solving the coupled linear evolution equations, we in effect include the contribution from 
the infinitely many tree-level Feynman diagrams coming from the quadratic $H^q$ term, 
and hence obtain a result correct to all orders in $\mathcal{I}$ (provided loop 
contributions are negligible). Therefore, the exact tree-level amplitude for the bispectrum, 
by standard application of Wick's theorem, is given by
\begin{eqnarray*}
\langle\zeta_{k_1}\zeta_{k_2}\zeta_{k_3}\rangle=&-&48\sqrt{2}\mathcal{I}^2\int ^{\eta }\frac{d\eta _1 }{\eta _1^4}\Im\left(\zeta^{a\ast}_{k_1}(\eta)\zeta^{b\ast}_{k_2}(\eta)\zeta^{c\ast}_{k_3}(\eta)\pi^a_{k_1}(\eta_1)\pi^b_{k_2}(\eta_1)\pi^c_{k_3}(\eta_1)\right)\nonumber\\
&+&48\mathcal{I}\int ^{\eta }d\eta _1\Im\left(\zeta^{a\ast}_{k_1}(\eta)\zeta^{b\ast}_{k_2}(\eta)\zeta^{c\ast}_{k_3}(\eta)\pi^a_{k_1}(\eta_1)\pi^b_{k_2}(\eta_1)\alpha_{k_3} ^{c\prime }(\eta_1)+{\rm 2 \ perms}\right)\nonumber\\
&-&12\sqrt{2}\int ^{\eta }d\eta _1\eta_1^4\Im\left(\zeta^{a\ast}_{k_1}(\eta)\zeta^{b\ast}_{k_2}(\eta)\zeta^{c\ast}_{k_3}(\eta)\pi^a_{k_1}(\eta_1)\alpha_{k_2} ^{b\prime }(\eta_1)\alpha_{k_3} ^{c\prime }(\eta_1)+{\rm 2 \ perms}\right)\,.
\end{eqnarray*}
In particular, we now only have to compute 1-vertex terms.

The evaluation of the integrand turns out to be a numerically unstable process sufficiently 
far outside the horizon, requiring a very precise cancellation of terms. It therefore becomes 
impractical to carry out the calculation with the numerically solved mode functions 
beyond a certain point. To overcome this difficulty, for $-k\eta<10^{-5}$ we switch to using 
the analytic superhorizon solution discussed in the previous section. The only difference is
in the matching of the analytic superhorizon solution; here we evaluate the numerical 
mode functions (and their time derivative) at $-k\eta=10^{-5}$ and use these as the matching 
conditions for the analytic superhorizon solution. Then, for $-k\eta<10^{-5}$ the time integrals 
in the above expression are computed analytically, therefore avoiding the problem of numerical 
instabilities. When performing the first stage of this computation (the numerical stage), 
we employ the technique recently developed in \cite{Funakoshi2012}.
As we will see later, the bispectrum (or more precisely, the shape function) is
peaked in the squeezed limit, and therefore to concentrate on the salient features we will restrict
most of our analysis to the squeezed limit.

\begin{figure}[htbp]
 \begin{center}
\includegraphics[width=0.3\linewidth]{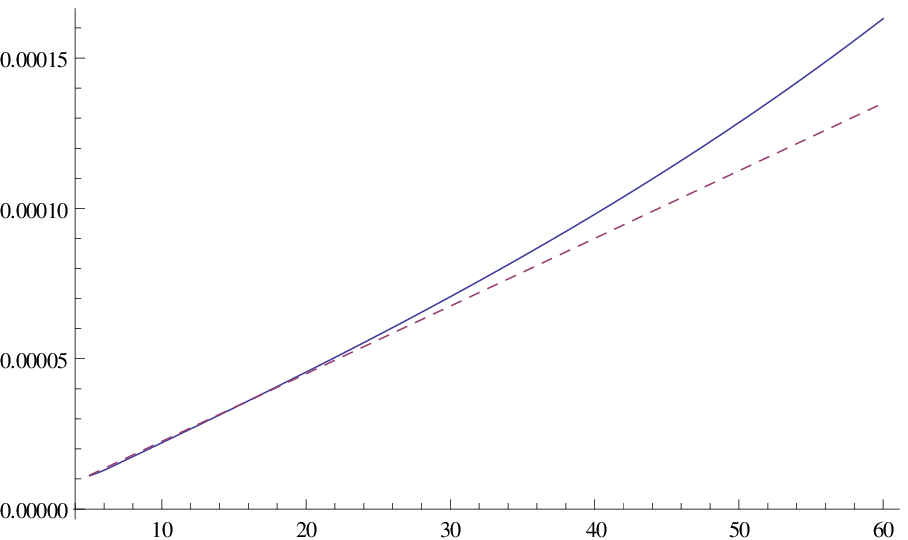}
\includegraphics[width=0.3\linewidth]{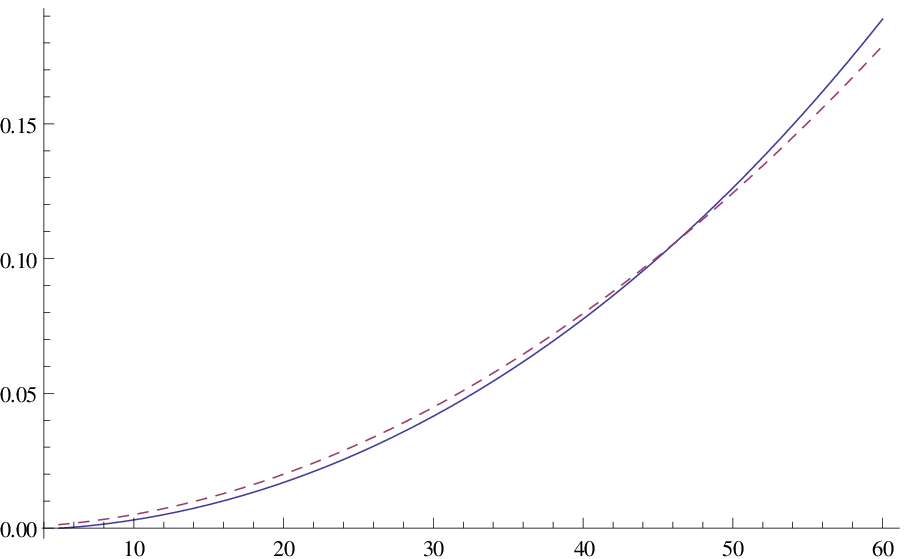}
\includegraphics[width=0.3\linewidth]{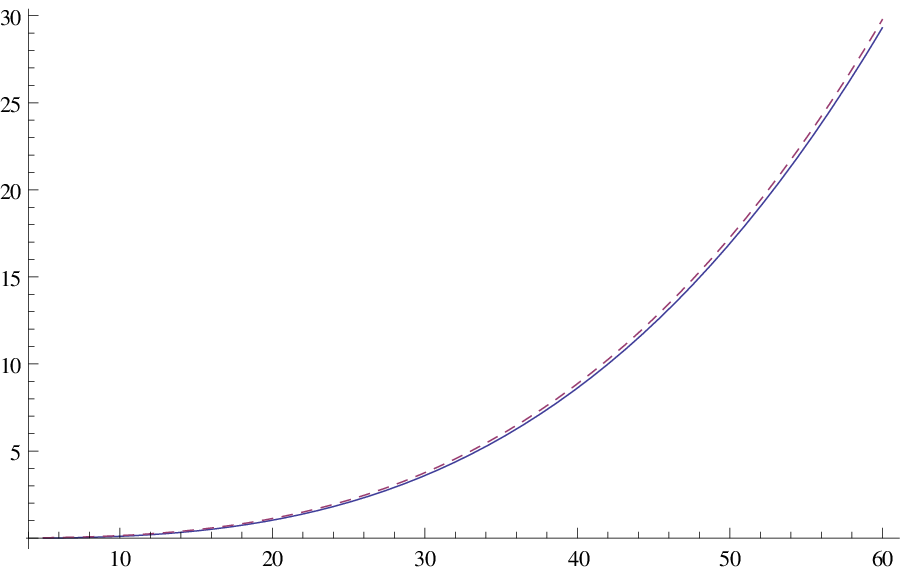}
\caption{Comparison between the analytic (perturbative) and numerical values of the bispectrum in the squeezed limit, for $\mathcal{I}=0.001$ just after horizon crossing, showing a very good agreement; from left to right are the contributions from $H^A$, $H^B$ and $H^C$ respectively, with the numerical value in solid lines and analytic in dashed lines. For all the results in this section, the squeezed limit is evaulated by taking the bispectrum in the configuation $k_1 \times 10^2=k_2 = k_3$.}\label{fig:smallscrifnl}
\includegraphics[width=0.29\linewidth]{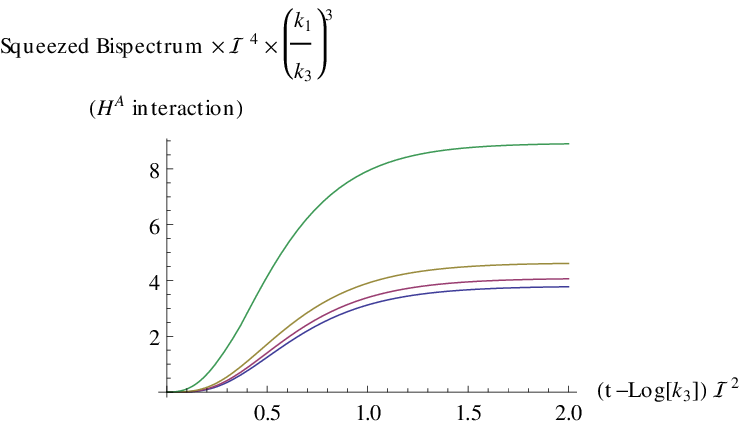}
\includegraphics[width=0.29\linewidth]{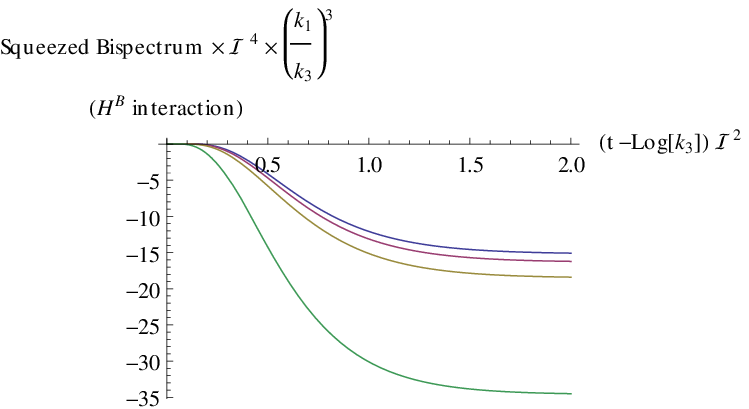}
\includegraphics[width=0.4\linewidth]{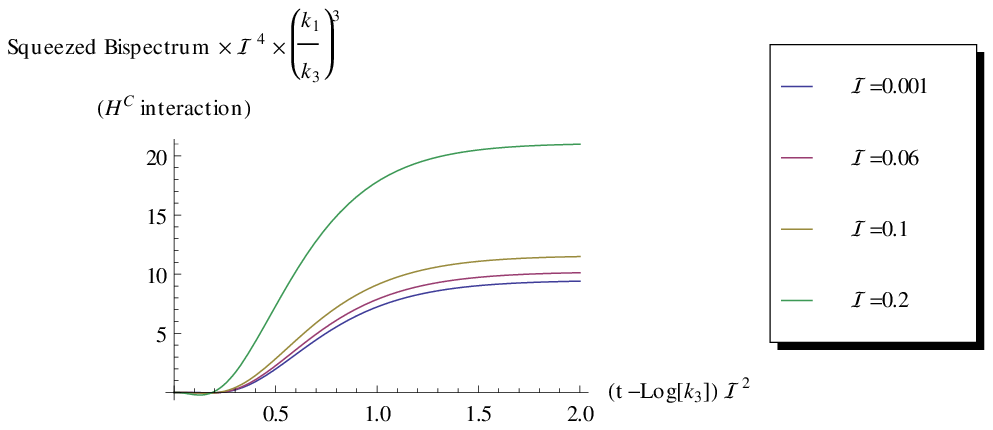}
\caption{Late-time evolution of the bispectrum in the squeezed limit. There are two properties here which cannot be seen from the perturbative calculation; the contribution to the bispectrum from the second vertex becomes negative, and the contribution from all three vertices become constant after some number of e-folds $\sim \mathcal{I}^{-2}$. For small $\mathcal{I}$, the final value of the squeezed bispectrum exhibits a $\mathcal{I}^{-4}$ scaling.}\label{fig:latebev}
\end{center}
\end{figure}

We start by crosschecking our numerical calculations against the perturbative results 
from section 3 (figure \ref{fig:smallscrifnl}), for small $\mathcal{I}$ ($=10^{-3}$) 
for sometime after horizon exit where the perturbative treatment of the quadratic vertex $H^q$ 
is justified. This underwrites the overall consistency between the analytical and numerical
methods.

In figure \ref{fig:latebev}, we confirm the convergence 
of the bispectrum generated by each cubic vertex. As one can see, while $H^C$ is the dominant
contribution in the perturbative regime as it grows the fastest $( \ln (-k\eta ) )^3$, it is overtaken
by $H^B$ when the perturbative approximation breaks down. It also exhibits 
an approximate $\mathcal{I}^{-4}$ scaling of the final value of bispectrum,
which is equivalent to the $\mathcal{I}$
independence of $f_{NL}$ that was inferred at the end of the previous section. The
characteristic timescale for the transition to constant is again shown to be 
$\mathcal{I}^{-2}$.

\begin{figure}
\begin{center}
\includegraphics[width=0.325\linewidth]{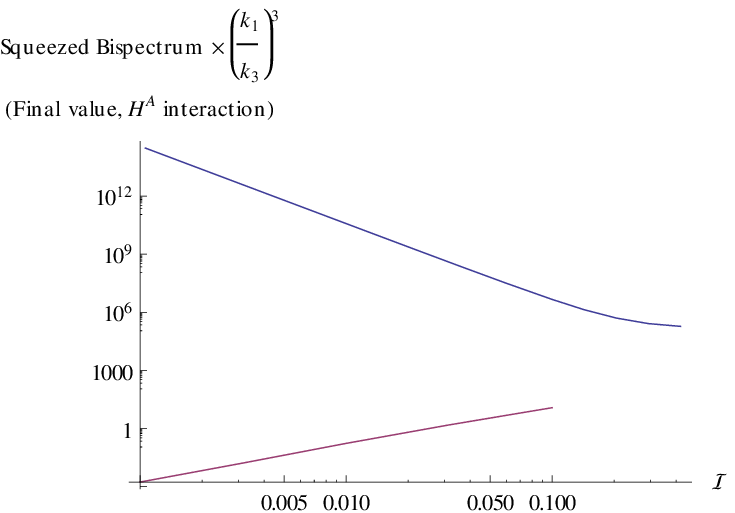}
\includegraphics[width=0.325\linewidth]{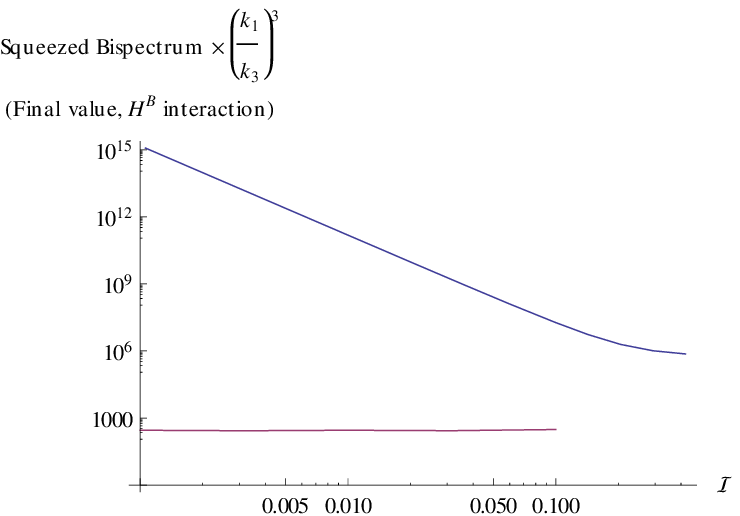}
\includegraphics[width=0.325\linewidth]{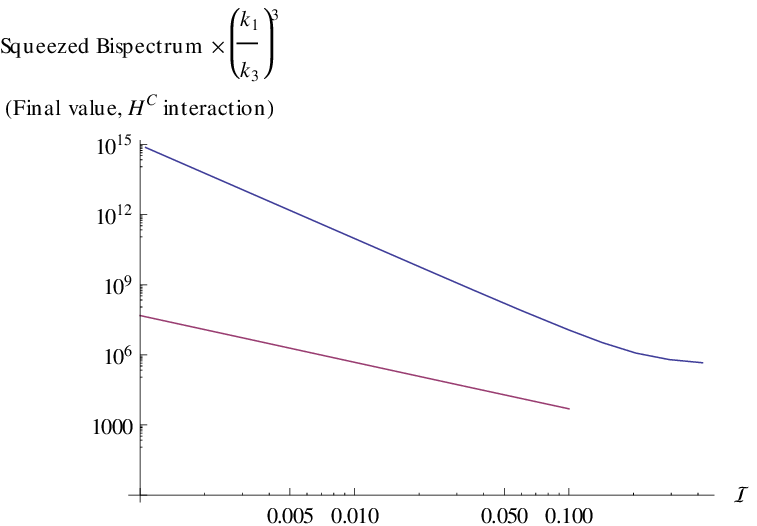}
\includegraphics[width=0.6\linewidth]{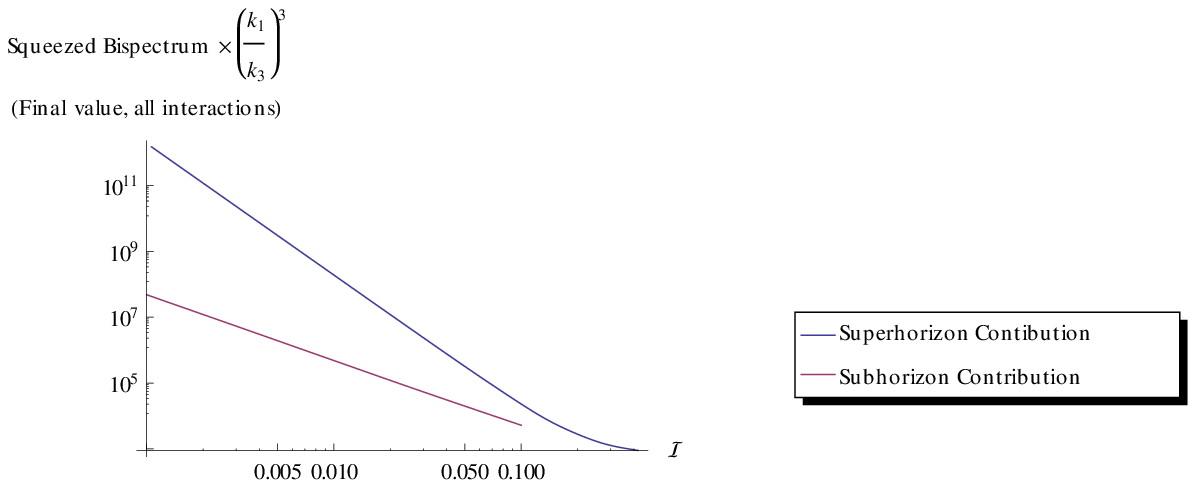}
\caption{Relative importance of the subhorizon and superhorizon contributions to the final value of the squeezed bispectrum. As $\mathcal{I}$ approaches 1, the subhorizon and superhorizon contributions become comparable. For the plots of the $H^B$ interaction and combined interactions, the negative of the bispectrum was taken for the purposes of taking a log plot. We were unable to obtain reliable results for the subhorizon contribution for $\mathcal{I}\gtrsim 0.1$, so they were not included in the above plots. }\label{fig:subsupercomp}
 \end{center}
\end{figure}

From the $(\ln(-k\eta))^3$ perturbative growth in the bispectrum, one may expect that the superhorizon 
contribution to the bispectrum dominates over the subhorizon contribution; in figure \ref{fig:subsupercomp} 
we verify that this is indeed the case for most values of $\mathcal{I}$. The bispectrum is evaluated
at $\eta=0$, and the $k\eta<-1$ (subhorizon) and $k\eta>-1$ (superhorizon) contributions to the
time integral are plotted separately as a function of $\mathcal{I}$. The two become comparable only as
$\mathcal{I}$ reaches order unity, and for $\mathcal{I}\lesssim0.1$ the subhorizon contribution
is negligible.

\begin{figure}[htbp]
\begin{center}
\includegraphics[width=0.45\linewidth]{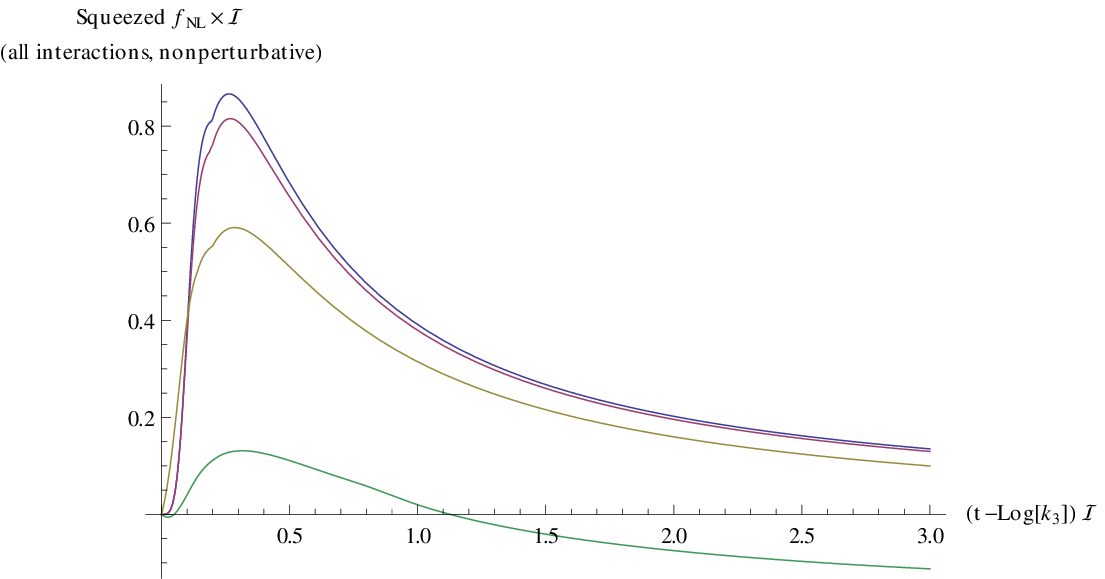}
\includegraphics[width=0.54\linewidth]{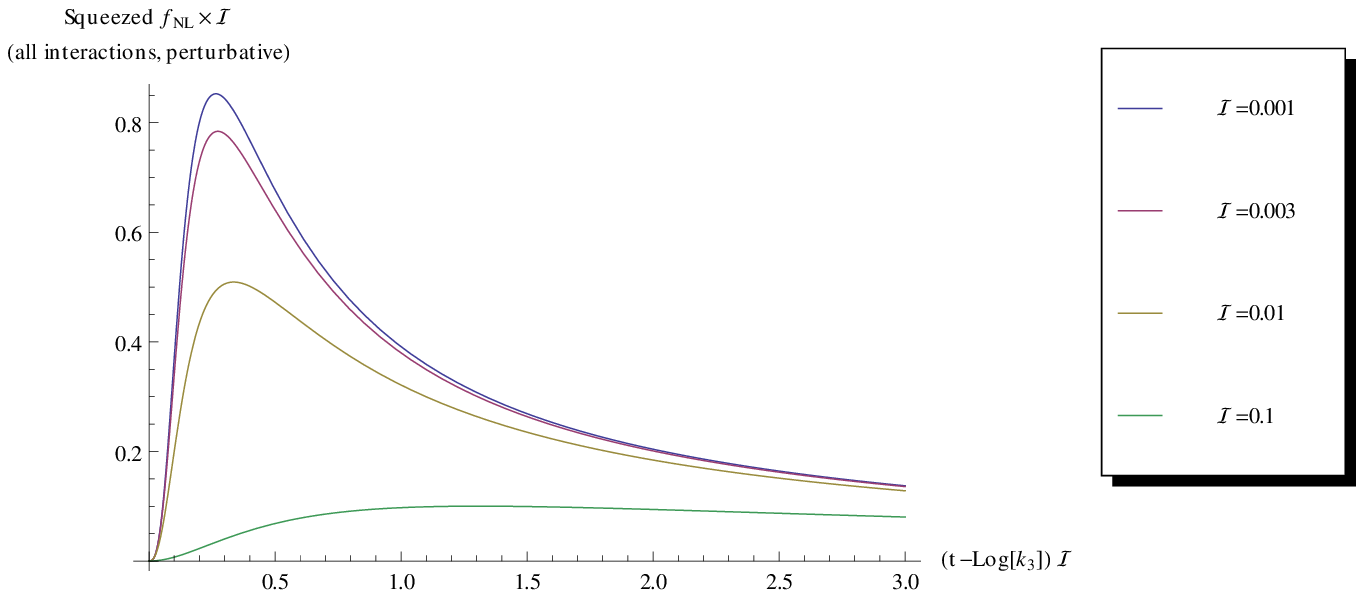}
\caption{Intermediate time evolution of $f_{NL}$ in the squeezed limit for 
$\mathcal{I} = 0.001, 0.003, 0.01$ and $0.1$. $f_{NL}$ first grows as $N_k^3$ and 
eventually peaks around $N_k \sim 0.3 / \mathcal{I}$. The peak value appears
to be $\sim \mathcal{I}^{-1}$. Then it monotonically decays until eventually 
settling down to a negative constant. The behaviour of the peak, at least for $\mathcal{I}\lesssim0.01$, can be understood perturbatively since the timescale for the peaking of $f_{NL}$ is smaller than that of the breakdown of perturbation theory.}
\label{fig:intermediateFNL}
\end{center}
\end{figure}

In figure \ref{fig:intermediateFNL}, we plot the intermediate time evolution of $f_{NL}$, 
with the numerical calculation on the left panel and perturbative result on the right.
For $N_k \ll \mathcal{I}^{-1}$, the power spectrum is essentially constant and $f_{NL}$ grows
as $N_k^3$. When $N_k \gtrsim 0.1 \mathcal{I}^{-1}$, the power spectrum starts to be overtaken
by the correction term and scale as $N_k^2$, which results in the peak around 
$N_k \sim 0.3 \mathcal{I}^{-1}$. The 
maximum value appears to scale as $\mathcal{I}^{-1}$, which means it may well be 
observable for a small value of $\mathcal{I}$. 
Since these peaks occur on timescales $\propto \mathcal{I}^{-1}$, 
the time dependence (and therefore scale dependence) of $f_{NL}$ around this
maximum can be understood by the perturbative results where analytical expressions
are available.

\begin{figure}
\begin{center}
\includegraphics[width=0.9\linewidth]{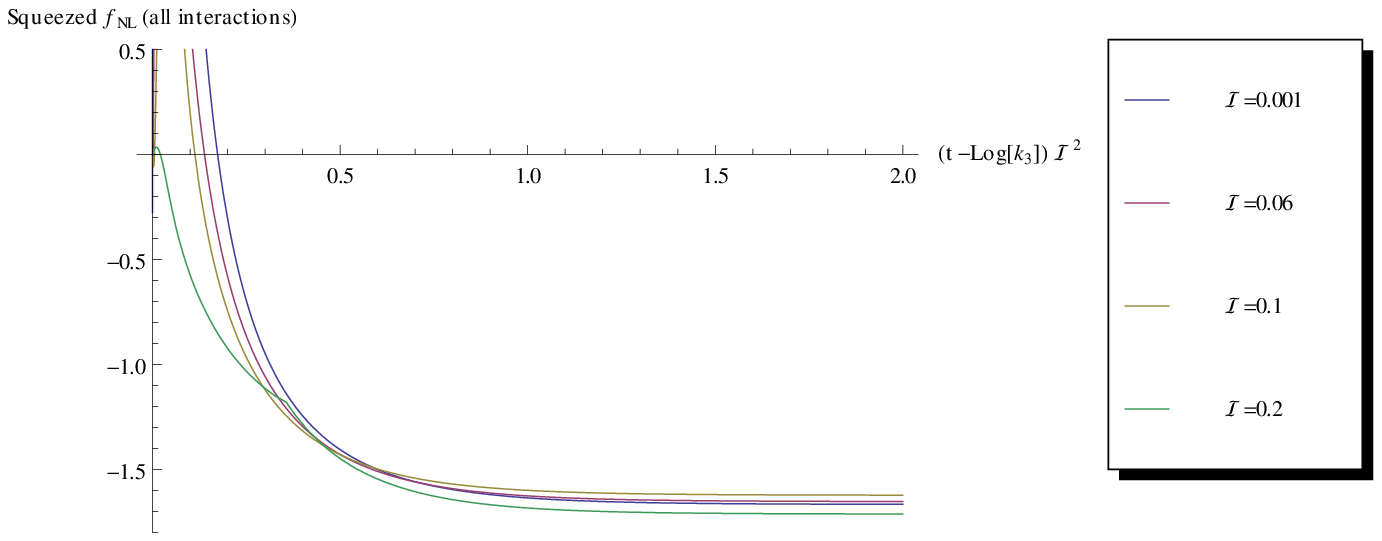}
\caption{Late time evolution of $f_{NL}$ in the squeezed limit for 
$\mathcal{I} = 0.001, 0.06, 0.1$ and $0.2$. The x-axis is e-folding number after horizon
crossing rescaled by $\mathcal{I}^{2}$, showing the characteristic timescale for $f_{NL}$ to become constant. The final value
is independent of $\mathcal{I}$.}
\label{fig:finalFNL}
\end{center}
\end{figure}
\begin{figure}
\begin{center}
\includegraphics[width=0.56\linewidth]{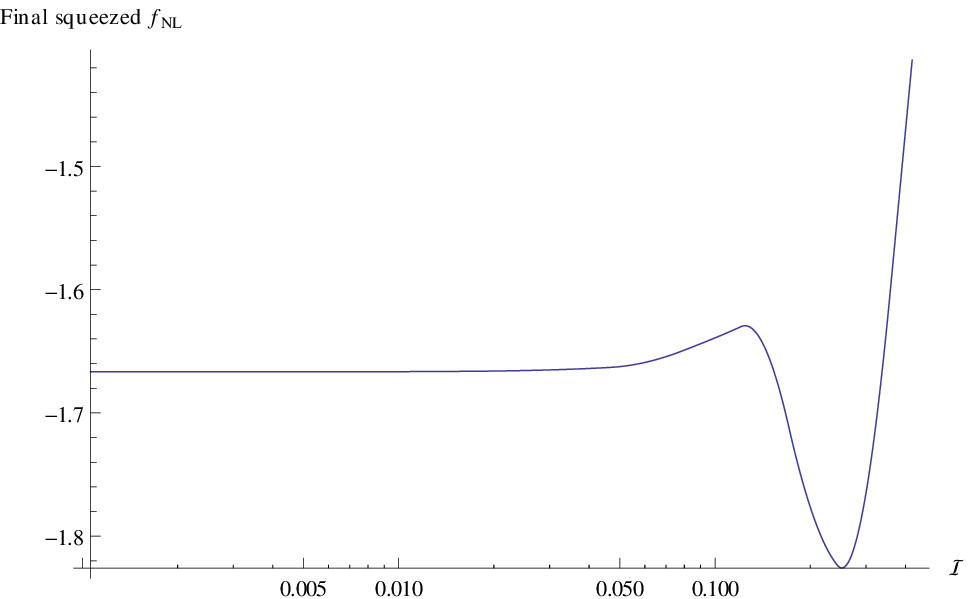}
\qquad
\includegraphics[width=0.36\linewidth]{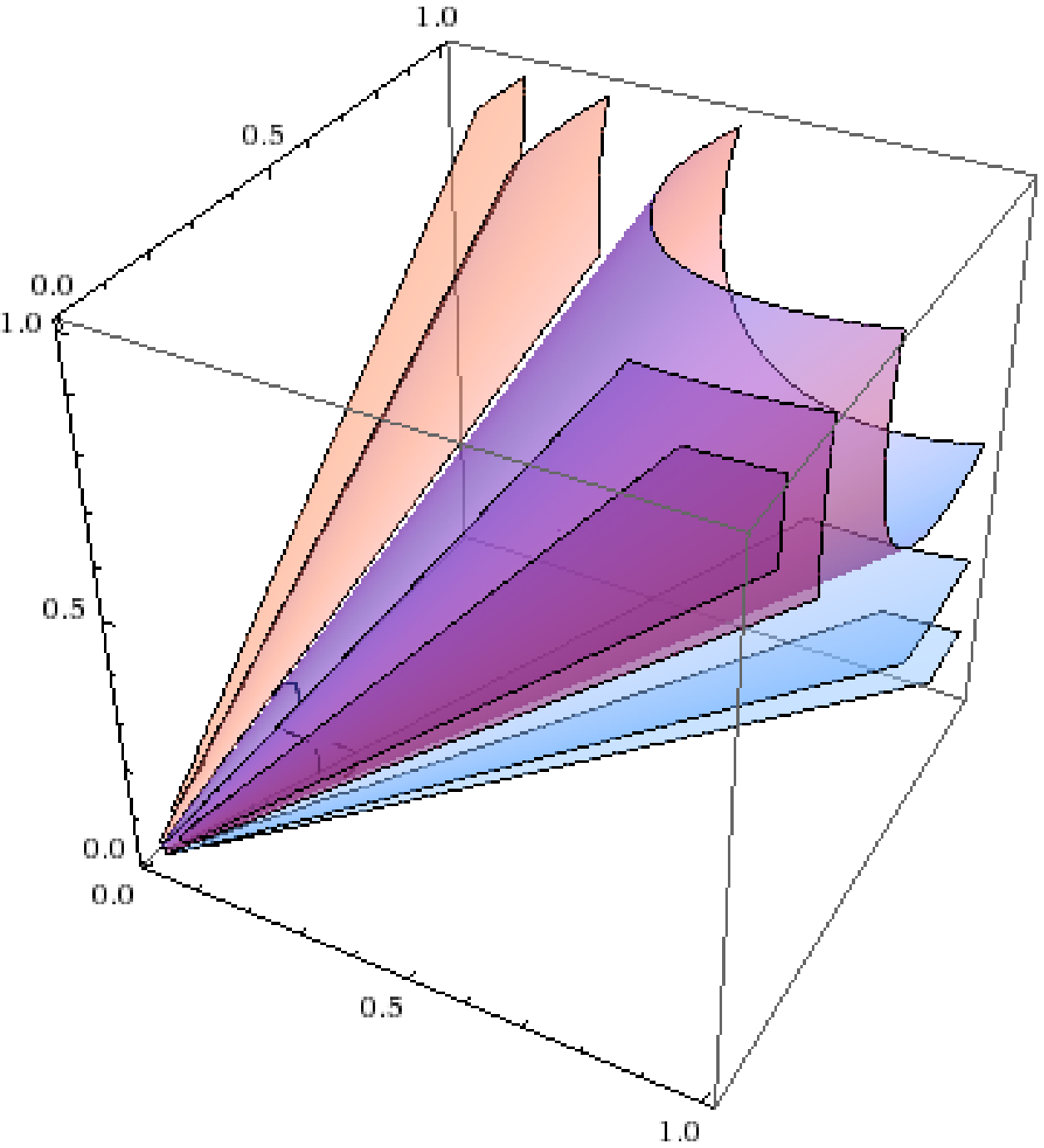}
\caption{Left: the final value of $f_{NL}$ in the squeezed limit, at least for $\mathcal{I}\lesssim0.1$, is independent of $\mathcal{I}$ appearing to take the value $-\frac{5}{3}$. Right: The shape of bispectrum for $\mathcal{I} = 0.01$, final value.
What is plotted here are contours for $(k_1k_2k_3)^2B(k_1,k_2,k_3)$, with the three axes
being $k_1$, $k_2$ and $k_3$. It is peaked in the squeezed limit and is scale-invariant.}
\label{fig:scridepend}
\end{center}
\end{figure}

However, for later times $\gtrsim\mathcal{I}^{-2}$ this is no longer the case, with $f_{NL}$ turning negative;
this behaviour is shown in figure \ref{fig:finalFNL}, where
we plot the late-time evolution of $f_{NL}$ in the squeezed limit 
for different values of $\mathcal{I}$. The final convergent value appears to be $-5/3$
independent of $\mathcal{I}$, in agreement with the results of the previous section. 
It is also seen in figure \ref{fig:scridepend} where 
the final value is presented as a function of $\mathcal{I}$. We suspect that the cause
of irregular behaviour for $\mathcal{I} \gtrsim 0.1$ is due to the significant contribution
from the subhorizon evolution. We also note that given that there are two scalar degrees of freedom here,
the single-field consistency relation \cite{Maldacena2002} does not hold.

We conclude this section by mentioning a few words about the shape of the bispectrum.
As can be seen from figure \ref{fig:scridepend}, the bispectrum is peaked in the squeezed
limit, which is expected given the fact that it is predominantly determined by the superhorizon
evolution which tends to generate local bispectra. Provided that we wait until all relevant modes 
have become constant (as done for the plot), it is perfectly scale-invariant too.  
This can be understood by noting that since the background
geometry is de-Sitter and the interaction terms are de-Sitter invariant,
the correlation functions for the perturbations which have become constant are scale-invariant;
in particular, for the modes which have settled down to the final value, both the power spectrum and
the bispectrum are scale-invariant. By similar arguments, where we have plotted the time evolution
of any quantity such as $f_{NL}$, they can be used to read off the scale dependence at any given time.



\section{Implications and concluding remarks}
We have studied the perturbation of a model of inflation where a stable isotropic phase
of inflation is realized by a scalar field coupled with a triplet of Abelian gauge fields. We
derived the general action for scalar perturbation up to cubic order and identified
all the relevant terms in the limit of vanishing slow-roll parameters. Using the standard
method of in-in formalism, we first treated both the quadratic and cubic vertices
perturbatively and computed the bispectrum at the leading order in the expansion
parameter $\mathcal{I}$. The resulting expression was consistent with the previous
studies and $f_{NL}$ in the squeezed limit was shown to be proportional to
$\mathcal{I}^2 N_k^3 $ where $N_k$ is the e-folding number after the relevant
modes exit the horizon. We then pointed out the limited applicability of this approach
even for $\mathcal{I} \ll 1$ and rectified it by introducing the exact linear mode
functions which take into account the effect of the infinite number of tree-level
diagrams generated by the quadratic vertices. Solving the linear evolution
equations analytically in the superhorizon limit, we proved that both the power spectrum
and the bispectrum are convergent in the limit $N_k \rightarrow \infty$, 
with the late-time bispectrum being local in shape.

In order to obtain a more quantitative estimate
of the bispectrum and $f_{NL}$, we carried out an extensive numerical analysis
employing in part the recently developed technique \cite{Funakoshi2012}. We confirmed the
analytical results and found a number of interesting features. In calculating
the time evolution of $f_{NL}$ in the squeezed limit, we find that it peaks at some characteristic time
after horizon crossing, with this peak value scaling as $\mathcal{I}^{-1}$.
After peaking, it settles down to the same value (independent of $\mathcal{I}$ for small $\mathcal{I}$) 
as was estimated analytically: $f_{NL}=-\frac{5}{3}$.


We now discuss the implications of our calculations. Recent Planck \cite{Ade2013} data suggest
$f_{NL}$ should be of order unity.
The modes observable in our Universe
typically experience around 50 e-folds after horizon crossing and we already argued that the perturbative
expression (\ref{eq:result1}) is valid as long as $\mathcal{I} \lesssim 0.1$. Thus excluding $f_{NL}>10$,
we can constrain $\mathcal{I}$ to satisfy either $\mathcal{I}^2 \lesssim 10^{-7}$ or $\mathcal{I}^2 \gtrsim 10^{-3}$. 
For $\mathcal{I} \gtrsim 0.1$, our numerical calculations suggest that $f_{NL}$ is in
the constant regime for $N_k \sim 50$ and its value is of order unity (left panel in figure
\ref{fig:scridepend}). 


We first emphasise that our analysis here is complete at tree-level. It indicates
the overall consistency of this model in the classical regime; any fluctuations present at horizon crossing
remain bounded and so do their correlation functions (at the very least at the 2-point and 3-point level).
For a certain range of values of $\mathcal{I}$, the model is ruled out by the latest observational
constraint on $f_{NL}$ by Planck. However, for large values of $\mathcal{I}$ approaching 1,
the length scales currently observable would come from the late-time stage where $f_{NL}$ is of order unity
and hence within Planck bounds. Similarly, for small $\mathcal{I}$, $f_{NL}$ grows sufficiently slowly after
horizon crossing and will remain within current observational constraints.  




One can expect that the same qualitative features will also apply to the anisotropic models
where the background is permeated by one or two gauge fields with non-vanishing
vacuum expectation values. The difference is that there the vector fields also contribute
to the spatial anisotropy and there is a strict upper bound for $\mathcal{I}$.
It is going to be difficult to repeat our analysis for anisotropic models since the
consistent perturbative expansion requires the inclusion of vector and tensor modes
which are coupled to the scalars through the background anisotropy. For this
purpose, it would be interesting to look into the relation between our results and the
delta-$N$ formalism \cite{Sasaki1996}. In fact, the isotropic case can be regarded as a particular
two-scalar model and the formalism should apply without any problem. Since
the convergence of the power spectrum and bispectrum is based on the
superhorizon evolution, the delta-$N$ formalism will reproduce them in
a more elegant manner. Since its mathematical basis resides in the equivalence
of the superhorizon curvature perturbation to the background evolution of the 
FLRW universe \cite{Sasaki1998}, an appropriate extension to anisotropic backgrounds sounds
plausible and can be a powerful tool to handle the complicated interactions
among different modes.

Another important theoretical issue is consistency of the quantum field theory
in the existence of background gauge fields. The authors of \cite{Bartolo2012} 
claimed that the infrared contribution of the one-loop diagrams can be interpreted
as the rescaling of the background vacuum expectation value of the gauge fields
so as to take into account the quantum mechanically created modes that froze
in outside the horizon. Although we have not discussed this issue in the present
article, it will be certainly an interesting direction of further research.

Finally, it is in principle straightforward to extend our analysis to inflationary models
with non-Abelian gauge fields, either the one based on gauge-kinetic coupling
\cite{Murata2011,Maeda2012} or Chern-Simons coupling \cite{Adshead2012}.
Given the qualitative similarity between Abelian and non-Abelian models
when all the vertices are treated perturbatively, it is natural to expect that a similar
convergent result in the limit of large e-folding can be established, although 
it is mathematically far from obvious. From a phenomenological point of view,
it would be important to clarify the difference among different scenarios so that one
is able to observationally distinguish between them.

\acknowledgments

We would like to thank Xingang Chen, Misao Sasaki and Jiro Soda for useful comments and
Federico Urban for interesting discussions. 
KY is also greateful to the support and hospitality of the Institute of Theoretical
Astrophysics in the University of Oslo where a part of this work was completed.

\appendix

\section{Details of the perturbative calculation of bispectrum}
Here, we give the details of the integrations and handling of second-order perturbations
necessary for determining the leading-order bispectrum (\ref{eq:bispectrum}). First of all,
let us introduce the following mode functions:
\begin{eqnarray*}
\xi (\tau ,\mathbf{x}) &=& \int \frac{d^3 k}{(2\pi )^3} \frac{H}{\sqrt{2k^3}\eta } \left( g_k (\eta ) a_{\mathbf{k}} e^{i\mathbf{k}\cdot \mathbf{x}} + g^{\ast }_k (\eta ) a^{\dagger }_{\mathbf{k}} e^{-i\mathbf{k}\cdot \mathbf{x}} \right) , \\
\alpha ^{\prime }(\eta ,\mathbf{x}) &=& \int \frac{d^3 k}{(2\pi )^3} \frac{1}{\sqrt{6c_s^3 k^3 }\eta ^4} \left( h_k (\eta ) b_{\mathbf{k}} e^{i\mathbf{k}\cdot \mathbf{x}} + h^{\ast }_k (\eta ) b^{\dagger }_{\mathbf{k}} e^{-i\mathbf{k}\cdot \mathbf{x}} \right) , \\
g_k (\eta ) &=& \left[ i (3+k^2 \eta ^2 ) -3k\eta  \right] e^{-ik\eta } , \\
h_k (\eta ) &=& \left[ i (3- c_s^2 k^2 \eta ^2 ) -3c_s k \eta \right] e^{-ic_s k\eta } .
\end{eqnarray*}
It will be useful later to note that
\begin{eqnarray}
g^{\ast }_k (\eta )u_k (\tilde{\eta }) &=& \left( 3 + k^2 \eta (\eta +3\tilde{\eta }) -3ik(\eta -\tilde{\eta }) +ik^3 \eta ^2 \tilde{\eta } \right) e^{ik(\eta - \tilde{\eta })} , \label{eq:propu} \\
h^{\ast }_k (\eta ) v_k (\tilde{\eta }) &=& \left( -3 +c_s^2 k^2 \eta (\eta -3\tilde{\eta }) +3ic_s k (\eta - \tilde{\eta }) +ic_s^3 k^3 \eta ^2 \tilde{\eta } \right) e^{ic_s k (\eta - \tilde{\eta })} .\label{eq:propv}
\end{eqnarray}

\subsection{1-vertex contributions}
Let us start from the single integration (\ref{eq:xi1}). One can easily go to Fourier space and derive
\begin{eqnarray*}
\langle \xi _k^3 \rangle &=& i\sqrt{\frac{2}{\epsilon _{\varphi }}} \frac{4\mathcal{I}^2}{H^2 \eta ^3 } \frac{H^6}{8k_1^3 k_2^3 k_3^3 } \int ^{\eta }\frac{d\eta _1 }{\eta _1^4} 6\left( u_{k_1} u_{k_2} u_{k_3} g^{\ast }_{k_1} g^{\ast }_{k_2} g^{\ast }_{k_3} - g_{k_1 }g_{k_2} g_{k_3} u^{\ast }_{k_1}u^{\ast }_{k_2}u^{\ast }_{k_3} \right) \\
&=& -\sqrt{\frac{2}{\epsilon _{\varphi }}} \frac{6\mathcal{I}^2 H^4}{k_1^3 k_2^3 k_3^3 \eta ^3} \int ^{\eta }\frac{d\eta _1}{\eta _1^4} \Im \left( g^{\ast }_{k_1}(\eta )g^{\ast }_{k_2}(\eta )g^{\ast }_{k_3}(\eta ) u_{k_1}(\eta _1) u_{k_2}(\eta _1) u_{k_3}(\eta _1) \right) .
\end{eqnarray*}
At a first glance, the integral looks divergent as $\eta \rightarrow 0$ even if the factor of $\eta ^3$ in (\ref{eq:zeta}) is taken into account. However, it is not the case since
we have
\begin{equation*}
\Im \left(  g^{\ast }_{k_1}(\eta )g^{\ast }_{k_2}(\eta ) g^{\ast }_{k_3}(\eta ) u_{k_1}(\eta ) u_{k_2}(\eta ) u_{k_3}(\eta ) \right) = 9 \left( k_1 ^3 + k_2^3 +k_3^3 \right) \eta ^3 + O(\eta ^5 ) .
\end{equation*}
Therefore, an integration by parts gives
\begin{eqnarray*}
\mathcal{A}_1 &=& \int ^{\eta }\frac{d\eta _1}{\eta _1^4} \Im \left( g^{\ast }_{k_1}(\eta )g^{\ast }_{k_2}(\eta )g^{\ast }_{k_3}(\eta ) u_{k_1}(\eta _1) u_{k_2}(\eta _1) u_{k_3}(\eta _1) \right) \\
&=& -3 (k_1^3 +k_2^3 +k_3^3 ) + O(\eta ^2)  +\frac{1}{3} \int ^{\eta }\frac{d\eta _1}{\eta _1^3} \Im \left( g^{\ast }_{k_1}g^{\ast }_{k_2}g^{\ast }_{k_3} \left( u_{k_1}u_{k_2}u_{k_3} \right) ^{\prime } \right) .
\end{eqnarray*}
The same type of cancelation of power holds for the remaining integrals. It is also helped by the fact that
\begin{equation*}
u_k^{\prime }(\eta ) = i k^2 \eta e^{-ik\eta } ,
\end{equation*}
which is a manifestation of the constancy after horizon exit of the de-Sitter mode functions. Repeating another integration by parts, we are left with
\begin{eqnarray*}
\mathcal{A}_1 &=& 6 (k_1^3 +k_2^3 + k_3^3 ) + O(\eta ^2)  \\
&& +\frac{1}{3} \int \frac{d\eta _1}{\eta _1} k_1^3 \Im \left( g^{\ast }_{k_1}(\eta )g^{\ast }_{k_2}(\eta )g^{\ast }_{k_3}(\eta ) e^{-ik_1 \eta _1}u_{k_2}(\eta _1) u_{k_3}(\eta _1 ) \right)  + {\rm 2 \ perms} \\
&& - \frac{1}{3} \int d\eta _1 \Im \left( g^{\ast }_{k_1} g^{\ast }_{k_2}g^{\ast }_{k_3} k_1^2 e^{-ik_1 \eta _1}  \left( k_2^2 e^{-ik_2 \eta _1}u_{k_3} + k_3^2 e^{-ik_3 \eta _1}u_{k_2} \right) \right) + {\rm 2 \ perms} .
\end{eqnarray*}
The third line is finite. The leading contribution is logarithmically divergent in $\eta $ as
\begin{equation*}
\langle \xi _k^3 \rangle  \rightarrow  \sqrt{\frac{2}{\epsilon _{\varphi }}} \frac{9H^4 \mathcal{I}^2 \left( k_1^3 +k_2^3 +k_3^3 \right)}{k_1^3 k_2^3 k_3^3 \eta ^3} \int ^{\eta } \frac{d\eta _1}{\eta _1} \cos \left[ \left( k_1 +k_2 +k_3 \right) \left( \eta -\eta _1 \right) \right] .
\end{equation*}

The cross correlations (\ref{eq:xia1}) and (\ref{eq:xiaa}) are in principle similar. We have
\begin{eqnarray*}
\langle \xi _{k_1} \xi _{k_2} \alpha ^{\prime }_{k_3} \rangle &=& \frac{24\mathcal{I}}{\sqrt{\epsilon _{\varphi }}H\eta ^6} \frac{H^4}{4k_1^2 k_2^2 } \frac{1}{6c_s^3 k_3^3}  \Im \left( g^{\ast }_{k_1}(\eta )g^{\ast }_{k_2}(\eta ) h_{k_3}^{\ast }(\eta )  T_{k_3}(\eta ) \right) , \\
T_{k_1} (\eta ) &=& \int ^{\eta }\frac{d\eta _1}{\eta _1^4} h_{k_1} (\eta _1 ) u_{k_2}(\eta _1 ) u_{k_3}(\eta _1 ) . 
\end{eqnarray*}
and
\begin{eqnarray*}
\langle \xi _{k_1} \alpha ^{\prime }_{k_2} \alpha ^{\prime }_{k_3} \rangle &=& -\sqrt{\frac{2}{\epsilon _{\varphi }}} \frac{6}{\eta ^9} \frac{H^2}{2k_1^3} \frac{1}{36 c_s^6 k_2^3 k_3^3}  \Im \left( g_{k_1}^{\ast }(\eta ) h^{\ast }_{k_2} (\eta ) h^{\ast }_{k_3} (\eta ) H_{k_1}(\eta  ) \right) , \\
H_{k_1}(\eta ) & = &  \int ^{\eta }\frac{d\eta _1}{\eta _1^4}u_{k_1}(\eta _1 ) h_{k_2}(\eta _1 ) h_{k_3} (\eta _1 ) .
\end{eqnarray*}
In carrying out these integrals, it is useful to note that
\begin{equation}
\frac{1}{\eta ^4}h_k = \left( \frac{v_k}{\eta ^3} \right) ^{\prime } .
\end{equation}
It will be later useful to derive the explicit forms of $T_{k}$ and $H_k$. Straightforward integrations by parts lead to
\begin{eqnarray*}
&& T_{k_1} (\eta ) = - \left( k_2^3 +k_3^3 \right) \int ^{\eta }\frac{d\eta _1}{\eta _1 }e^{-i(c_s k_1 +k_2 +k_3 )\eta _1} \\
&& +\left( \frac{i}{\eta ^3} - \frac{c_s k_1 + k_2 + k_3}{\eta ^2 } -i \frac{c_s k_1 \left( k_2 + k_3 \right) -k_2^2 + k_2 k_3 -k_3^2 }{\eta } + \frac{c_s^2 k_1^2 k_2 k_3}{c_s k_1 +k_2 +k_3} \right) e^{-i(c_s k_1 +k_2 +k_3 )\eta } ,
\end{eqnarray*}
and
\begin{eqnarray*}
&& H_{k_3} (\eta ) = -3k_3^3 \int ^{\eta }\frac{d\eta _1}{\eta _1} e^{-i(c_s k_1 +c_s k_2 +k_3 )\eta _1}  + i \frac{c_s^4 k_1^2 k_2^2 k_3 }{c_s k_1 +c_s k_2 +k_3}\eta e^{-i(c_s k_1 +c_s k_2 +k_3)\eta } \\
&&  +\frac{c_s^3 k_1 k_2 \left( c_s^2 k_1 k_2 (k_1 +k_2 ) +c_s k_3 (3k_1^2 +8k_1 k_2 +3k_2^2 ) + 3(k_1 +k_2 ) k_3^2 \right)}{\left( c_s k_1 +c_s k_2 +k_3 \right) ^2 } e^{-i(c_s k_1 +c_s k_2 +k_3)\eta } \\
&& +\left( \frac{3i}{\eta ^3} - \frac{3\left( c_s k_1 +c_s k_2 +k_3 \right) }{\eta ^2} - i \frac{3\left( c_s^2 k_1 k_2 +c_s k_3 (k_1 +k_2 ) -k_3^2 \right)}{\eta } \right) e^{-i(c_s k_1 +c_s k_2 +k_3)\eta } .
\end{eqnarray*}
All the terms with negative powers of $\eta $ cancel when taking the imaginary parts, due to the rapid decay of the imaginary part of the propagators beyond the Hubble horizon;
\begin{equation*}
\Im \left( g^{\ast }_k (\eta ) u_k (\eta  ) \right) = k^3 \eta ^3 +O(\eta ^5 )  \ \ \ {\rm and} \ \ \ \Im \left( h^{\ast }_k (\eta ) v_k (\eta  ) \right) = c_s^3 k^3 \eta ^3 + O(\eta ^5 ) .
\end{equation*}
The end results are again logarithmic dependences on $\eta $;
\begin{eqnarray*}
 \langle \xi _{k_1}\xi _{k_2}\alpha ^{\prime }_{k_3} \rangle &\rightarrow & -\frac{27 H^3 \mathcal{I} (k_1^3 +k_2^3 )}{\sqrt{\epsilon _{\varphi }}\eta ^6 c_s^3 k_1^3 k_2^3 k_3^3 } \int ^{\eta }\frac{d\eta _1}{\eta _1 }\cos \left[ \left( k_1 +k_2 +c_s k_3 \right) \left( \eta -\eta _1 \right) \right]  , \\
 \langle \xi _{k_1}\alpha ^{\prime }_{k_2}\alpha ^{\prime }_{k_3} \rangle  &\rightarrow & \frac{27 H^2 k_1^3 }{2\sqrt{2\epsilon _{\varphi }}\eta ^9 c_s^6 k_1^3 k_2^3 k_3^3 } \int ^{\eta }\frac{d\eta _1}{\eta _1 } \cos \left[ \left( k_1 +   c_s k_2 +c_s k_3 \right) \left( \eta - \eta _1 \right) \right]  .
\end{eqnarray*}

\subsection{2-vertex contributions}
Let us start from the term (\ref{eq:xi2A}). By definition, the connected tree-level contribution is given by
\begin{eqnarray*}
&&- \langle H^B (\eta _2 ) H^q (\eta _1 ) \xi ^3 \rangle = \frac{12\mathcal{I}}{\sqrt{\epsilon _{\varphi }}H} \int d^3 w_2 \frac{6\sqrt{2}\mathcal{I}}{H} \int d^3 w_1 \langle \pi (\mathbf{w}_2 )^2 \alpha ^{\prime }(\mathbf{w}_2 ) \pi (\mathbf{w}_1 )\alpha ^{\prime }(\mathbf{w}_1 ) \xi (\mathbf{x})\xi (\mathbf{y}) \xi (\mathbf{z}) \rangle \\
&=& \sqrt{\frac{2}{\epsilon _{\varphi }}} \frac{144 \mathcal{I}^2 }{H^2} \iint d^3 w_2 d^3 w_1 \langle \alpha ^{\prime }(\mathbf{w}_2) \alpha ^{\prime }(\mathbf{w}_1) \rangle  \left( \langle \pi (\mathbf{w}_2 ) \xi (\mathbf{y} \rangle \langle \pi (\mathbf{w}_2 )\xi (\mathbf{z}) \rangle \langle \pi (\mathbf{w}_1)\xi (\mathbf{x})\rangle  + {\rm 2 \ perms} \right) .
\end{eqnarray*}
Given
\begin{eqnarray*}
\langle \pi (\eta _1 ,\mathbf{w}_1 ) \xi (\eta ,\mathbf{x}) \rangle &=& \int \frac{d^3 k_1}{(2\pi )^3} \frac{H^2}{2k_1^3 \eta } u_{k_1}(\eta _1) g^{\ast }_{k_1}(\eta ) e^{-i\mathbf{k}_1 \cdot (\mathbf{x} - \mathbf{w}_1)}, \\
\langle \alpha ^{\prime } (\eta _2 ,\mathbf{w}_2) \alpha ^{\prime }(\eta _1 ,\mathbf{w}_1 ) \rangle &=& \int \frac{d^3 p}{(2\pi )^3} \frac{1}{6c_s^3 p^3 \eta _1^4 \eta _2^4}h_p (\eta _2) h^{\ast }_p (\eta _1) e^{-i\mathbf{p}\cdot (\mathbf{w}_1 -\mathbf{w}_2)} ,
\end{eqnarray*}
we can write down its Fourier transform as
\begin{eqnarray*}
- \langle H^B (\eta _2 )H^q (\eta _1) \xi _k^3 \rangle &=& \sqrt{\frac{2}{\epsilon _{\varphi }}} \frac{3 H^4 \mathcal{I}^2 }{c_s^3 k_1^6 k_2^3 k_3^3 \eta ^3} g_{k_1}^{\ast }(\eta ) g^{\ast }_{k_2} (\eta ) g^{\ast }_{k_3} (\eta ) u_{k_1}(\eta _1 ) \frac{h_{k_1}^{\ast }(\eta _1)}{\eta _1^4} u_{k_2} (\eta _2) u_{k_3}(\eta _2 ) \frac{h_{k_1}(\eta _2)}{\eta _2^4} \\
&& + {\rm 2 \ pemrs} .
\end{eqnarray*}
Taking the commutators, it becomes
\begin{eqnarray*}
- \langle \left[ H^B (\eta _2) ,\left[ H^q (\eta _1) , \xi _k^3 \right] \right] \rangle  &=& -\sqrt{\frac{2}{\epsilon _{\varphi }}} \frac{12 H^4 \mathcal{I}^2}{c_s^3 k_1^6 k_2^3 k_3^3 \eta ^3} \Im \left( g^{\ast }_{k_1}(\eta )u_{k_1} (\eta _1) \right) \\
&& \times \Im \left( g^{\ast }_{k_2}(\eta ) g^{\ast }_{k_3}(\eta ) \frac{h^{\ast }_{k_1}(\eta _1)}{\eta _1^4} u_{k_2}(\eta _2 ) u_{k_3} (\eta _2 ) \frac{h_{k_1}(\eta _2)}{\eta _2^4} \right) + {\rm 2 \ perms} .
\end{eqnarray*}
Now our task is to carry out the integral
\begin{equation*}
\mathcal{A}_{2a} = \int ^{\eta }\frac{d\eta _1}{\eta _1^4} \Im \left( g^{\ast }_{k_1} u_{k_1} (\eta _1) \right) \Im \left( g_{k_2}^{\ast }g^{\ast }_{k_3} h_{k_1}^{\ast }(\eta _1) T_{k_1}(\eta _1 ) \right) .
\end{equation*}
Substituting the integrated expression for $T_{k_1}(\eta _1 )$, the single integral arising 
from its second line gives at most $\propto \ln (-\eta )$ in the limit $\eta \rightarrow 0$ 
since any power divergence disappears after taking the imaginary 
part as demonstrated for the 1-vertex cases.\footnote{In fact, one can directly show there is no power
divergence by Taylor-expanding the propagators (\ref{eq:propu}) and (\ref{eq:propv}). 
Namely, the integrand does not contain any power of $\eta , \eta _1 $ and $\eta _2 $ lower than $-1$
when the exponential is written as a power series. Since the calculation relies on Mathematica, 
however, we stick to the integrations by parts here.} We then only
have to check if the remaining term gives rise to similar logarithmic contributions. 
Evaluating the double integral, we obtain
\begin{eqnarray*}
&& \int ^{\eta }\frac{d\eta _1}{\eta _1^4} \Im \left( g_{k_1}^{\ast }u_{k_1}(\eta _1 ) \right) \Im \left( g_{k_2}^{\ast }g_{k_3}^{\ast }h_{k_1}^{\ast }(\eta _1 ) \int ^{\eta _1}\frac{d\eta _2}{\eta _2 }e^{-i(c_s k_1 +k_2 +k_3 )\eta _2} \right) \\
&=& -k_1^3 \int ^{\eta }\frac{d\eta _1}{\eta _1} \Im \left( g_{k_1}^{\ast }e^{-ik_1 \eta _1} \right) \Im \left( g_{k_2}^{\ast }g_{k_3}^{\ast } v_{k_1}^{\ast }(\eta _1 ) \int ^{\eta _1}\frac{d\eta _2}{\eta _2} e^{-i(c_s k_1 +k_2 +k_3 )\eta _2} \right) \\
&& + \frac{1}{\eta ^3} \Im \left( g_{k_1}^{\ast } \left[ u_{k_1} + ik_1^2 \eta ^2 e^{-ik_1 \eta } \right] \right) \Im \left( g_{k_2}^{\ast }g_{k_3}^{\ast } v_{k_1}^{\ast } \int ^{\eta }\frac{d\eta _1}{\eta _1}e^{-i(c_s k_1 +k_2 +k_3)\eta _1} \right) \\
&& -c_s^2 k_1^4 \int ^{\eta }d\eta _1 \Im \left( ig_{k_1}^{\ast } e^{-ik_1 \eta _1} \right) \Im \left( i g_{k_2}^{\ast }g_{k_3}^{\ast } e^{ic_s k_1 \eta _1}\int ^{\eta _1} \frac{d\eta _2}{\eta _2} e^{-i(c_s k_1 +k_2 +k_3)\eta _2} \right) \\
&& - \int ^{\eta } \frac{d\eta _1}{\eta _1^4} \Im \left( g_{k_1}^{\ast }\left[ u_{k_1}(\eta _1 ) +ik_1^2 \eta _1^2 u^{-ik_1 \eta _1} \right] \right) \Im \left( g_{k_2}^{\ast }g_{k_3}^{\ast } v_{k_1}^{\ast }(\eta _1 )e^{-i(c_s k_1 +k_2 + k_3 )\eta _1} \right)   .
\end{eqnarray*}
All the integrations are at most of order $\ln (-\eta )$ as $\eta \rightarrow 0$ except for the first line whose leading term yields
\begin{equation*}
\mathcal{A}_{2a} \rightarrow 27 k_1^3 (k_2^3 +k_3^3 ) \int ^{\eta }\frac{d\eta _1}{\eta _1} \int ^{\eta _1}\frac{d\eta _2}{\eta _2} \cos \left( k_1 (\eta - \eta _1 ) \right) \cos \left(  c_s k_1 (\eta _1 -\eta _2 )+(k_2 +k_3 )(\eta -\eta _2 ) \right)   
\end{equation*}
and behaves as $\left( \ln (-\eta ) \right) ^2 $.
The leading-order behaviors for the other contributions are essentially the same. For (\ref{eq:xi2B}), the amplitude reads
\begin{eqnarray*}
-\langle \left[ H^q (\eta _2 ) , \left[ H^B (\eta _1 ) , \xi _k^3 \right] \right] \rangle &=& - \sqrt{\frac{2}{\epsilon _{\varphi }}} \frac{12 H^4 \mathcal{I}^2}{c_s^3 k_1^6 k_2^3 k_3^3 \eta ^3} \Im \left( g_{k_2}^{\ast }(\eta ) g^{\ast }_{k_3} (\eta ) u_{k_2}(\eta _1 ) u_{k_3}(\eta _1) \right) \\
&& \times \Im \left( g_{k_1}^{\ast }(\eta ) \frac{h^{\ast }_{k_1} (\eta _1 ) }{\eta _1^4} u_{k_1}(\eta _2 ) \frac{h_{k_1}(\eta _2 )}{\eta _2^4} \right) + {\rm 2 \ perms } .
\end{eqnarray*}
Defining
\begin{eqnarray*}
F_{k_1} (\eta _1 ) &=& \int ^{\eta _1}\frac{d\eta _2}{\eta _2^4} u_{k_1} (\eta _2 ) h_{k_1} (\eta _2 ) \\
&=& \frac{1}{\eta _1^3} u_{k_1}(\eta _1 )v_{k_1}(\eta _1 ) + \frac{k_1^2 }{\eta _1 } e^{-ik_1 (1+c_s ) \eta _1} + ik_1^3 \int ^{\eta _1}\frac{d\eta _2}{\eta _2} e^{-ik_1 (1+c_s ) \eta _2} , 
\end{eqnarray*}
we retain only the most divergent term to obtain
\begin{eqnarray*}
\mathcal{A}_{2b} &=& \int ^{\eta }\frac{d\eta _1}{\eta _1^4 } \Im \left( g_{k_2}^{\ast }g_{k_3}^{\ast } u_{k_2}(\eta _1 ) u_{k_3}(\eta _1 ) \right) \Im \left( g_{k_1}^{\ast } h_{k_1}^{\ast }(\eta _1 ) F_{k_1}(\eta _1 ) \right)  \\
&\sim & -k_1^3 \int \frac{d\eta _1}{\eta _1} \Im \left( g_{k_2}^{\ast }g_{k_3}^{\ast } \left[ k_2^3 e^{-ik_2 \eta _1}u_{k_3} +k_3^3 e^{-ik_3 \tau _1 }u_{k_2} \right] \right) \Im \left( ig_{k_1}^{\ast } v_{k_1}^{\ast } \int ^{\tau _1}\frac{d\eta _2}{\eta _2} e^{-ik_1 (1+c_s )\eta _2} \right) \\
& \sim &  27 k_1^3 (k_2^3 +k_3^3 ) \int ^{\eta } \frac{d\eta _1}{\eta _1}\int ^{\eta _1}\frac{d\eta _2}{\eta _2} \cos \left( (k_1 +k_2 )(\eta -\eta _1 ) \right)  \cos \left( k_1 (\eta - \eta _2 ) +c_s k_1 (\eta _1 -\eta _2 ) \right) .
\end{eqnarray*}
For (\ref{eq:xia2A}), 
\begin{eqnarray*}
 - \langle H^C (\eta _2 ) H^q (\eta _1 ) \xi ( \mathbf{x} ) \xi (\mathbf{y} ) \alpha ^{\prime } (\mathbf{z} ) \rangle &=& -\frac{72 \mathcal{I}}{\sqrt{\epsilon _{\varphi }}H} \tau _2^4 \iint d^3 w_1 d^3 w_2 \langle \alpha ^{\prime }(\mathbf{w}_2 ) \alpha ^{\prime }(\mathbf{w}_1 ) \rangle \langle \alpha ^{\prime } (\mathbf{w}_2 ) \alpha ^{\prime } (\mathbf{z} ) \rangle \\
 && \times \left( \langle \pi (\mathbf{w}_2 ) \xi (\mathbf{y} ) \rangle \langle \pi (\mathbf{w}_1 ) \xi (\mathbf{x} ) \rangle \langle \pi (\mathbf{w}_2 ) \xi (\mathbf{x} ) \rangle \langle \pi (\mathbf{w}_1 ) \xi (\mathbf{y} ) \rangle \right) ,
 \end{eqnarray*}
  and in Fourier space, it becomes
  \begin{eqnarray*}
  - \langle H^C (\eta _2 ) H^q (\eta _1 ) \xi _{k_1 }\xi _{k_2} \alpha ^{\prime }_{k_3} \rangle &=& -\frac{H^3 \mathcal{I}}{2\sqrt{\epsilon _{\varphi }}\eta ^6 c_s^6 k_1^6 k_2^3 k_3^3 }g^{\ast }_{k_1} (\eta ) g^{\ast }_{k_2}(\eta ) h_{k_3}^{\ast }(\eta ) \\
  && \times u_{k_1}(\eta _1 ) \frac{h^{\ast }_{k_1} (\eta _1 )}{\eta _1 ^4 } h_{k_1}(\eta _2 ) u_{k_2 }(\eta _2 ) \frac{h_{k_3}(\eta _2 )}{\eta _2^4 } + (1 \leftrightarrow 2) .
  \end{eqnarray*}
 Then, we derive
 \begin{eqnarray*}
 -\langle \left[ H^C (\eta _2 ) , \left[ H^q (\eta _1 ) , \xi _{k_1} \xi _{k_2} \alpha ^{\prime }_{k_3 } \right] \right] &=& \frac{2H^3 \mathcal{I}}{\sqrt{\epsilon _{\varphi }}c_s^6 k_1^6 k_2^3 k_3^3 \eta ^6} \Im \left( g_{k_1}^{\ast }(\eta ) u_{k_1} (\eta _1 ) \right) \\
 && \times \Im \left( g_{k_2}^{\ast }(\eta ) h^{\ast }_{k_3}(\eta ) \frac{h_{k_1}^{\ast } (\eta _1 )}{\eta _1^4} h_{k_1}(\eta _2 ) u_{k_2} (\eta _2 ) \frac{h_{k_3}(\eta _2 )}{\eta _2^4} \right) + (1 \leftrightarrow 2 ).
 \end{eqnarray*}
  As before, integration goes as
  \begin{eqnarray*}
  \mathcal{B}_{2a} &=& \int ^{\eta }\frac{d\eta _1}{\eta _1^4} \Im \left( g_{k_1}^{\ast }u_{k_1}(\eta _1 ) \right) \Im \left( g_{k_2 }^{\ast }h_{k_3}^{\ast } h_{k_1}^{\ast }(\eta _1 ) H_{k_2} (\eta _1 ) \right) \\
  &\sim & -3k_2^3 \int ^{\eta }\frac{d\eta _1}{\eta _1^4} \Im \left( g_{k_1}^{\ast }u_{k_1} \right) \Im \left( g_{k_2}^{\ast } h_{k_3}^{\ast } h_{k_1}^{\ast } \int ^{\eta _1}\frac{d\eta _2}{\eta _2 }e^{-i(c_s k_1 +k_2 +c_s k_3 )\eta _2} \right) \\
  & \sim & 3k_1^3 k_2^3 \int ^{\eta } \frac{d\eta _1}{\eta _1} \Im \left( g_{k_1}^{\ast }u^{-ik_1 \eta _1} \right) \Im \left( g_{k_2}^{\ast }h_{k_3}^{\ast } v_{k_1}^{\ast }(\eta _1 )  \int ^{\eta _1}\frac{d\eta _2}{\eta _2}e^{-i(c_s k_1 +k_2 +c_s k_3 )\eta _2} \right) \\
  &\sim & 81 k_1^3 k_2^3 \int ^{\eta }\frac{d\eta _1}{\eta _1} \int ^{\eta _1}\frac{d\eta _2}{\eta _2 } \cos \left( k_1 (\eta  -\eta _1 ) \right) \cos \left( c_s k_1 (\eta _1 -\eta _2 ) + (k_2 + c_s k_3 ) (\eta - \eta _2 ) \right) .
  \end{eqnarray*}
Finally, (\ref{eq:xia2B}) yields
\begin{eqnarray*}
-\langle \left[ H^q (\eta _2 ) , \left[ H^C (\eta _1 ) , \xi _{k_1}\xi _{k_2}\alpha ^{\prime }_{k_3} \right] \right] &=& \frac{2H^3 \mathcal{I}}{\sqrt{\epsilon _{\varphi }}c_s^6 k_1^6 k_2^3 k_3^3 \eta ^6}\Im \left( g^{\ast }_{k_2} (\eta ) h_{k_3}^{\ast }(\eta )u_{k_2}(\eta _1 ) \frac{h_{k_3}(\eta _1 )}{\tau _1^4} \right) \\
&& \times \Im \left( g_{k_1}^{\ast }(\eta ) h_{k_1}^{\ast }(\eta _1 )\frac{h_{k_1}(\eta _2 )}{\eta _2^4} u_{k_1}(\eta _2 ) \right)  + ( 1\leftrightarrow 2 ) .
\end{eqnarray*}
The leading-order contribution is
\begin{eqnarray*}
\mathcal{B}_{2b} &=& \int ^{\eta }\frac{d\eta _1}{\eta _1^4} \Im \left( g_{k_2}^{\ast }h_{k_3}^{\ast } u_{k_2}(\eta _1 ) h_{k_3}(\eta _1 )\right) \Im \left( g_{k_1 }^{\ast } h_{k_1}^{\ast }(\eta _1 ) F_{k_1 }(\eta _1 ) \right) \\
& \sim & -k_1^3 k_2^3 \int ^{\eta }\frac{d\eta _1}{\eta _1 }\Im \left( g_{k_2}^{\ast } h_{k_3}^{\ast } e^{-ik_2 \eta _1}h_{k_3} (\eta _1 ) \right) \Im \left( i g_{k_1}^{\ast }v_{k_1}^{\ast }(\eta _1 ) \int ^{\eta _1}\frac{d\eta _2 }{\eta _2 } e^{-i(1+c_s )k_1 \eta _2 } \right) \\
&\sim & 81 k_1^3 k_2^3 \int ^{\eta }\frac{d\eta _1}{\eta _1} \int ^{\eta _1}\frac{d\eta _2 }{\eta _2 } \cos \left( (k_2 +c_s k_3 ) (\eta -\eta _1 ) \right) \cos \left( c_s k_1 (\eta _1 -\eta _2 ) + k_1 (\eta -\eta _2 ) \right) .
\end{eqnarray*}

\subsection{3-vertex contributions}
We saw that the 1-vertex terms that involve only single time integrals resulted in $\propto \ln (-\eta )$
while the leading contributions from 2-vertex terms come from double integrals and proportional to 
$\left( \ln (-\eta ) \right) ^2 $. Hence, one expects that 3-vertex contributions behave like 
$\left( \ln (-\eta ) \right) ^3$ and dominate the tree-level amplitude at the order $\mathcal{I}^2$.
This was also the result of \cite{Bartolo2012}.
We explicitly prove it and derive the coefficients in front. The principle of the calculations is 
the same as the previous sections although the algebra gets increasingly complicated. 
First of all, we write
\begin{eqnarray*}
&& -i \langle H^C (\eta _3 ) H^q (\eta _2 ) H^q (\eta _1 ) \xi (\mathbf{x})\xi (\mathbf{y} ) \xi (\mathbf{z}) \rangle = -i 3 \eta _3^4 \sqrt{\frac{2}{\epsilon _{\varphi }}} \frac{144\mathcal{I}^2}{H^2} \iiint d^3 w_1 ^3 w_2 d^3 w_3  \\
&& \times \langle \alpha ^{\prime } (\mathbf{w}_3 ) \alpha ^{\prime } (\mathbf{w}_2 ) \rangle \langle \alpha ^{\prime } (\mathbf{w}_3 ) \alpha ^{\prime } (\mathbf{w}_1 ) \rangle \left( \langle \pi (\mathbf{w}_3 ) \xi (\mathbf{z}) \rangle \langle \pi (\mathbf{w}_2 ) \xi (\mathbf{y} ) \rangle \langle \pi (\mathbf{w}_1 ) \xi (\mathbf{x}) \rangle  + {\rm 5 \ perms} \right) 
\end{eqnarray*}
and
\begin{eqnarray*}
\langle H^C (\eta _3 ) H^q (\eta _2 ) H^q (\eta _1 ) \xi _{k_1} \xi _{k_2} \xi _{k_3} \rangle &=& \frac{3H^4 \mathcal{I}^2 }{\sqrt{2\epsilon _{\varphi }}\eta ^3 \left( \eta _1 \eta _2 \eta _3 \right) ^4 } \frac{1}{c_s^6 k_1^6 k_2^6 k_3^3 } g_{k_1}^{\ast } (\eta ) g_{k_2}^{\ast } (\eta ) g_{k_3}^{\ast } (\eta ) \\
&& \times u_{k_1}(\eta _1 ) h^{\ast }_{k_1}(\eta _1 ) u_{k_2}(\eta _2 ) h^{\ast }_{k_2}(\eta _2 )h_{k_1}(\eta _3 ) h_{k_2}(\eta _3 ) u_{k_3}(\eta _3 ) + {\rm 5 \ perms} .
\end{eqnarray*}
In the end, our integrand is
\begin{eqnarray*}
&& -i \langle \left[ H^C (\eta _3 ) , \left[ H^q (\eta _2 ) , \left[ H^q (\eta _1 ) , \xi _k^3 \right] \right] \right] \rangle = - \sqrt{\frac{2}{\epsilon _{\varphi }}}\frac{12H^4 \mathcal{I}^2 }{\eta ^3 \left( \eta _1 \eta _2 \eta _3 \right) ^4 } \frac{1}{c_s^6 k_1^6 k_2^6 k_3^3 } \\
&& \times \Im \left( g_{k_1}^{\ast } u_{k_1}(\eta _1 )  \right) \Im \left( g_{k_2}^{\ast } u_{k_2}(\eta _2 ) \right) \Im \left( g_{k_3}^{\ast } h_{k_1}^{\ast } (\eta _1 ) h_{k_2}^{\ast }(\eta _2 ) h_{k_1}(\eta _3 ) h_{k_2 }(\eta _3 ) u_{k_3}(\eta _3 ) \right) + {\rm 5 \ perms} .
\end{eqnarray*}
Anticipating the cancellation of terms with negative powers of $\eta $, we seek the expected $\left( \ln (-\eta )\right) ^3 $ contribution. It can only come from
\begin{eqnarray*}
\mathcal{A}_{3a} &=& \int ^{\eta }\frac{d\eta _1}{\eta _1^4 } \Im \left( g_{k_1}^{\ast }u_{k_1}(\eta _1 )  \right)  \int ^{\eta _1}\frac{d\eta _2}{\eta _2^4} \Im \left( g_{k_2}^{\ast }u_{k_2}(\eta _2 ) \right) \Im \left( g_{k_3}^{\ast } h^{\ast }_{k_1}(\eta _1 ) h_{k_2}^{\ast }(\eta _2 ) H_{k_3}(\eta _2 ) \right) \\
&\sim & 3k_2^3 k_3^3 \int ^{\eta }\frac{d\eta _1}{\eta _1^4} \Im \left( g_{k_1}^{\ast }u_{k_1}(\eta _1 ) \right) \int ^{\eta _1} \frac{d\eta _2}{\eta _2 } \Im \left( g_{k_2}^{\ast } e^{-ik_2 \eta _2 } \right) \\
&& \times  \Im \left( g_{k_3}^{\ast } h_{k_1}^{\ast }(\eta _1 ) v_{k_2}^{\ast }(\eta _2 ) \int ^{\eta _2}\frac{d\eta _3}{\eta _3} e^{-i(c_s k_1 +c_s k_2 + k_3 )\eta _3} \right) 
\end{eqnarray*}
where we integrated by parts for $\eta _2 $. We perform another integration by parts with $\eta _1$ as follows:
\begin{eqnarray*}
&& \int ^{\eta }d\eta _1 \Im \left( g_{k_1}^{\ast }u_{k_1}(\eta _1 ) \right) \int ^{\eta _1}\frac{d\eta _2}{\eta _2} \Im \left( g_{k_2}^{\ast } e^{-ik_2 \eta _2 } \right) \Im \left( g_{k_3}^{\ast } \left( \frac{v_{k_1}^{\ast }(\eta _1 )}{\eta _1^3} \right) ^{\prime } v_{k_2}^{\ast }(\eta _2 ) \int ^{\eta _2}\frac{d\eta _3}{\eta _3} e^{-i(c_s k_1 +c_s k_2 + c_s k_3 )\eta _3} \right) \\
&& = \frac{1}{\eta ^3 } \Im \left( g^{\ast }_{k_1} u_{k_1} \right) \int ^{\eta }\frac{d\eta _2}{\eta _2} \Im \left( g_{k_2}^{\ast } e^{-ik_2 \eta _2} \right) \Im \left( g_{k_3}^{\ast } v_{k_1}^{\ast } (\eta ) v_{k_2}^{\ast }(\eta _2 ) \int ^{\eta _2 } \frac{d\eta _3}{\eta _3} e^{-i(c_s k_1 +c_s k_2 +k_3 )\eta _3} \right)  \\
&& - \int ^{\eta }\frac{d\eta _1}{\eta _1^2 } \Im \left( ik_1^2 g_{k_1}e^{-ik_1 \eta _1} \right) \int ^{\eta _1}\frac{d\eta _2}{\eta _2} \Im \left( g_{k_2}^{\ast } e^{-ik_2 \eta _2 } \right) \Im \left( g_{k_3}^{\ast } v_{k_1}^{\ast } (\eta _1 ) v_{k_2}^{\ast } (\eta _2 ) \int ^{\eta _2}\frac{d\eta _3}{\eta _3} e^{-i(c_s k_1 +c_s k_2 +k_3)\eta _3} \right) \\
&& - \int ^{\eta }\frac{d\eta _1}{\eta _1^4} \Im \left( g_{k_1}^{\ast } u_{k_1}(\eta _1 ) \right) \Im \left( g_{k_1}^{\ast } e^{-ik_2 \eta _1 } \right) \Im \left( g_{k_3}^{\ast } v_{k_1}^{\ast }(\eta _1 ) v_{k_2}^{\ast }(\eta _1 ) \int ^{\eta _1} \frac{d\eta _3}{\eta _3} e^{-i(c_s k_1 +c_s k_2 +k_3 )\eta _3 } \right)  .
\end{eqnarray*}
Only the second term can give rise to the sought dependence on $\eta $. We derive
\begin{eqnarray*}
\mathcal{A}_{3a} & \sim  & -3k_1^3 k_2^3 k_3^3 \int ^{\eta } \frac{d\eta _1}{\eta _1} \Im \left( g_{k_1}^{\ast }e^{-ik_1 \eta _1}\right)  \int ^{\eta _1} \frac{d\eta _2}{\eta _2} \Im \left( g_{k_2}^{\ast } e^{-ik_2 \eta _2} \right) \\
&& \times  \Im \left( g_{k_3}^{\ast } v_{k_1}^{\ast }(\eta _1 ) v_{k_2}^{\ast }(\eta _2 ) \int ^{\eta _2} \frac{d\eta _3}{\eta _3} e^{-i(c_s k_1 +c_s k_2 +k_3 )\eta _3} \right) \\
& \sim & -81 k_1^3 k_2^3 k_3^3 \int ^{\eta }\frac{d\eta _1}{\eta _1} \cos \left( k_1 (\eta -\eta _1) \right) \int ^{\eta _1}\frac{d\eta _2}{\eta _2} \cos \left( k_2 (\eta - \eta _2 ) \right) \\
&& \times  \int ^{\eta _2}\frac{d\eta _3}{\eta _3} \cos \left( c_s k_1 (\eta _1 -\eta _3 ) +c_s k_2 (\eta _2 -\eta _3 ) +k_3 (\eta - \eta _3 ) \right) .
\end{eqnarray*}

For (\ref{eq:xi3B}), we have
\begin{eqnarray*}
&& -i \langle H^q (\eta _3 ) H^C (\eta _2 ) H^q (\eta _1 ) \xi (\mathbf{x} ) \xi (\mathbf{y} ) \xi (\mathbf{z}) \rangle = -i3\eta _2^4 \sqrt{\frac{2}{\epsilon _{\varphi }}} \frac{144\mathcal{I}^2}{H^2} \iiint d^3 w_1 d^3 w_2 d^3 w_3 \\
&& \times \langle \alpha ^{\prime }(\mathbf{w}_3 ) \alpha ^{\prime } (\mathbf{w}_2 ) \rangle \langle \alpha ^{\prime } (\mathbf{w}_2 ) \alpha ^{\prime } (\mathbf{w}_1 ) \rangle \left( \langle \pi (\mathbf{w}_3 )\xi (\mathbf{z}) \rangle \langle \pi (\mathbf{w}_2 ) \xi (\mathbf{y} ) \rangle \langle \pi (\mathbf{w}_1 ) \xi (\mathbf{x}) \rangle + {\rm 5 \ perms} \right) ,
\end{eqnarray*}
and
\begin{eqnarray*}
&& -i \langle \left[ H^q (\eta _3 ) , \left[ H^C (\eta _2 ) , \left[ H^q (\eta _1 ) ,\xi _k^3 \right] \right] \right] = -\sqrt{\frac{2}{\epsilon _{\varphi }}}\frac{12H^4 \mathcal{I}^2}{\eta ^3 \left( \eta _1 \eta _2 \eta _3 \right) ^4}\frac{1}{c_s^6 k_1^6 k_2^3 k_3^6 } \\
&& \times \Im \left( g_{k_1}^{\ast } u_{k_1}(\eta _1 ) \right) \Im \left( g_{k_2}^{\ast } h_{k_1}^{\ast }(\eta _1 ) u_{k_2}(\eta _2 ) h_{k_1}(\eta _2 ) \right) \Im \left( g_{k_3}^{\ast } h_{k_3}^{\ast } (\eta _2 ) u_{k_3}(\eta _3 ) h_{k_3} (\eta _3 ) \right) + {\rm 5 \ perms} .
\end{eqnarray*}
The integration results in 
\begin{eqnarray*}
\mathcal{A}_{3b} &=& \int ^{\eta }\frac{d\eta _1}{\eta _1^4} \Im \left( g_{k_1}^{\ast }u_{k_1}(\eta _1 ) \right) \int ^{\eta _1}\frac{d\eta _2}{\eta _2^4} \Im \left( g_{k_2}^{\ast } h_{k_1}^{\ast }(\eta _1 ) u_{k_2}(\eta _2 ) h_{k_1}(\eta _2 ) \right)\Im \left( g_{k_3}^{\ast } h_{k_3}^{\ast }(\eta _2 ) F_{k_3}(\eta _2 ) \right) \\
&\sim & k_1^3 k_2^3 k_3^3 \int ^{\eta }\frac{d\eta _1}{\eta _1} \Im \left( g_{k_1}^{\ast } e^{-ik_1 \eta _1} \right) \int ^{\eta _1}\frac{d\eta _2}{\eta _2} \Im \left( g_{k_2}^{\ast } v_{k_1}^{\ast }(\eta _1 ) e^{-ik_2 \eta _2 } h_{k_1}(\eta _2 ) \right) \\
&& \times \Im \left( ig_{k_3}^{\ast } v_{k_3}^{\ast }(\eta _2 ) \int ^{\eta _2 } \frac{d\eta _3}{\eta _3} e^{-i(1+c_s )k_3 \eta _3} \right) \\
& \sim & -81 k_1^3 k_2^3 k_3^3 \int ^{\eta }\frac{d\eta _1}{\eta _1 } \cos \left( k_1 (\eta -\eta _1 ) \right) \int ^{\eta _1}\frac{d\eta _2}{\eta _2} \cos \left( c_s k_1 (\eta _1 -\eta _2 ) + k_2 (\eta - \eta _2 ) \right) \\
&& \times \int ^{\eta _2 }\frac{d\eta _3}{\eta _3} \cos \left( k_3 (\eta - \eta _3 ) +c_s k_3 (\eta _2 -\eta _3 ) \right) .
\end{eqnarray*}

For (\ref{eq:xi3C}), we have
\begin{eqnarray*}
&& -i \langle H^q (\eta _3 ) H^q (\eta _2 ) H^C (\eta _1 ) \xi (\mathbf{x}) \xi (\mathbf{y} ) \xi (\mathbf{z} ) \rangle  = -i3\eta _2^4 \sqrt{\frac{2}{\epsilon _{\varphi }}} \frac{144\mathcal{I}^2}{H^2} \iiint d^3 w_1 d^3 w_2 d^3 w_3 \\
&& \times \langle \alpha ^{\prime }(\mathbf{w}_3 ) \alpha ^{\prime } (\mathbf{w}_1 ) \rangle \langle \alpha ^{\prime } (\mathbf{w}_2 ) \alpha ^{\prime } (\mathbf{w}_1 ) \rangle \left( \langle \pi (\mathbf{w}_3 )\xi (\mathbf{z}) \rangle \langle \pi (\mathbf{w}_2 ) \xi (\mathbf{y} ) \rangle \langle \pi (\mathbf{w}_1 ) \xi (\mathbf{x}) \rangle + {\rm 5 \ perms} \right) ,
\end{eqnarray*}
and 
\begin{eqnarray*}
&& -i \langle \left[ H^q (\eta _3 ) , \left[ H^q (\eta _2 ) , \left[ H^C (\eta _1 ) ,\xi _k^3 \right] \right] \right] = -\sqrt{\frac{2}{\epsilon _{\varphi }}}\frac{12H^4 \mathcal{I}^2}{\eta ^3 \left( \eta _1 \eta _2 \eta _3 \right) ^4}\frac{1}{c_s^6 k_1^6 k_2^3 k_3^6 } \\
&& \times \Im \left( g_{k_1}^{\ast } u_{k_1}(\eta _1 ) \right) \Im \left( g_{k_2}^{\ast } h_{k_1}^{\ast }(\eta _1 ) u_{k_2}(\eta _2 ) h_{k_1}(\eta _2 ) \right) \Im \left( g_{k_3}^{\ast } h_{k_3}^{\ast } (\eta _1 ) u_{k_3}(\eta _3 ) h_{k_3} (\eta _3 ) \right) + {\rm 5 \ perms} .
\end{eqnarray*}
Similar to the other two, we obtain
\begin{eqnarray*}
\mathcal{A}_{3c} &=& \int ^{\eta }\frac{d\eta _1}{\eta _1^4} \Im \left( g_{k_1}^{\ast } u_{k_1}(\eta _1 ) \right) \int ^{\eta _1} \frac{d\eta _2 }{\eta _2^4 } \Im \left( g_{k_2}^{\ast } h_{k_1}^{\ast } (\eta _1 ) u_{k_2}(\eta _2 ) h_{k_1}(\eta _2 )\right) \Im \left( g_{k_3}^{\ast } h_{k_3}^{\ast } (\eta _1 ) F_{k_3}(\eta _2 ) \right) \\
&\sim & -k_2^3 k_3^3 \int ^{\eta }\frac{d\eta _1}{\eta _1^4} \Im \left( g_{k_1}^{\ast } u_{k_1}(\eta _1 ) \right) \int ^{\eta _1} \frac{d\eta _2}{\eta _2} \Im \left( g_{k_2}^{\ast } h_{k_1}^{\ast }(\eta _1 ) e^{-ik_2 \eta _2 }v_{k_1}(\eta _2 ) \right) \\
&& \times \Im \left( ig_{k_3}^{\ast }h_{k_3}^{\ast }(\eta _1 )  \int ^{\eta _2}\frac{d\eta _3}{\eta _3} e^{-i(1+c_s )k_3 \eta _3 } \right) \\
&\sim & k_2^3 k_3^3 \int ^{\eta }\frac{d\eta _1}{\eta _1^2 } \Im \left( ik_1^2 g_{k_1}^{\ast } e^{-ik_1 \eta _1} \right) \int ^{\eta _1}\frac{d\eta _2}{\eta _2} \Im \left( g_{k_2}^{\ast } v_{k_1}^{\ast }(\eta _1 ) e^{-ik_2 \eta _2}v_{k_1}(\eta _2 ) \right) \\
&& \times \Im \left( ig_{k_3}^{\ast }h_{k_3}^{\ast } (\eta _1 ) \int ^{\eta _2}\frac{d\eta _3}{\eta _3} e^{-i(1+c_s )k_3 \eta _3 } \right) \\
&\sim & k_1^3 k_2^3 k_3^3 \int ^{\eta }\frac{d\eta _1}{\eta _1} \Im \left( g_{k_1}^{\ast } e^{-ik_1 \eta _1} \right) \int ^{\eta _1} \frac{d\eta _2}{\eta _2} \Im \left( g_{k_2}^{\ast }v_{k_1}^{\ast }(\eta _1) e^{-ik_2 \eta _2 } v_{k_1}(\eta _2 ) \right) \\
&& \times \Im \left( i g_{k_2}^{\ast } h_{k_3}^{\ast } (\eta _1 ) \int ^{\eta _2} \frac{d\eta _3}{\eta _3 } e^{-i(1+c_s )k_3 \eta _3 } \right) \\
&\sim & -81 k_1^3 k_2^3 k_3^3 \int ^{\eta }\frac{d\eta _1}{\eta _1} \cos \left( k_1 (\eta -\eta _1 ) \right) \int ^{\eta _1}\frac{d\eta _2}{\eta _2} \cos \left( c_s k_1 (\eta _1 -\eta _2 ) +k_2 (\eta - \eta _2 ) \right) \\
&& \times \int ^{\eta _2}\frac{d\eta _3}{\eta _3}\cos \left( k_3 (\eta - \eta _3 ) +c_s k_3 (\eta _1 -\eta _3 ) \right) .
\end{eqnarray*}

\subsection{Second order curvature perturbation and summary}
So far, we have only discussed the linear part of the curvature perturbation since it is the only term
that picks up contributions from cubic vertices at tree level. The second- and higher order terms 
in $\zeta $ also contribute to the bispectrum, however, through the combinations such as 
$\langle \zeta _{(1)} \zeta _{(1)} \zeta _{(2)} \rangle$. Since it is impossible to examine at
all orders if they give any contribution within the order $\mathcal{I}^2 $, here we just look
at the second-order term and check that they do not become dominant over the 3-vertex
contributions derived in the previous subsection.

First, we note that ignoring higher order corrections in $\epsilon _{H,\varphi } , \eta _{H,\varphi }$ and the terms with spatial derivatives and using (\ref{eq:curvature1}), we can
rewrite $\Xi _{ij}$ as
\begin{equation*}
\Xi _{ij} = \left( 4\zeta ^2_{(1)} - \frac{2}{\bar{\rho }^{\prime }}\delta \rho _{(1)}^{\prime } \zeta _{(1)} \right) \delta _{ij} -\frac{2}{\mathcal{H}} \left( \zeta _{(1),i}B^{(1)}_{,j} + \zeta _{(1),j}B^{(1)}_{,i} \right) .
\end{equation*}
Comparing equation (\ref{eq:zeta1}) with
\begin{equation*}
\nabla ^2 B^{(1)} = -\sqrt{\frac{\epsilon _{\varphi }}{2}} \left[ \pi ^{\prime } + \frac{6\mathcal{I}}{\eta }\pi -\frac{3}{\sqrt{2}}H\eta ^3 \left( \alpha ^{\prime } -\frac{2}{\eta }\alpha \right) \right] ,
\end{equation*}
we see the second term is suppressed by a factor of $\epsilon _H$. Throwing it away, 
equation (\ref{eq:curvature2}) yields
\begin{equation*}
-\zeta _{(2)} = \frac{\mathcal{H}}{\bar{\rho }^{\prime }} \delta \rho _{(2)} +2\frac{\delta \rho _{(1)}^{\prime }}{\bar{\rho }^{\prime }}\zeta _{(1)}- 2\zeta _{(1)}^2 .
\end{equation*}
Expanding the energy-momentum tensor up to second order, we find
\begin{eqnarray*}
\delta \rho _{(2)} &=& -2T^0_{(2)0} +\frac{2}{\bar{\rho }+\bar{p}} T^0_{(1)i}T^i_{(1)0} \\
&=& \frac{1}{a^2}\pi ^{\prime 2} + \frac{1}{a^2}\pi _{,i}\pi _{,i} + \left( V_{,\varphi \varphi } + \frac{3c^2}{a^2 f^2}\frac{(f^2)_{,\varphi \varphi }}{f^2} \frac{1}{a^2} \right) \pi ^2 + \frac{1}{a^2}\left( \bar{\varphi }^{\prime 2} + \frac{3c^2}{a^2 f^2} \right) \left( 3\phi _{(1)}^2 -\phi _{(2)} \right) \\
&& -\frac{4\bar{\varphi }^{\prime }}{a^2}\phi _{(1)}\pi ^{\prime } -\frac{2c^2}{a^2 f^2} \frac{(f^2)_{,\varphi }}{f^2}\frac{1}{a^2} \phi _{(1)}\pi +\frac{6c}{af}\frac{(f^2)_{,\varphi }}{f^2} \frac{f}{a^3}\pi \alpha ^{\prime } - \frac{12c}{af} \frac{f}{a^3}\phi _{(1)}\alpha ^{\prime } +\frac{f^2}{a^4} \left( 3\alpha ^{\prime 2} +2\alpha _{,i}\alpha _{,i} \right) .
\end{eqnarray*}
The appearance of $\phi _{(2)}$ forces us to look into the constraint equations at
the second order. In fact, they are not too bad for the scalar perturbations in the flat gauge.
The relevant equation is obtained from variation of $N_i$ in the ADM formalism and reads
\begin{equation*}
-\frac{2\mathcal{H} \delta _{ij} + B_{,ij} -\nabla ^2 B \delta _{ij}}{ 1 +\phi } \phi _{,i} = -\varphi ^{\prime } \varphi _{,j} + \varphi _{,i}B_{,i} \varphi _{,j} + f^2 L^a_i F^{ai}_{\ \ j} .
\end{equation*}
Expanding it to the second order, we find
\begin{eqnarray}
2\mathcal{H}\phi _{(2),j} &=& \left( 2\mathcal{H} \phi _{(1)} + \nabla ^2 B^{(1)} \right) \phi _{(1),j} - B^{(1)}_{,ij} \phi _{(1),i}  \\
&& + \pi ^{\prime } \pi _{,j} + \frac{f^2}{a^2} \left( 2\alpha ^{\prime }\alpha _{,j} + \tau ^{\prime }_{,i} \tau _{,ij} + \tau ^{\prime }_{,j}\nabla ^2 \tau \right) . \nonumber
\end{eqnarray}
In the end, its contribution is subdominant. Keeping the leading-order terms in slow roll and discarding
higher spatial derivatives, we obtain
\begin{equation}
\zeta _{(2)} = \frac{\eta ^2}{6\epsilon _H}\pi ^{\prime 2} +\frac{1}{6\epsilon _H}\left( 24\mathcal{I}^2 \pi ^2 -12 H\mathcal{I}\eta ^4 \pi \alpha ^{\prime } + 3H^2 \eta ^8 \alpha ^{\prime 2} \right) +2\zeta _{(1)}^2 -2\eta \zeta _{(1)}^{\prime } \zeta _{(1)} .  \label{eq:secondorder}
\end{equation}
The contribution to the bispectrum is 
\begin{equation*}
\langle \zeta (\mathbf{x} ) \zeta (\mathbf{y} )\zeta (\mathbf{z}) \rangle \sim \langle \zeta _{(1)} ({\bf x}) \zeta _{(1)} ({\bf y}) \zeta _{(2)} ({\bf z}) \rangle + (2\ {\rm perms}) .
\end{equation*}
The first and the last terms in (\ref{eq:secondorder}) are subdominant. The term quadratic in $\zeta _{(1)}$
gives contributions such as
\begin{equation*}
\langle \zeta _{(1)}(\mathbf{x} )\zeta _{(1)}(\mathbf{z}) \rangle \langle \zeta _{(1)}(\mathbf{y}) \zeta _{(1)}(\mathbf{z}) \rangle ,
\end{equation*}
which exist regardless of the dynamics and give $|f_{NL} | \lesssim 1$. The rest are the generic
effects of the background gauge fields. Looking at (\ref{eq:zetag}), we see that the leading order
contributions in $\mathcal{I}$ are quadratic, which involve terms such as 
\begin{eqnarray*}
 \frac{\epsilon _{\varphi }}{\epsilon _H^3}\mathcal{I}^2 \left(  2\langle \pi (\mathbf{x} )\pi (\mathbf{z} ) \rangle \langle \pi (\mathbf{y}) \pi (\mathbf{z}) \rangle +\frac{1}{8}H^4 \eta ^{16} \langle \alpha ^{\prime }(\mathbf{x})\alpha ^{\prime }(\mathbf{z}) \rangle \langle \alpha ^{\prime }(\mathbf{y}) \alpha ^{\prime }(\mathbf{z}) \rangle \right)  
 \end{eqnarray*}
and
\begin{eqnarray*}
\frac{\epsilon _{\varphi }}{\sqrt{2}\epsilon _H^3}H^2 \mathcal{I}^2 \eta ^8 \left( \langle \pi (\mathbf{x})\pi (\mathbf{z}) \rangle \langle \alpha ^{\prime }(\mathbf{y}) \alpha ^{\prime }(\mathbf{z}) \rangle + \langle \pi (\mathbf{y})\pi (\mathbf{z}) \rangle \langle \alpha ^{\prime }(\mathbf{x}) \alpha ^{\prime }(\mathbf{z}) \rangle \right) .
\end{eqnarray*}
At the leading order, $\pi $ and $\alpha ^{\prime }$ are essentially just $u_k (\eta )$ and $h_k(\eta )/\eta ^4$, therefore their contribution will be constant of $|f_{NL}| \sim O(\mathcal{I}^2 )$.

\bibliographystyle{JHEP}
\bibliography{refVB}

\end{document}